\documentclass[11pt,english]{elsarticle}
\usepackage{mathptmx}
\usepackage[T1]{fontenc}
\usepackage[latin9]{inputenc}
\usepackage[letterpaper]{geometry}
\usepackage[active]{srcltx}
\usepackage{color}
\usepackage{float}
\usepackage{bm}
\usepackage{amsmath}
\usepackage{amssymb}
\usepackage{graphicx}

\makeatletter

\newcommand{\lyxdot}{.}

\floatstyle{ruled}
\newfloat{algorithm}{tbp}{loa}
\providecommand{\algorithmname}{Algorithm}
\floatname{algorithm}{\protect\algorithmname}

\AtBeginDocument{}

\DeclareMathAlphabet{\mathsfsl}{OT1}{cmss}{m}{sl}
\newcommand{\tensor}[1]{\stackrel{\leftrightarrow}{\mathsfsl{#1}}}
\renewcommand{\vec}[1]{\mathbf{#1}}

\newcommand{\ftilde}{{\widetilde f}}
\usepackage[english]{babel}
\addto\captionsenglish{}

\begin{document}

\title{An Eulerian Vlasov-Fokker-Planck Algorithm for Spherical Implosion
Simulations of Inertial Confinement Fusion Capsules}

\author[lanl]{W. T. Taitano\corref{cor1}}

\ead{taitano@lanl.gov}

\author[lanl2]{B. D. Keenan}

\author[lanl]{L. Chac{\'o}n}

\author[lanl]{S. E. Anderson}

\author[lanl4]{H. R. Hammer}

\author[lanl3]{A. N. Simakov}

\cortext[cor1]{Corresponding author}

\address[lanl]{T-5 Applied Mathematics and Plasma Physics Group, Theoretical Division
Los Alamos National Laboratory, Los Alamos, NM 87545}

\address[lanl2]{XCP-6 Plasma Theory and Applications Group, Computational Physics
Division, Los Alamos National Laboratory, Los Alamos, NM 87545}

\address[lanl3]{Theoretical Design Division, Los Alamos National Laboratory, Los
Alamos, NM 87545}

\address[lanl4]{T-3 Fluid Dynamics and Solid Mechanics Group, Theoretical Division
Los Alamos National Laboratory, Los Alamos, NM 87545}
\begin{abstract}
We present a numerical algorithm that enables a phase-space adaptive
Eulerian Vlasov-Fokker-Planck (VFP) simulation of inertial confinement
fusion (ICF) capsule implosions. The approach relies on extending
a recent mass, momentum, and energy conserving phase-space moving-mesh
adaptivity strategy to spherical geometry. In configuration space,
we employ a mesh motion partial differential equation (MMPDE) strategy
while, in velocity space, the mesh is expanded/contracted and shifted
with the plasma's evolving temperature and drift velocity. The mesh
motion is dealt with by transforming the underlying VFP equations
into a computational (logical) coordinate, with the resulting inertial
terms carefully discretized to ensure conservation. To deal with the
spatial and temporally varying dynamics in a spherically imploding
system, we have developed a novel nonlinear stabilization strategy
for MMPDE in the configuration space. The strategy relies on a nonlinear
optimization procedure that optimizes between mesh quality and the
volumetric rate change of the mesh to ensure both accuracy and stability
of the solution. Implosions of ICF capsules are driven by several
boundary conditions: 1) an elastic moving wall boundary; 2) a time-dependent
Maxwellian Dirichlet boundary; and 3) a pressure-driven Lagrangian
boundary. Of these, the pressure-driven Lagrangian boundary driver
is new to our knowledge. The implementation of our strategy is verified
through a set of test problems, including the Guderley and Van-Dyke
implosion problems --the first-ever reported using a Vlasov-Fokker-Planck
model. 
\end{abstract}

\maketitle

\section{Introduction\label{sec:introduction}}

In many high-energy-density (HED) laboratory experiments --including
inertial confinement fusion (ICF) ones-- experimental design employs
single-fluid radiation hydrodynamic models (rad-hydro for short),
where the constituent ions are modeled as a single-average fluid with
an additional electron energy equation (for a comprehensive description
overview, see Refs. \citep{zeldovich_raizer_rad_hydro_textbook,paul_drake_higher_energy_density_physics,atzenni_icf_textbook}).
Contrary to the rad-hydro predictions, however, the National Ignition
Facility (NIF) has not achieved ignition of ICF targets to date. Various
authors have proposed the importance and observations of kinetic effects
\citep{petschek_nuclear_fusion_1979_high_energy_ion_loss,rygg_pop_2006,welser-sherrill-hedp_2009_icf_mix_using_exp_data_and_theo_model,hermann_pop_2009,amendt_pop_2011_baro_diffusion,thomas_JCP_2012_review_of_vfp_in_icf,ma_prl_2013_mix,bellei_pop_2013_species_separation_in_icf,baumgaertel_pop_2014_ti_tracer_omega,rinderknecht_2014_prl_first_mix_nonhydro,rosenberg_prl_2014_hydro_kinetic_transition,kagan_pra_2014_thermo_diffusion_icf,rinderknecht_2015_prl_ion_thermal_decoupling,yin_pop_2016_interfacial_mix,simakov_pop_2016_multi_ion_braginskii,Le_pop_2016_simulation_of_direct_drive_capsule_implosion,sio_rsi_2016_pxtd,keenan_pop_2017_ion_species_stratification_wh_strong_shocks,keenan_pre_2017_deciphering_kinetic_structure_of_plasma_shocks,simakov_pop_2017_plasma_ion_stratification_by_planar_shocks,joshi_2017_species_separation_via_xray,vold_pop_2018_diffusion_fluid_dynamics_in_plasmas,rinderknecht_ppcf_2018_present_understanding_and_future_directions,sio_prl_2019_use_of_pxtd,sadler_pre_2019_kin_fusion_ignition_wh-hotspot_ablator_mix,vold_pop_2019_multispecies_plasma_transport_in_1d_icf_implosion,tikhonchuk_nuclear_fusion_2019_lpi_icp_review,young_2019,haines_pop_2020}
and their potential role in altering the rad-hydro predictions. Some
of these effects include, but are not limited to: 1) plasma ion stratification
due to differential motion of multiple ions; 2) long-mean-free path
modification to transport physics (e.g., heat flux and viscosity);
and 3) laser-plasma interactions.

The Vlasov-Fokker-Planck (VFP) kinetic equation, coupled with the
Maxwell equations is regarded as the first-principles model for describing
the dynamic evolution of weakly collisional (i.e., classical) plasma
particle distribution function in 6-dimensional phase-space (3-dimensional
in configuration and 3-dimensional in velocity). Consequently, there
is wide applicability of the VFP-Maxwell equation set to laboratory
(e.g., fusion experiments, space propulsion) and astrophysical systems.
However, VFP simulations of ICF capsules are challenged by the disparate
length and velocity scales supported, which must be efficiently and
accurately dealt with. In configuration space, there exist over 9
orders of magnitude of separation between the Debye length ($\sim10^{-12}$
m), and capsule size ($10^{-3}$ m). In velocity space, there exists
more than 3 orders of magnitude separation between the thermal speed,
$v_{th}=\sqrt{2T/m}$ (where $T$ is the temperature, and $m$ is
the particle mass) of the initially preheated fuel, the compressed
heated fuel, cold shell, electrons, and fusion-produced $\alpha$-particles.
Even if one considers fluid electrons (thus analytically removing
kinetic electron length and velocity scales), more than 6 orders of
magnitude of separation remains in space between the ion mean free
path and the capsule size.

As far as we are aware, in the literature there are only three codes
capable of performing Eulerian and Semi-Lagrangian VFP implosion simulation
of ICF capsules: 1) The FPion and FUSE code by the Commissariat {à}
l'{è}nergie atomique et aux {è}nergies alternatives (CEA) group
\citep{casanova_1991_prl_kin_sim_of_col_shock_plasma,larroche_1993_kin_sim_of_plasma_collision_exp,Vidal_PoP_1993_ion_kin_sim_of_plan_col_shock,vidal_pre_1995_spherical_ion_kinetic_simulations_of_DT_imposions,Vidal_PoP_1995_nonlocal_electron_heat_flow,larroche_EPJ_2003_icf_fuel_ion_implosion_sim,larroche_2003_kin_sim_fuel_ion_transp_in_icf,larroche_2007_lsse_jcp,larroche_pop_2012_Dhe_3_icf_sim,peigney_JCP_2014_fp_kinetic_modeling_of_alpha,inglebert_epl_2014_species_separation_kinetic_effect_neutron_diagnostics,larroche_pre_2018_kinetic_exploding_pusher},
which includes $\alpha$-particle transport and burn physics capabilities;
2) the HIMICO code by the Institute for Laser Energetics (ILE) group
\citep{nishiguchi_pop_1996_himico_code,honda_pop_1996_himico_code,mima_aip_1996_himico_code,honda_pop_1999_nonlocal_rayleigh_taylor},
which in addition to burn physics, includes the treatment of multi-group
radiation transport; and 3) the iFP code at the Los Alamos National
Laboratory (LANL) \citep{Taitano_2015_rfp_0d2v_implicit,yin_pop_2016_interfacial_mix,Taitano_2016_rfp_0d2v_implicit,keenan_pre_2017_deciphering_kinetic_structure_of_plasma_shocks,keenan_pop_2017_ion_species_stratification_wh_strong_shocks,Taitano_JCP_2017_FP_equil_disc,taitano_pop_2018,Taitano_2018_vrfp_1d2v_implicit,vold_pop_2018_diffusion_fluid_dynamics_in_plasmas,keenan_pop_2020_revolver_physics,keenan_pop_2020_shock_driven_high_z_mix,taitano_cpc_2020_1d2v_cartesian_phase_space,anderson_jcp_2020_vlasov_ampere_adaptive}.
The approach taken by the CEA group relies on a 1D-2V semi-Lagrangian
formalism with a periodic remapping of solution in the phase-space
to track heating/cooling of plasmas and implosion of the capsule,
while that by the ILE group relies, effectively, on a spherically
symmetric (1D), multi-energy group P1 approximation, where the energy
space is discretized in a finite difference fashion while a first
order Legendre polynomial expansion is used in the pitch-angle coordinate.
The CEA approach can handle arbitrary distribution functions in phase-space
and can thus investigate capsules that are effectively collisionless
such as exploding pushers. However, the formalism (which relies on
remapping) is not conservative. On the other hand, the ILE approach
relies on an Eulerian discretization --on a static energy grid--
hence conservation symmetries related to the Vlasov spatial advection
operator are trivially satisfied. However, their approach assumes
linear deviations from isotropic plasmas, and is thus limited in the
distributions that it can describe. Further, the ILE approach relies
on a Lagrangian formalism for the configuration space grid, where
the grid is evolved according to the single-fluid hydrodynamic velocity,
and is thus neither able to selectively adapt the grid near regions
where strong kinetic features may evolve (i.e., the shock front),
nor handle multiple species.

In this study, we present a multiscale 1D2V phase-space moving-grid
kinetic-ion/fluid-electron hybrid algorithm, used in iFP to simulate
spherical implosions of ICF capsules. The algorithm addresses the
limitations of the previous approaches by FPION/FUSE and HIMICO by
allowing the description of arbitrary distribution functions for an
arbitrary number of plasma species while enforcing conservation of
mass and energy (note that, global linear momentum conservation is
enforced automatically by spherical symmetry), and moving the grid
to resolve evolving sharp gradient structures. In the proposed algorithm,
we extend to spherical geometry a recent moving phase-space grid strategy
developed for Cartesian geometry \citep{taitano_cpc_2020_1d2v_cartesian_phase_space}.
The hybrid equations are formulated in Cartesian logical coordinates
in phase-space, resulting in several additional inertial terms --representing
mesh motion-- that are discretized so as to enforce the underlying
continuum conservation symmetries discretely. In velocity space, the
grid is normalized and shifted based on the local and instantaneous
plasma thermal speed and drift velocity, while in configuration space
a mesh-motion partial differential equation (MMPDE) formalism \citep{huang_1994_jcp_mmpde_ep,huang_1994_siam_J_numer_anal_mmpdes_ep,budd_huang_russell_2009}
is used to track evolving macroscopic moment structures. The MMPDE
strategy employed in our previous study, Ref. \citep{taitano_cpc_2020_1d2v_cartesian_phase_space},
defines a fixed grid relaxation time scale to provide a temporal smoothing
mechanism for grid evolution. However, in realistic problems where
different parts of the domain evolve at different rates, a constant
relaxation time scale often leads to disparate cell volume evolution
rate that can cause numerical brittleness and accuracy degradation
\citep{li_1998_siam_jci_comp_stability_of_mmpde}. To address these
limitations, we develop the nonlinearly stabilized (NS)-MMPDE algorithm,
which nonlinearly optimizes the MMPDE grid with the local rate of
change of volume. The algorithm's advertised capabilities are demonstrated
with known Guderley and Van-Dyke asymptotic hydrodynamic solutions
as well as published implosion results. For reproducibility purposes,
we also provide details on the numerical implementations of several
implosion driving boundary conditions, namely: 1) elastic moving wall;
2) time-dependent Maxwellian Dirichlet; and 3) pressure-driven Lagrangian
boundaries.

The rest of paper is organized as follows. Section \ref{sec:vfp_and_fluid_electrons}
introduces hybrid ion-VFP and fluid-electron equations in a 1D2V spherical
geometry. In Sec. \ref{sec:transformed_equations_in_spherical_geometry_with_moving_Grid},
we introduce the transformed VFP ion and the fluid-electron equations
in terms of the logical phase-space coordinates. In Sec. \ref{sec:NS-MMPDE},
we discuss our NS-MMPDE algorithm. In Sec. \ref{sec:numerical_implementations},
we discuss in detail the numerical implementation of the proposed
scheme in the following order: 1) the conservative finite differencing
discretization of the ion VFP equation; 2) the discretization of the
fluid-electron energy equation; 3) the discretization of the NS-MMPDE
algorithm ; 4) the various implosion driving boundary conditions;
5) our adaptive time-stepping strategy; and 6) a brief description
of the solver and overall algorithm combining the various components.
The numerical performances and properties of the new algorithm are
demonstrated with various implosion tests of varying degrees of complexity
in Sec. \ref{sec:numerical_results}. Finally, we conclude in Sec.
\ref{sec:conclusions_and_future_work}.

\section{Hybrid Vlasov-Fokker-Planck Ion and Fluid Electron Formulation\label{sec:vfp_and_fluid_electrons}}

In this study, we employ an electrostatic hybrid formulation with
a Vlasov-Fokker-Planck kinetic description for the ions and a fluid
description for electrons. The Vlasov-Fokker-Planck (VFP) equation
describes the particle distribution function for ion species $\alpha$,
$f_{\alpha}=f_{\alpha}\left(\vec{x},\vec{v},t\right)$, interacting
with the electric field, $\vec{E}$, and colliding with $\beta$ species
(which includes both ions and electrons), given as:

\begin{equation}
\partial_{t}f_{\alpha}+\vec{v}\cdot\nabla_{x}f_{\alpha}+\frac{q_{\alpha}}{m_{\alpha}}\vec{E}\cdot\nabla_{v}f_{\alpha}=\sum_{\beta}^{N_{s}+1}C_{\alpha\beta}\left(f_{\beta},f_{\alpha}\right).\label{eq:vfp_3d3v_continuum}
\end{equation}
Here, $m_{\alpha}$ is the mass, $q_{\alpha}$ is the charge, $\vec{v}$
is the particle velocity, $\nabla_{x}=\left\{ \partial_{x},\partial_{y},\partial_{z}\right\} $
$\left(\nabla_{v}=\left\{ \partial_{v_{x}},\partial_{v_{y}},\partial_{v_{z}}\right\} \right)$
is the configuration- (velocity-) space gradient operator, $C_{\alpha\beta}$
is the bilinear Fokker-Planck collision operator of species $\alpha$
colliding with species $\beta$, 
\begin{equation}
C_{\alpha\beta}=\Gamma_{\alpha\beta}\nabla_{v}\cdot\left[\tensor D_{\beta}\cdot\nabla_{v}f_{\alpha}-\vec{A}_{\beta}f_{\alpha}\right],\label{eq:fokker_planck}
\end{equation}
$N_{s}$ is the total number of ion species, $\Gamma_{\alpha\beta}=\frac{q_{\alpha}^{2}q_{\beta}^{2}\Lambda_{\alpha\beta}}{8\pi\epsilon_{0}^{2}m_{\alpha}^{2}}$,
$\Lambda_{\alpha\beta}$ is the Coulomb logarithm, $\epsilon_{0}$
is the permittivity of vacuum, $\tensor D_{\beta}$ and $\vec{A}_{\beta}$
are the collisional velocity space tensor diffusion and friction coefficients
of species $\beta$, and the detailed numerical treatments are discussed
in Refs. \citep{taitano-jcp-15-vfp,Taitano_2016_rfp_0d2v_implicit,Taitano_2018_vrfp_1d2v_implicit}.
Electrons are described by quasi-neutrality, 
\begin{equation}
n_{e}=-\frac{\sum_{\alpha}^{N_{s}}q_{\alpha}n_{\alpha}}{q_{e}},\label{eq:electron_number_density}
\end{equation}
ambipolarity, 
\begin{equation}
\vec{u}_{e}=-\frac{\sum_{\alpha}^{N_{s}}q_{\alpha}n_{\alpha}\vec{u}_{\alpha}}{q_{e}n_{e}},\label{eq:electron_drift_velocity}
\end{equation}
in conjunction with a temperature equation: 
\begin{equation}
\frac{3}{2}\partial_{t}\left(n_{e}T_{e}\right)+\frac{5}{2}\nabla_{x}\cdot\left(\vec{u}_{e}n_{e}T_{e}\right)+\nabla_{x}\cdot\vec{Q}_{e}-q_{e}n_{e}\vec{u}_{e}\cdot\vec{E}=\sum_{\alpha}^{N_{s}}W_{e\alpha}.\label{eq:fluid_electron}
\end{equation}
Here, $n_{e}$ ($n_{\alpha}$) is the electron ($\alpha$-ion species)
number density, $\vec{u}_{e}$ $(\vec{u}_{\alpha})$ is the electron
($\alpha$-ion species) drift velocity, $T_{e}$ is the electron temperature,
$\vec{Q}_{e}$ is the electron heat flux, $W_{e\alpha}$ is the electron-ion
energy exchange, and $\vec{E}$ is computed from the generalized Ohm's
law which includes the friction contributions with ions. For completeness,
we provide the expressions for $\vec{Q}_{e}$, $W_{e\alpha}$, and
$\vec{E}$ in (\ref{app:fluid_electron_model}). For detailed descriptions,
we refer the readers to Ref. \citep{simakov_PoP_2014_e_transp_wh_multi_ion},
and to Ref. \citep{Taitano_2018_vrfp_1d2v_implicit} for numerical
implementations in the context of our hybrid VFP ions/fluid electrons.

For spherical implosion problems, we adopt a spherically symmetric
radial coordinate system in the configuration space and an azimuthally
symmetric cylindrical velocity coordinate system. The transformed
ion VFP equation reads:

\begin{eqnarray}
\partial_{t}\left(J_{S}f_{\alpha}\right)+\partial_{r}\left(J_{S}v_{||}f_{\alpha}\right)+\frac{q_{\alpha}}{m_{\alpha}}E_{||}\partial_{v_{||}}\left(J_{S}f_{\alpha}\right)+r\partial_{\vec{v}}\cdot\left(\vec{\widetilde{a}}f_{\alpha}\right)=J_{S}\left[\sum_{\beta}^{N_{s}}C\left(f_{\beta},f_{\alpha}\right)+C_{\alpha e}\left(n_{e},u_{||,e}T_{e},f_{\alpha}\right)\right]\label{eq:1d2v_vfp_spherical_geometry}
\end{eqnarray}
and the fluid electron equation reads: 
\begin{equation}
\frac{3}{2}\partial_{t}\left(J_{S}n_{e}T_{e}\right)+\frac{5}{2}\partial_{r}\left(J_{S}u_{||,e}n_{e}T_{e}\right)+\partial_{r}\left(J_{S}Q_{||,e}\right)-J_{S}q_{e}n_{e}u_{||,e}E_{||}=J_{S}\sum_{\alpha}^{N_{s}}W_{e\alpha}.\label{eq:1d_fluid_electron_spherical_geometry}
\end{equation}
Here, $J_{S}=r^{2}$ is the Jacobian of transformation going from
Cartesian to a spherical coordinate system; $v_{||}$ and $v_{\perp}$
are the parallel (to the radial coordinate) and perpendicular particle
velocity; $r$ is the radial coordinate; $u_{||,e}$ is the parallel
electron bulk velocity; $T_{e}\leftarrow k_{B}T_{e}$ is the electron
temperature in units of energy and $k_{B}$ is the Boltzmann constant;
$Q_{||,e}$ is the parallel electron heat flux; $E_{||}$ is the parallel
electric field; and $C_{\alpha e}$ is the reduced ion-electron collision
operator (for details, refer to Ref. \citep{Taitano_2018_vrfp_1d2v_implicit}).
The fourth term on the left-hand-side of Eq. (\ref{eq:1d2v_vfp_spherical_geometry})
is the inertial term that arises from the coordinate transformation,
with $\widetilde{\vec{a}}=\left[v_{\perp}^{2},-v_{||}v_{\perp}\right]^{T}$
the associated pseudo-acceleration vector and $\partial_{\vec{v}}\cdot\vec{A}=\partial_{v_{||}}A_{||}+v_{\perp}^{-1}\partial_{v_{\perp}}\left(v_{\perp}A_{\perp}\right)$
the velocity-space divergence operator on a vector $\vec{A}=\left[A_{||},A_{\perp}\right]^{T}$.
The details of the transformation are provided in \ref{app:CC_2_SC_transformation}.

\section{Transformation of the Governing Equations with Moving Phase-Space
Grid\label{sec:transformed_equations_in_spherical_geometry_with_moving_Grid}}

To accommodate the disparate spatio-temporal variation in the thermal
speed and drift velocity encountered during the course of ICF capsule
implosions, we adopt a moving phase-space grid strategy similar to
that proposed in Refs. \citep{taitano_cpc_2020_1d2v_cartesian_phase_space}.
For this, we transform the governing equations on a static-computational
grid, $\left(\xi,\widetilde{v}_{||},\widetilde{v}_{\perp}\right)$,
where $\xi\in[0,1]$ is the logical coordinate in configuration space
with a mapping to the physical coordinate given as

\begin{equation}
r\left(\xi,t\right)=r_{min}+\int_{0}^{\xi}J_{\xi}\left(\xi',t\right)d\xi',\label{eq:logial_2_physical_configuration_transformation}
\end{equation}
and $\widetilde{v}_{||}$ and $\widetilde{v}_{\perp}$ are the transformed
velocity-space coordinates related to the physical coordinates \citep{Filbet_JCP_2013_rescaling_velocity_method,Taitano_2016_rfp_0d2v_implicit,Chertock_KRM_2018_AP_moving_grid,Taitano_2018_vrfp_1d2v_implicit,taitano_cpc_2020_1d2v_cartesian_phase_space}
as: 
\begin{equation}
v_{||}=v_{\alpha}^{*}\widetilde{v}_{||}+u_{||,\alpha}^{*}\label{eq:para_logical_2_physical}
\end{equation}
and 
\begin{equation}
v_{\perp}=v_{\alpha}^{*}\widetilde{v}_{\perp}.\label{eq:perp_logical_2_physical}
\end{equation}
Here, $r_{min}$ is the inner-radial domain in the configuration space
($r_{min}=0$ for a sphere), $J_{\xi}=\partial_{\xi}r$ is the Jacobian
of transformation from physical to logical coordinates in the configuration
space, $v_{\alpha}^{*}$ is the normalization speed and $u_{||,\alpha}^{*}$
is the transformation velocity. Accordingly, the VFP equation in the
transformed computational-coordinate system is given as (details shown
in \ref{app:physical_2_logical_transformation}):

\begin{eqnarray}
\partial_{t}\left(J_{S\xi}\widetilde{f}_{\alpha}\right)+\partial_{\xi}\left[\left(J_{S}v_{||}-\dot{J}_{r^{3}}\right)\widetilde{f}_{\alpha}\right]-v_{\alpha}^{*^{-1}}\partial_{\vec{\widetilde{v}}}\cdot\left\{ \left[J_{S\xi}\partial_{t}\vec{v}+\left(J_{r^{2}}v_{||}-\dot{J}_{r^{3}}\right)\partial_{\xi}\vec{v}\right]\widetilde{f}_{\alpha}\right\} \nonumber \\
+rv_{\alpha}^{*^{-1}}J_{\xi}\partial_{\vec{\widetilde{v}}}\cdot\left(\widetilde{\vec{a}}\widetilde{f}_{\alpha}\right)-\frac{q_{\alpha}}{m_{\alpha}}E_{||}J_{S\xi}v_{\alpha}^{*^{-1}}\partial_{\widetilde{v}_{||}}\widetilde{f}_{\alpha}=J_{S\xi}\left[\sum_{\beta}^{N_{s}}\widetilde{C}_{\alpha\beta}+\widetilde{C}_{\alpha e}\right].\label{eq:composite_transformed_vfp}
\end{eqnarray}
Here, $\widetilde{f}_{\alpha}=\left(v_{\alpha}^{*}\right)^{3}f_{\alpha}$
is the normalized distribution function, $J_{S\xi}=J_{S}J_{\xi}$
is the composite Jacobian, $\dot{J}_{r^{3}}=\frac{1}{3}\partial_{t}r^{3}$
is the local volumetric rate of change in configuration space; $\partial_{\vec{\widetilde{v}}}\cdot\vec{A}=\partial_{\widetilde{v}_{||}}A_{||}+\widetilde{v}_{\perp}^{-1}\partial_{\widetilde{v}_{\perp}}\left(\widetilde{v}_{\perp}A_{\perp}\right)$
is the velocity-space divergence operator on a vector, $\vec{A}=\left[A_{||},A_{\perp}\right]^{T}$,
in the transformed coordinate system; and $\vec{v}=v_{\alpha}^{*}\vec{\widetilde{v}}+u_{||,\alpha}^{*}\vec{e}_{||}$,
where $\vec{e}_{||}=[1,0]^{T}$ is the unit vector along the parallel
velocity. For convenience, we define commonly used velocity moments
of the normalized ion distribution function such as the number density,
\begin{equation}
n_{\alpha}=\left\langle 1,\widetilde{f}_{\alpha}\right\rangle _{\vec{v}},
\end{equation}
the drift velocity 
\begin{equation}
\vec{u}_{\alpha}=u_{||,\alpha}\vec{e}_{||}=\frac{\left\langle v_{||},\widetilde{f}_{\alpha}\right\rangle _{\vec{v}}}{\left\langle 1,\widetilde{f}_{\alpha}\right\rangle _{\vec{v}}}\vec{e}_{||},
\end{equation}
and the temperature, 
\begin{equation}
T_{\alpha}=\frac{m_{\alpha}}{3}\frac{\left\langle \left(\vec{v}-\vec{u}_{\alpha}\right)^{2},\widetilde{f}_{\alpha}\right\rangle _{\vec{v}}}{\left\langle 1,\widetilde{f}_{\alpha}\right\rangle _{\vec{v}}},
\end{equation}
where $\left\langle A\left(\vec{v}\right),B\left(\vec{v}\right)\right\rangle _{\vec{v}}=2\pi\int_{0}^{\infty}d\widetilde{v}_{\perp}\int_{-\infty}^{\infty}\widetilde{v}_{\perp}A\left(\vec{\widetilde{v}}\right)B\left(\vec{\widetilde{v}}\right)d\widetilde{v}_{||}$
denotes the transformed velocity-space inner product operation between
$A$ and $B$, and $\vec{e}_{||}=[1,0]$ is the unit vector along
the radial axis. Similarly, the fluid electron equation is transformed
as: 
\begin{equation}
\frac{3}{2}\partial_{t}\left(J_{S\xi}n_{e}T_{e}\right)+\partial_{\xi}\left[\left(\frac{5}{2}J_{S}u_{||,e}-\frac{3}{2}\dot{J}_{r^{3}}\right)n_{e}T_{e}\right]+\partial_{\xi}\left(J_{S}Q_{||,e}\right)-J_{S\xi}q_{e}n_{e}u_{||,e}E_{||}=J_{S\xi}\sum_{\alpha}^{N_{s}}W_{e\alpha}.\label{eq:composite_transformed_fluid_electron}
\end{equation}

\section{\textcolor{black}{Nonlinearly-Stabilized Moving Mesh Partial Differential
Equation (NS-MMPDE)\label{sec:NS-MMPDE}}}

We evolve the configuration-space grid by a predictor-corrector strategy.
In the predictor stage, we numerically solve the MMPDE \citep{huang_1994_jcp_mmpde_ep,huang_1994_siam_J_numer_anal_mmpdes_ep}:

\begin{equation}
\partial_{t}r^{*}=\tau_{r}^{-1}\partial_{\xi}\left[\omega\partial_{\xi}r^{*}\right],\label{eq:mmpde_original}
\end{equation}
\textcolor{black}{where $r^{*}$ is the }\textcolor{black}{\emph{predictor}}\textcolor{black}{{}
grid location (to be discussed shortly) in configuration space with
boundary conditions, $\left.r^{*}\right|_{\xi=0}=r_{min}=0$ and $\left.r^{*}\right|_{\xi=1}=R$},
and $\tau_{r}$ is the grid-equilibration time-scale. The grid evolves
according to a given monitor function, $\omega=\omega\left(\xi,t\right)$\textcolor{black}{.
}\textcolor{black}{The monitor function depends on inverse gradient-length
scales of number density, $n$, drift velocity, $u_{||}$, thermal
speed, $v_{th}$, and electron temperature, $T_{e}$, as:}

\textcolor{black}{
\begin{equation}
\omega^{*}=\sqrt{\frac{1}{2}\sum_{\alpha}^{N_{s}}\left[\left(L_{n,\alpha}^{-1}\right)^{2}+\left(L_{v_{th},\alpha}^{-1}\right)^{2}+\left(L_{u_{||},\alpha}^{-1}\right)^{2}\right]+\frac{1}{2}\left(L_{T_{e}}^{-1}\right)^{2}},\label{eq:inverse_grad_scale_length_monitor_function}
\end{equation}
where the ($*$) superscript denotes the pre-smoothed (to be discussed
shortly) monitor function. The inverse-gradient-length scales, $L^{-1}$,
are computed as:}

\textcolor{black}{
\begin{eqnarray}
L_{n,\alpha}^{-1}=\left|\delta_{\xi}\ln n_{\alpha}\right|+\delta_{min},\;\;\;\;L_{v_{th},\alpha}^{-1}=\left|\delta_{\xi}\ln v_{th,\alpha}\right|+\delta_{min},\label{eq:inverse_gradient_length_scale}\\
L_{u_{||},\alpha}^{-1}=\left|\frac{\delta_{\xi}u_{||,\alpha}}{v_{th,\alpha}}\right|+\delta_{min},\;\;\;\;L_{T_{e}}^{-1}=\left|\delta_{\xi}\ln T_{e}\right|+\delta_{min},\nonumber 
\end{eqnarray}
where $\delta_{min}$ is the user-specified floor for the relative
variation ($\delta_{min}=0.01$ unless otherwise specified).}\textcolor{black}{{}
}\textcolor{black}{To avoid an arbitrarily fine mesh, we limit the
minimum-to-maximum ratio of the monitor function, $\omega_{min}^{*}/\omega_{max}^{*}$,
to a cutoff value, $\eta$, by modifying the monitor function by a
constant, $a$,}

\textcolor{black}{
\begin{equation}
\omega^{*}:=\omega^{*}+a,\label{eq:flooring_of_monitor_function}
\end{equation}
where $a=\left(\eta\omega_{max}^{*}-\omega_{min}^{*}\right)/\left(1-\eta\right)$.
Note that this limiting is performed only when $\omega_{min}^{*}/\omega_{max}^{*}\le\eta$.
Unless otherwise stated, $\eta=10^{-2}$ is used in this study. Further,
during singular events (e.g., shock collapse), to avoid the grid from
being unnecessarily concentrated near $r=0$ --where the volume vanishes--
we modify the monitor function as:
\begin{equation}
\omega^{*}:=\theta\omega_{avg}^{*}+\left(1-\theta\right)\omega^{*}.\label{eq:monitor_fn_singularity_smoothing}
\end{equation}
Here $\omega_{avg}^{*}=\int_{0}^{1}d\xi\omega^{*}$ is the average
monitor function, the $\theta=\exp\left[-(r^{(p)}/\Delta r_{avg}^{(p)})^{2}\right]$
weights the monitor function between the average and local value,
and $r^{(p)}$ and $\Delta r^{(p)}$ are the cell-center location
and size (to be discussed in Sec. \ref{subsec:discretization_vfp_in_spherical_geo})
at the previous time, $p$, respectively. The intuition behind Eq.
\eqref{eq:monitor_fn_singularity_smoothing} is, if the distance between
the previous-time cell location and $r=0$ is within the average cell
size, $\omega^{*}$ will be floored to $\omega_{avg}$ and otherwise
retained. Finally}, we employ a Winslow smoothing \textcolor{black}{operation
\citep{winslow_jcp_winslow_smoothing} for the monitor function to
ensure the smoothness of the mesh:}

\textcolor{black}{
\begin{equation}
\left[1-\lambda_{\omega}\partial_{\xi\xi}^{2}\right]\omega=\omega^{*}.\label{eq:winslow_smoothing}
\end{equation}
Here, $\omega$ is the smoothed monitor function, and $\lambda_{\omega}$
is an empirically chosen constant smoothing coefficient ($10^{-3}$
in this study). }

In problems where the boundary deforms (i.e., imploding ICF capsules),
we prevent the grid from \emph{tangling} at the boundary by evolving
the MMPDE equation in terms of the normalized radial coordinate, 
\begin{equation}
\partial_{t}\widehat{r}^{*}=\tau_{r}^{-1}\partial_{\xi}\left[\omega\left(\xi,t\right)\partial_{\xi}\widehat{r}^{*}\right],\label{eq:mmpde_modified}
\end{equation}
where $\widehat{r}^{*}\left(\xi,t\right)=r^{*}\left(\xi,t\right)/R\left(t\right)$
is the normalized radial coordinate with boundary conditions, $\left.\widehat{r}^{*}\right|_{\xi=0}=0$
and $\left.\widehat{r}^{*}\right|_{\xi=1}=1$, and $R(t)$ is the
instantaneous radial domain size (determined by the implosion boundary
conditions -- to be discussed in Sec. \ref{subsec:implosion_bc}).
Since $\widehat{r}^{*}\in\left[0,1\right]$ at all times, the regularity
of the renormalized grid, $r_{i+1/2}^{*}=\widehat{r}_{i+1/2}^{*}R$,
is trivially ensured.

In ICF capsule implosions, where a variety of time-scales exist, it
is often not possible to choose a single value for $\tau_{r}$ that
simultaneously ensures 1) a stable grid variation that does not excite
numerical instabilities, and 2) a fast enough grid velocity that tracks
important physical features (e.g., shocks). To address these limitations
with the classic MMPDE scheme, we apply a corrector step where the
predicted grid obtained from solving Eq. (\ref{eq:mmpde_modified})
is \emph{optimized} by minimizing the following cost function:

\begin{eqnarray}
{\cal F}=\frac{1}{2}\int_{0}^{1}d\xi\left\{ \left(\textnormal{ln}\zeta\right)^{2}+\psi_{vol}^{2}\left[\left(1-\frac{J_{S\xi}^{(p)}}{\zeta J_{S\xi}^{*}}\right)^{2}+\left(1-\frac{\zeta J_{S\xi}^{*}}{J_{S\xi}^{(p)}}\right)^{2}\right]\right\} -\lambda\left[\int_{0}^{1}d\xi\zeta J_{S\xi}^{*}-\int_{0}^{1}d\xi J_{S\xi}^{*}\right],\label{eq:grid_cost_function}
\end{eqnarray}
for the modifier function, $\zeta\left(\xi\right)$, and the Lagrange
multiplier, $\lambda$, for the total volume preservation constraint
(both discussed shortly). This is achieved by solving the following
system of nonlinear equations, 
\begin{equation}
\left[\begin{array}{c}
\partial_{\vec{\zeta}}{\cal F}\\
\partial_{\lambda}{\cal F}
\end{array}\right]=\vec{0},\label{eq:grid_cost_function_minimization_equation_continuum}
\end{equation}
as described later in this study (Sec. \ref{subsec:nonlinear_mmpde}).
Here, the function $\zeta\in\left(0,\infty\right)$ is the modification
function for the Jacobian of the predicted grid (i.e., $J_{S\xi}=\zeta J_{S\xi}^{*}$),
$\psi_{vol}=0.05$ is the empirically chosen penalty factor for the
cost function based on the volumetric-rate-of-change of the grid with
respect to the previous time step, $p$, and $J_{r^{2}\xi}^{*}=\left(r^{*}\right)^{2}\partial_{\xi}r^{*}$
is the composite Jacobian for the \emph{predicted grid}. We remark
that the form of the cost function includes $\textnormal{ln}\zeta$
to ensure the positivity of $\zeta$, and the last term in the Eq.
(\ref{eq:grid_cost_function}) is a constraint that ensures that the
optimized grid preserves the total volume of the domain determined
by the predicted grid (inspired by similar considerations by Ref.
\citep{chacon_jcp_2015_mesh_motion_kantorovich}), which is dictated
by the boundary conditions for $r^{*}$. The optimized grid is finally
computed from, 
\begin{equation}
r\left(\xi\right)=r_{min}+\int_{0}^{\xi}\zeta\left(\xi'\right)J_{\xi}^{*}\left(\xi'\right)d\xi.'
\end{equation}
\textcolor{black}{We close the section by noting that, although the
introduction of the optimization procedure may tempt one to remove
$\tau_{r}$ entirely, we have discovered that a small amount of temporal
smoothing provides enhanced robustness for grid evolution. The NS-MMPDE
allows one to choose $\tau_{r}$ to be closer to the dynamical time-scale
of the problem (as will be shown in Sec. \ref{subsec:fuel_only_compression_yield}).
Additionally, due to the need to use several hyperparameters for the
NS-MMPDE scheme ($\delta_{min}$, $\eta$, $\lambda_{\omega}$), we
provide in \ref{app:intuition_on_ns_mmpde} visualizations and intuitions
on how these parameters and the normalization procedure affect the
resulting grid.}

\section{Numerical Implementation\label{sec:numerical_implementations}}

\subsection{Finite Difference Discretization of the Vlasov-Fokker-Planck Equation\label{subsec:discretization_vfp_in_spherical_geo}}

We employ a conservative finite-difference discretization in the 1D2V
phase space, where the phase-space cell volume is defined on the grid
$\left(\xi_{i},\widetilde{v}_{||,j},\widetilde{v}_{\perp,k}\right)$
as:

\begin{equation}
\Delta V_{i,j,k}=2\pi J_{S\xi,i}\widetilde{v}_{\perp,k}\Delta\widetilde{v}_{||}\Delta\widetilde{v}_{\perp},
\end{equation}
with, $\Delta\widetilde{v}_{||}=\frac{\widetilde{L}_{||}}{N_{||}}$
$\left(\Delta v_{\perp}=\frac{\widetilde{L}_{\perp}}{N_{\perp}}\right)$
the parallel- (perpendicular-) velocity cell size, $\widetilde{L}_{||}=\text{\ensuremath{\widetilde{v}_{||,max}-\widetilde{v}_{||,min}}}$
($\widetilde{L}_{\perp}=\widetilde{v}_{\perp,max}$) is the transformed
parallel- (perpendicular-) velocity domain size, $\widetilde{v}_{||,max}$
($\widetilde{v}_{\perp,max}$) is the maximum transformed parallel-
(perpendicular-) velocity bound, $\widetilde{v}_{||,min}$ is the
minimum transformed parallel velocity bound, $N_{||}$ ($N_{\perp}$)
is the number of cells in the parallel (perpendicular) velocity space,
$J_{S\xi,i}=J_{S,i}J_{\xi,i}$, $J_{S,i}=r_{i}^{2}$, and $J_{\xi,i}=\frac{\Delta r_{i}}{\Delta\xi}$.
Here, $\Delta r_{i}=r_{i+1/2}-r_{i-1/2}$, is the discrete configuration-space
cell size, and 
\begin{equation}
r_{i}=\sqrt{\frac{1}{3}\frac{r_{i+1/2}^{3}-r_{i-1/2}^{3}}{r_{i+1/2}-r_{i-1/2}}}\label{eq:vol_pres_cell_loc}
\end{equation}
is the cell-center location defined such that the discrete-volume
integral preserves the exact volume of a sphere, i.e., $r_{i}^{2}\Delta r_{i}=\int_{r_{i-1/2}}^{r_{i+1/2}}r^{2}dr=\frac{r_{i+1/2}^{3}-r_{i-1/2}^{3}}{3}$
. We define the discrete velocity-space moment of function $B$ as,

\begin{equation}
\left\langle A\left(\vec{v}\right),B\left(\vec{v}\right)\right\rangle _{\vec{v}}\approx\left\langle A\left(\vec{v}\right),B\left(\vec{v}\right)\right\rangle _{\delta\vec{v}}=2\pi\sum_{j=1}^{N_{||}}\Delta\widetilde{v}_{||}\sum_{k=1}^{N_{\perp}}\widetilde{v}_{\perp,k}\Delta\widetilde{v}_{\perp}A_{j,k}B{}_{j,k}.
\end{equation}
Additionally, with the distribution function defined on cell centers,
fluxes defined on cell faces, \textcolor{black}{and the requirement
for implicit time integration (to deal with the stiffness in both
the Fokker-Planck collision operator and the $r\rightarrow0$ singularity
in spherical geometry), the disc}retized VFP equation in the transformed
coordinate system reads:

\begin{eqnarray}
\left[\delta_{t}\left(J_{r^{2}\xi}\widetilde{f}_{\alpha}\right)\right]_{i,j,k}^{(p+1)}+\underbrace{\left[\delta_{\xi}F_{r,\alpha}\right]_{i,j,k}^{(p+1)}}_{\textcircled a}+\underbrace{\left[\delta_{\xi}F_{\dot{r},\alpha}\right]_{i,j,k}^{(p+1)}}_{\textcircled b}+\underbrace{\left[\delta_{\widetilde{v}_{||}}J_{acc,\alpha}\right]_{i,j,k}^{(p+1)}}_{\textcircled c}\nonumber \\
+\underbrace{\left[\delta_{\vec{\widetilde{v}}}\cdot\left(\gamma_{t,\alpha}\vec{J}_{t,\alpha}\right)\right]_{i,j,k}^{(p+1)}}_{\textcircled d}+\underbrace{\frac{1}{2}\left[\delta_{\vec{\widetilde{v}}}\cdot\left(\gamma_{r,\alpha,i+1/2}\vec{J}_{r,\alpha}^{-}+\gamma_{r,\alpha,i-1/2}\vec{J}_{r,\alpha}^{+}\right)\right]_{i,j,k}^{(p+1)}}_{\textcircled e}+\nonumber \\
\underbrace{\left[\delta_{\vec{\widetilde{v}}}\cdot\left(\gamma_{S,\alpha}\vec{J}_{S,\alpha}\right)\right]_{i,j,k}^{(p+1)}}_{\textcircled f}=J_{r^{2}\xi,i}^{(p+1)}\left[\sum_{\beta}^{N_{s}}\widetilde{C}_{\alpha\beta}+\widetilde{C}_{\alpha e}\right]_{i,j,k}^{(p+1)}.\label{eq:vfp_discrete_eqn}
\end{eqnarray}
Here, we introduce a shorthand notation for discrete-time derivatives,

\begin{equation}
\left[\delta_{t}\left(J_{S\xi}\widetilde{f}_{\alpha}\right)\right]_{i,j,k}^{(p+1)}=\frac{c^{(p+1)}J_{S\xi,i}^{(p+1)}\widetilde{f}_{\alpha,i,j,k}^{(p+1)}+c^{(p)}J_{S\xi,i}^{(p)}\widetilde{f}_{\alpha,i,j,k}^{(p)}+c^{(p-1)}J_{S\xi,i}^{(p-1)}\widetilde{f}_{\alpha,i,j,k}^{(p-1)}}{\Delta t^{(p)}},\label{eq:discrete_time_derivative}
\end{equation}
where $c^{(p+1)}$, $c^{(p)}$, and $c^{(p-1)}$ are the coefficients
for the second-order backward-differencing implicit time differencing
scheme (BDF2) \citep{byrne-acmtms-75-bdf} and superscripts $(p)$
indicate the discrete time index. The discrete spatial divergence
operator is defined as, 
\begin{equation}
\left[\delta_{\xi}F\right]_{i,j,k}^{(p+1)}=\frac{F_{i+1/2,j,k}^{(p+1)}-F_{i-1/2,j,k}^{(p+1)}}{\Delta\xi},\label{eq:discrete_spatial_divergence}
\end{equation}
and velocity-space divergence operator as, 
\begin{eqnarray}
\left[\delta_{\vec{\widetilde{v}}}\cdot\vec{J}^{(p+1)}\right]_{i,j,k}=\left[\delta_{\widetilde{v}_{||}}J_{||}+\widetilde{v}_{\perp}^{-1}\delta_{\widetilde{v}_{\perp}}\left(\widetilde{v}_{\perp}J_{\perp}\right)\right]_{i,j,k}^{(p+1)}=\nonumber \\
\frac{J_{||,i,j+1/2,k}^{(p+1)}-J_{||,i,j-1/2,k}^{(p+1)}}{\Delta\widetilde{v}_{||}}+\frac{\widetilde{v}_{\perp,k+1/2}J_{\perp,i,j,k+1/2}^{(p+1)}-\widetilde{v}_{\perp,k-1/2}J_{\perp,i,j,k-1/2}^{(p+1)}}{\widetilde{v}_{\perp,k}\Delta\widetilde{v}_{\perp}}.\label{eq:discrete_velocity_divergence}
\end{eqnarray}
\textcolor{black}{Here, the subscripts $i\pm1/2$, $j\pm1/2$, $k+1/2$
denote cell-face interpolated quantities. The term $\textcircled a$
corresponds to the discrete representation of the spatial streaming
term, $\textcircled b$ corresponds to the inertial term in the configuration
space (arising from the moving grid), $\textcircled c$ corresponds
to the electrostatic-acceleration term, $\textcircled d$ corresponds
to the velocity-space inertial terms due to temporal variation of
the velocity-space metrics (i.e., $v_{\alpha}^{*}$ and $\widehat{u}_{||,\alpha}^{*}$),
$\textcircled e$ corresponds to the velocity-space inertial terms
due to the spatial variation of the metrics, and $\textcircled f$
corresponds to the inertial term arising from transformation from
Cartesian phase-space to a spherical-configuration and cylindrical-velocity-space
coordinate system. Additionally, $\gamma_{t}\left(\xi,\widetilde{\vec{v}},t\right)$,
$\gamma_{r}\left(\xi,\widetilde{\vec{v}},t\right)$, and $\gamma_{S}\left(\xi,\widetilde{\vec{v}},t\right)$
in terms $\textcircled d$, $\textcircled e$, and $\textcircled f$
are discrete-nonlinear-constraint functions, which enforce the underlying
integral-discrete-conservation symmetries, and therefore the discrete
conservation of energy of the system. The functional forms of $\gamma_{t}$,
$\gamma_{r}$, and $\gamma_{S}$ are discussed in \ref{app:symmetries_for_transformed_coordinates}.
For reproducibility purposes, the detailed treatment of fluxes for
individual terms are discussed in \ref{app:discretization_of_fluxes}.
Finally, the right-hand-side of Eq. (\ref{eq:vfp_discrete_eqn}) corresponds
to the discrete collision operator, for which we follow Refs. \citep{Taitano_2015_rfp_0d2v_implicit,Taitano_2016_rfp_0d2v_implicit,DOV_CPC_effective_advection_tensor_diffusion}.
Finally, all velocity space fluxes, $\vec{J}$, at the boundaries
are set to zero, $\vec{J}\cdot\vec{n}_{v,B}=0$ (i.e., a particle-conserving
boundary is employed), where $\vec{n}_{v,B}$ is the velocity space
boundary outward normal vector. }

\subsection{Discretization of the fluid electron equation\label{subsec:discretization_fluid_e}}

The electron temperature equation in the transformed coordinate system,
Eq. (\ref{eq:composite_transformed_fluid_electron}), is discretized
using a finite-difference scheme in space and BDF2 in time: 
\begin{eqnarray}
\frac{3}{2}\left[\delta_{t}\left(J_{S\xi}n_{e}T_{e}\right)\right]_{i}^{(p+1)}+\frac{5}{2}\left[\delta_{\xi}\left(J_{S}u_{||,e}\widehat{n_{e}T_{e}}\right)\right]_{i}^{(p+1)}-\frac{3}{2}\left[\delta_{\xi}\left(\dot{J}_{r^{3}}\widehat{n_{e}T_{e}}\right)\right]_{i}^{(p+1)}\nonumber \\
+\left[\delta_{\xi}\left(J_{S}Q_{||,e}\right)\right]_{i}^{(p+1)}-J_{S\xi,i}^{(p+1)}q_{e}\left[n_{e}u_{||,e}E_{||}\right]_{i}^{(p+1)}=J_{S\xi,i}^{(p+1)}\sum_{\alpha}^{N_{s}}W_{e\alpha,i}^{(p+1)}.\label{eq:discrete_electron_eqn}
\end{eqnarray}
Here, the hat denotes cell-face interpolated quantities, for which
we use a SMART discretization \citep{smart_limiter_gaskel_lau_1988}.
Following Ref. \citep{Taitano_2018_vrfp_1d2v_implicit}, we discretize
the Joule heating term (last term on the left hand side) to ensure
exact energy balance with the ion electrostatic acceleration term.

\subsection{Nonlinearly stabilized (NS)-MMPDE and geometric conservation law
preserving grid velocity\label{subsec:nonlinear_mmpde}}

The modified MMPDE equation, Eq. (\ref{eq:mmpde_modified}), is discretized
on a staggered grid as:

\begin{equation}
\frac{c^{(p+1)}\widehat{r}_{i+1/2}^{*}+c^{(p)}\widehat{r}_{i+1/2}^{(p)}+c^{(p-1)}\widehat{r}_{i+1/2}^{(p-1)}}{\Delta t^{(p)}}-\frac{1}{\tau_{r}}\frac{\omega_{i+1}^{(p)}\left(\widehat{r}_{i+3/2}^{*}-\widehat{r}_{i+1/2}^{*}\right)-\omega_{i}^{(p)}\left(\widehat{r}_{i+1/2}^{*}-\widehat{r}_{i-1/2}^{*}\right)}{\Delta\xi^{2}}=0.\label{eq:mmpde_discrete}
\end{equation}
To reduce the nonlinearity between the solution and the grid for a
given time step, the monitor function is evaluated from the previous
time solution. The cost function for the MMPDE optimization in Eq.
(\ref{eq:grid_cost_function}) is discretized as:

\begin{equation}
{\cal F}=\frac{1}{2}\sum_{i=1}^{N_{\xi}}\Delta\xi\left\{ \left(\textnormal{ln}\zeta_{i}\right)^{2}+\psi_{vol}^{2}\left[\left(1-\frac{J_{S\xi,i}^{(p)}}{\zeta_{i}J_{S\xi,i}^{*}}\right)^{2}+\left(1-\frac{\zeta_{i}J_{S\xi,i}^{*}}{J_{S\xi,i}^{(p)}}\right)^{2}\right]\right\} -\lambda\left[\sum_{i=1}^{N_{\xi}}\Delta\xi\zeta_{i}J_{S\xi,i}^{*}-\sum_{i=1}^{N_{\xi}}\Delta\xi J_{S\xi,i}^{*}\right].\label{eq:grid_cost_function_discrete}
\end{equation}
Here, $J_{r^{2}\xi,i}^{*}=\frac{\left(r_{i}^{*}\right)^{2}\left[r_{i+1/2}^{*}-r_{i-1/2}^{*}\right]}{\Delta\xi}$
is the discrete composite Jacobian for the predicted grid, and the
cell-center grid location is defined by Eq. (\ref{eq:vol_pres_cell_loc}).
The discrete cost function is minimized with respect to $\zeta_{i}$
and $\lambda$ by solving : 
\begin{equation}
\left[\begin{array}{c}
\partial_{\zeta_{1}}{\cal F}\\
\vdots\\
\partial_{\zeta_{N_{\xi}}}{\cal F}\\
\partial_{\lambda}{\cal F}
\end{array}\right]=\left[\begin{array}{c}
\Delta\xi\left\{ \frac{\textnormal{ln}\zeta_{1}}{\zeta_{1}}+\psi_{vol}^{2}\left[\left(1-\frac{J_{S\xi,1}^{(p)}}{\zeta_{1}J_{S\xi,1}^{*}}\right)\frac{J_{S\xi,1}^{(p)}}{\zeta_{1}^{2}J_{S\xi,1}^{*}}-\left(1-\frac{\zeta_{1}J_{S\xi,1}^{*}}{J_{S\xi,1}^{(p)}}\right)\frac{J_{S\xi,1}^{*}}{J_{S\xi,1}^{(p)}}\right]-\lambda J_{S\xi,1}^{*}\right\} \\
\vdots\\
\Delta\xi\left\{ \frac{\textnormal{ln}\zeta_{N_{\xi}}}{\zeta_{N_{\xi}}}+\psi_{vol}^{2}\left[\left(1-\frac{J_{S\xi,N_{\xi}}^{(p)}}{\zeta_{N_{\xi}}J_{S\xi,N_{\xi}}^{*}}\right)\frac{J_{S\xi,N_{\xi}}^{(p)}}{\zeta_{N_{\xi}}^{2}J_{S\xi,N_{\xi}}^{*}}-\left(1-\frac{\zeta_{N_{\xi}}J_{S\xi,N_{\xi}}^{*}}{J_{S\xi,N_{\xi}}^{(p)}}\right)\frac{J_{S\xi,N_{\xi}}^{*}}{J_{S\xi,N_{\xi}}^{(p)}}\right]-\lambda J_{S\xi,N_{\xi}}^{*}\right\} \\
-\sum_{i=1}^{N_{\xi}}\Delta\xi\zeta_{i}J_{S\xi,i}^{*}+\sum_{i=1}^{N_{\xi}}\Delta\xi J_{S\xi,i}^{*}
\end{array}\right]=\vec{0},\label{eq:ns_mmpde_discrete_nonlinear_system}
\end{equation}
which is a nonlinearly coupled system of equations and must be solved
iteratively. We employ a Newton solver, where the solution is iterated
as:

\begin{equation}
\vec{x}^{(k+1)}=\vec{x}^{(k)}+\delta\vec{x}^{(k)},\label{eq:newton_nonlinear_iteration_ns_mmpde}
\end{equation}
where, the superscript, $k$, denotes the Newton iteration index,
$\vec{x}=\left[\bm{\zeta}^{T},\lambda\right]^{T}$ is the solution
vector, 
\begin{equation}
\delta\vec{x}^{(k)}=-\mathbb{J}^{(k),-1}\vec{R}^{(k)}\label{eq:nonlinear_update}
\end{equation}
is the Newton update, $\mathbb{J}=\frac{\partial\vec{R}}{\partial\vec{x}}$
is the Jacobian matrix, and $\vec{R}=\left[\partial_{\bm{\zeta}}{\cal F}^{T},\partial_{\lambda}{\cal F}\right]^{T}$
is the nonlinear residual vector. Due to the trivial 1D nature of
the problem, Eq. (\ref{eq:nonlinear_update}) is solved using a direct
inversion of $\mathbb{J}^{(k)}$. The iteration is continued until
$|\vec{R}^{(k)}|_{2}\le\epsilon_{grid}|\vec{R}^{(0)}|_{2}$ , where
$\epsilon_{grid}=10^{-10}$ is the nonlinear convergence tolerance.
Finally, the new-time grid is updated as:

\begin{equation}
r_{i+1/2}^{(p+1)}=r_{i-1/2}^{(p+1)}+\sum_{j=1}^{i}\zeta_{j}J_{\xi,j}^{*}.\label{eq:corrector_grid_update}
\end{equation}
The procedure for the NS-MMPDE algorithm is given in Alg. \ref{alg:grid_update}.
We note that, relative to the overall nonlinear solve of the coupled
hybrid equations (to be discussed in Sec. \ref{subsec:solvers}),
the cost of the nonlinear solve for Eq. (\ref{eq:ns_mmpde_discrete_nonlinear_system})
is negligible\textcolor{black}{{} (typically, between 1 and 3 percent
of the overall CPU cost). }
\begin{algorithm}[h]
1. Compute the non-smoothed monitor function, $\omega^{*,(p)}$, based
on the previous time solution.

2. Limit the monitor function via $\omega^{*}:=\omega^{*}+\left(\eta\omega_{max}^{*}-\omega_{min}^{*}\right)/\left(1-\eta\right)$
if $\omega_{min}^{*}/\omega_{max}^{*}\le\eta$.

\textcolor{black}{3. Floor the monitor function near the singularity
via $\omega^{*}:=\theta\omega_{avg}^{*}+\left(1-\theta\right)\omega^{*}$
with $\omega_{avg}^{*}=\int_{0}^{1}d\xi\omega^{*}$ and $\theta=\exp\left[-\left(r^{(p)}/\Delta r_{avg}^{(p)}\right)^{2}\right]$.}

4. Smooth the monitor function by solving $\left[1-\lambda_{\omega}\partial_{\xi\xi}^{2}\right]\omega=\omega^{*}$
for $\omega$, where $\lambda_{\omega}=5\times10^{-3}$ in this study.

5. Solve the normalized predictor grid, $\widehat{r}_{i+1/2}^{*}$,
from Eq. (\ref{eq:mmpde_discrete}).

6. Compute the renormalized predictor grid, $r_{i+1/2}^{*}=\widehat{r}_{i+1/2}^{*}R^{(p+1)}$,
where $R^{(p+1)}$ is the instantaneous domain size, determined from
implosion-driving boundary conditions (to be discussed in Sec \ref{subsec:implosion_bc}).

7. Compute the predictor Jacobian, $J_{S\xi,i}^{*}=\frac{\left(r_{i}^{*}\right)^{2}\left[r_{i+1/2}^{*}-r_{i-1/2}^{*}\right]}{\Delta\xi}$.

8. Solve for $\vec{\zeta}$ and $\lambda$ from Eqs. (\ref{eq:ns_mmpde_discrete_nonlinear_system})-(\ref{eq:nonlinear_update}).

9. Update the new time grid from Eq. (\ref{eq:corrector_grid_update}).

\caption{Grid update procedure with the NS-MMPDE algorithm.\label{alg:grid_update}}
\end{algorithm}

Once the new-time grid is obtained, the volumetric rate of change
of the cell $\dot{J}_{r^{3}}$ (which is related to the grid speed
as $\partial_{t}r=\dot{r}=\dot{J}_{r^{3}}/J_{S}$), is obtained in
such a way as to discretely ensure the geometric conservation law,
\begin{equation}
\partial_{t}J_{S\xi}=\partial_{\xi}\dot{J}_{r^{3}}.
\end{equation}
Discretizing and using Eq. (\ref{eq:vol_pres_cell_loc}), we find:

\begin{eqnarray}
\frac{c^{(p+1)}\left[\left(r_{i+1/2}^{(p+1)}\right)^{3}-\left(r_{i-1/2}^{(p+1)}\right)^{3}\right]+c^{(p)}\left[\left(r_{i+1/2}^{(p)}\right)^{3}-\left(r_{i-1/2}^{(p)}\right)^{3}\right]+c^{(p-1)}\left[\left(r_{i+1/2}^{(p-1)}\right)^{3}-\left(r_{i-1/2}^{(p-1)}\right)^{3}\right]}{3\Delta t^{(p)}\Delta\xi}\nonumber \\
-\frac{\dot{J}_{r^{3},i+1/2}^{(p+1)}-\dot{J}_{r^{3},i-1/2}^{(p+1)}}{\Delta\xi}=0.\label{eq:discrete_null_space_preserving_constraint}
\end{eqnarray}
Equating the terms with common cell indices gives the definition of
$\dot{J}_{r^{3}}$ as:

\begin{equation}
\dot{J}_{r^{3},i+1/2}^{(p+1)}=\frac{c^{(p+1)}\left(r_{i+1/2}^{(p+1)}\right)^{3}+c^{(p)}\left(r_{i+1/2}^{(p)}\right)^{3}+c^{(p-1)}\left(r_{i+1/2}^{(p-1)}\right)^{3}}{3\Delta t^{(p)}}.\label{eq:discrete_volume_rate_change}
\end{equation}

\subsection{Implosion Boundary Conditions\label{subsec:implosion_bc}}

We discuss next the various ICF implosion-driving boundary conditions.
Unless otherwise specified, for the spatial advection terms, we employ
an upwind discretization at the configuration-space boundary face
to compute the flux, i.e., 
\[
F_{B,j,k}=v_{||,j}\left\{ \begin{array}{ccc}
\widetilde{f}_{B-1/2,j,k} & \textnormal{if} & v_{||,j}\ge0\\
\widetilde{f}_{B+1/2,j,k} & \textnormal{otherwise}
\end{array}\right\} .
\]
Here, the subscript $B$ denotes the boundary surface and $B\pm1/2$
denotes the adjacent configuration-space computational/ghost cell-center
locations.

\subsubsection{Elastic moving wall and symmetry boundary conditions\label{subsubsec:symmetry_elastic_bc}}

\begin{figure}[h]
\begin{centering}
\includegraphics[scale=0.6]{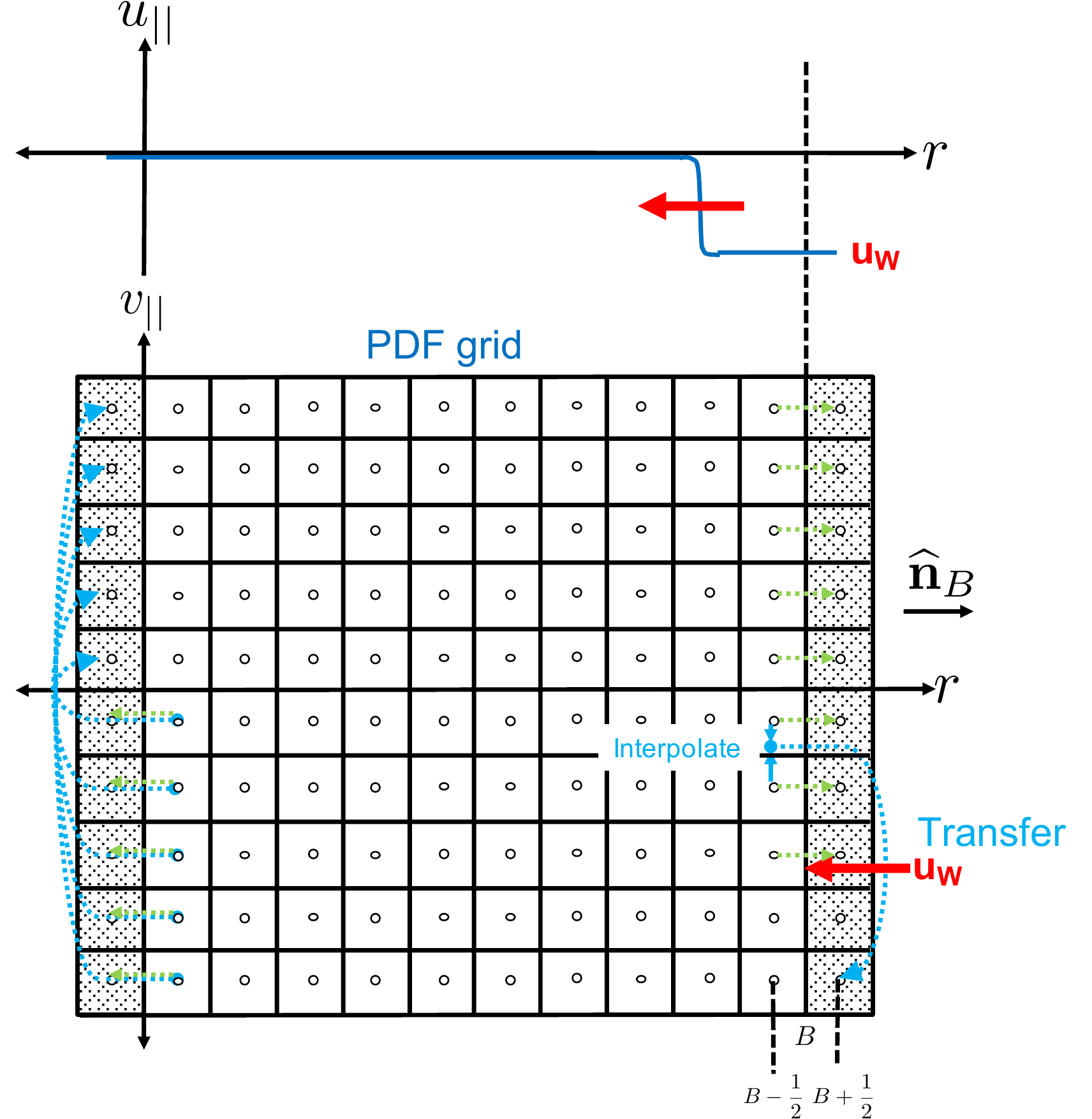} 
\par\end{centering}
\caption{Illustration of the elastic moving wall. The green arrows denote the
values that are ``copied'' to the ghost cell, while the blue arrows
denote quantities that are interpolated and transferred to the ghost
cell. The shaded regions denote the ghost-cell regions in which the
values of the adjacent computational cell quantities are transferred
to. \label{fig:illustration_of_moving_wall_bc}}
\end{figure}

To model an infinitely massive piston compressing a plasma, we consider
the elastic scattering of a collection of particles off a moving wall.
Consider a particle with velocity, $\vec{v}$, reflecting off a moving
wall with velocity, $\vec{u}_{W}$. One can derive the exact kinematic
relation between the pre- and post-scattered particle velocities,
$\left(\vec{v},\vec{v}_{s}\right)$ by transforming the velocity in
the reference of the wall as:

\begin{equation}
\text{\ensuremath{\vec{v}'=\vec{v}-\vec{u_{W}}},}\label{eq:v_prime}
\end{equation}
and 
\begin{equation}
\vec{v}'_{s}=\vec{v}_{s}-\vec{u}_{W}.\label{eq:vs_prime}
\end{equation}
In the frame of reference of the moving wall, an elastic scatter will
reflect the particle's parallel (to wall normal vector) velocity component,
\begin{equation}
\vec{v}'_{||}=\vec{v}'\cdot\vec{n}_{W}=v_{||}-u_{W,||},\label{eq:v_parallel_prime}
\end{equation}
in the equal and opposite direction while the perpendicular/tangent
component, 
\begin{equation}
\vec{v}'_{\perp}=\vec{v}'-\vec{v}'_{||},\label{eq:v_perpendicular_prime}
\end{equation}
is unchanged, i.e., 
\begin{equation}
\vec{v}'_{||,s}=-\vec{v}'_{||},\label{eq:vs_parallel_prime}
\end{equation}
\begin{equation}
\vec{v}'_{\perp,s}=\vec{v}'_{\perp}.\label{eq:vs_perpendicular_prime}
\end{equation}
Here, $\vec{n}_{W}$ is the wall normal vector (in 1D spherically
symmetric geometry, simply -1/1 at inner/outer radial boundary). Finally,
substituting Eq. (\ref{eq:vs_parallel_prime}) into Eq. (\ref{eq:vs_prime})
and solving for $\vec{v}_{s}$ (i.e., in the laboratory frame), we
obtain for the scattered-particle parallel velocity as

\[
v_{||,s}=-v_{||}+2u_{W,||}.
\]
Given a distribution function of particles, this kinematic relationship
can be expressed equivalently in the moving frame as:

\[
\left.\widetilde{f}\left(v'_{||}\right)\right|_{B}=\left\{ \begin{array}{cc}
\left.\widetilde{f}\left(-v'_{||}\right)\right|_{B} & \textnormal{if}\;\;v'_{||}n_{W,||}\le0\\
\left.\widetilde{f}\left(v'_{||}\right)\right|_{B} & \textnormal{else}
\end{array}\right..
\]
Numerically, the boundary condition in the laboratory frame is ensured
accordingly for the distribution function:

\begin{equation}
\widetilde{f}\left(\xi_{B},v_{||}\right)=\left\{ \begin{array}{cc}
C\widetilde{f}\left(\xi_{B\pm1/2},-v_{||}+2u_{W,||}\right) & \textnormal{if}\;\;\left(v_{||}-u_{W,||}\right)n_{W,||}\le0\\
\widetilde{f}\left(\xi_{B\pm1/2},v_{||}\right) & \textnormal{else}
\end{array}\right.,\label{eq:moving_wall_bc_dist_func}
\end{equation}
where $C$ is the scaling factor that ensures discrete mass conservation
at the boundary (discussed shortly). We note that, in general, it
is not possible to reconstruct numerically an exact one-to-one map
between the pre- and post-scattered particle distribution function
due to the discrete nature of velocity-space grid (i.e., there may
be no grid point at the specified reflection velocity). For numerical
purposes, we reconstruct the scattered distribution function by splining
the incoming velocity distribution - in this study, we consider a
second-order spline to ensure monotonicity of the reflected distribution
function. Further, because of the discrete violation of symmetry in
the reconstruction procedure, the mass is not automatically conserved
for a finite-wall velocity. In order to discretely conserve mass,
we rescale the reflected distribution function to ensure the following
constraint: 
\begin{equation}
\left\langle 1,\left(J_{S}v_{||}\widetilde{f}-\dot{J}_{r^{3}}\widetilde{f}\right)_{B}\right\rangle _{\delta\vec{v}}=C\left\langle 1,\left(J_{S}v_{||}\widetilde{f}-\dot{J}_{r^{3}}\widetilde{f}\right)_{B}\right\rangle _{\delta\vec{v}}^{v_{||}\ge2u_{W,||}}+\left\langle 1,\left(J_{S}v_{||}\widetilde{f}-\dot{J}_{r^{3}}\widetilde{f}\right)_{B}\right\rangle _{\delta\vec{v}}^{v_{||}<2u_{W,||}},\label{eq:wall_mass_conservation_constraint}
\end{equation}
where $J_{r^{2}}v_{||}\widetilde{f}-\dot{J}_{r^{3}}\widetilde{f}$
is the effective configuration-space flux including the boundary mesh
motion, and: 
\begin{equation}
\left\langle 1,\left(\cdots\right)\right\rangle _{\delta\vec{v}}^{v_{||}\ge2u_{W,||}}=2\pi\sum_{k=1}^{N_{\perp}}\sum_{j=j'}^{N_{||}}\widetilde{v}_{\perp,k}\Delta\tilde{v}_{\perp}\Delta\tilde{v}_{||}\left(\cdots\right)_{j,k},
\end{equation}
\begin{equation}
\left\langle 1,\left(\cdots\right)\right\rangle _{\delta\vec{v}}^{v_{||}<2u_{W,||}}=2\pi\sum_{k=1}^{N_{\perp}}\sum_{j=1}^{j'-1}\widetilde{v}_{\perp,k}\Delta\tilde{v}_{\perp}\Delta\tilde{v}_{||}\left(\cdots\right)_{j,k},
\end{equation}
where $j'$ corresponds to the discrete parallel velocity grid point
nearest to the $2u_{W,||}n_{W}$ value, 
\begin{equation}
j'=\left\{ \begin{array}{ccc}
\textnormal{floor}\left(\frac{2u_{W,||}-u_{||}^{*}-v^{*}\widetilde{v}_{||,min}}{v^{*}\Delta\widetilde{v}_{||}}\right) & \textnormal{if} & 1\le\textnormal{floor}\left(\frac{2u_{W,||}-u_{||}^{*}-v^{*}\widetilde{v}_{||,min}}{v^{*}\Delta\widetilde{v}_{||}}\right)\le N_{||}\\
1 & \textnormal{if} & \textnormal{floor}\left(\frac{2u_{W,||}-u_{||}^{*}-v^{*}\widetilde{v}_{||,min}}{v^{*}\Delta\widetilde{v}_{||}}\right)<1\\
N_{||} & \textnormal{otherwise}
\end{array}\right\} .\label{eq:j_prime_for_elastic_wall}
\end{equation}
We note that the symmetry boundary condition --identical to the \emph{specular
reflection boundary condition} considered in Ref. \citep{vogman_jcp_2018}--
is a special case of the treatment above in the limit of vanishing
wall velocity and is used for the $r=0$ boundary. An illustration
of the elastic moving wall boundary condition is shown in Fig. \ref{fig:illustration_of_moving_wall_bc}.

\subsubsection{Time dependent Maxwellian Dirichlet conditions with total mass conserving
boundary mesh speed\label{subsubsec:hydro_bc}}

In a full ICF capsule implosion, kinetic plasma physics is not sufficient
for fidelity. For a predictive capability, additional physics are
required such as radiation transport, $\alpha$ particle generation
and transport, laser ray tracing, as well as non-ideal equations of
state for cryogenic and high-Z materials. However, in certain experiments
(at the Omega facility for example), these effects become important
only in regions away from the fuel-pusher interface. Thus, in these
instances, it is possible to model only part of the capsule by decoupling
the outer pusher dynamics from the fuel region. In the literature,
Ref. \citep{larroche_2003_kin_sim_fuel_ion_transp_in_icf} was amongst
the first to explore this coupling with rad-hydro codes such as XRAGE
\citep{gittings_arxiv_2008}, HYDRA \citep{marinak_pop_1996}, and
LILAC \citep{delettrez_1976_lle_report}. In their approach, the authors
used the rad-hydro simulations to drive the kinetic simulation by
sourcing the boundary condition with a time-dependent Maxwellian distribution
function, in which the state variables ($n,u,T$) were provided by
rad-hydro simulations as a function of time. The particle distribution
function at this VFP-rad-hydro interface is defined as, 
\begin{equation}
\widetilde{f}_{B}=\begin{cases}
\widetilde{f}_{M,H}\left(n_{H,B}\left(t\right),u_{||,H}\left(t\right),T_{H}\left(t\right),\vec{v}\right) & \textnormal{for}\;\;\;\;\vec{v}\cdot\vec{n}_{B}\le0\\
\widetilde{f}_{B-1/2}\left(\vec{v}\right) & \textnormal{otherwise}
\end{cases},\label{eq:boundary_pdf}
\end{equation}
where, the subscript \emph{H} denotes rad-hydro simulation quantities.

In a Lagrangian formulation of the hydrodynamic equations, grid velocities
are chosen to be the center-of-mass velocity and consequently, lead
to a local mass conservation theorem of the associated computational
cell. In order to ensure that the boundary conserves total mass, we
evolve the boundary velocity to ensure a zero total mass flux. A total
mass conservation theorem can be derived by taking the $m_{\alpha}v^{0}$
moment of Eq. (\ref{eq:composite_transformed_vfp}), integrating over
$\xi$, and summing over all species, 
\begin{equation}
\sum_{\alpha}^{N_{s}}\int_{0}^{1}d\xi\left[\partial_{t}\left(J_{S\xi}\rho_{\alpha}+\partial_{\xi}\left[J_{S}\rho u_{||,\alpha}-\dot{J}_{r^{3}}\rho_{\alpha}\right]\right)\right]=0,\label{eq:reduced_continuity_for_bc_mass_conservation}
\end{equation}
where $\rho_{\alpha}=m_{\alpha}\left\langle 1,\widetilde{f}_{\alpha}\right\rangle _{\delta\vec{v}}$
and $\rho u_{||,\alpha}=m_{\alpha}\left\langle v_{||},\widetilde{f}_{\alpha}\right\rangle _{\delta\vec{v}}$.
We can reduce the above equation to:

\begin{equation}
\partial_{t}M+\left[J_{S}\rho u_{||}-\dot{J}_{r^{3}}\rho\right]_{\xi=0}^{\xi=1}=0,\label{eq:total_mass_conservation_bc}
\end{equation}
where $\partial_{t}M=0$ with $M=\sum_{\alpha}^{N_{s}}\int_{0}^{1}d\xi J_{S\xi}\rho_{\alpha}$,
$\rho=\sum_{\alpha}^{N_{s}}\rho_{\alpha}$, and $\rho u_{||}=\sum_{\alpha}^{N_{s}}\rho u_{||,\alpha}$.
We note that, with $r\left(\xi=0,t\right)=0$ for all time, $\left[J_{S}\rho u_{||}-\dot{J}_{r^{3}}\rho\right]_{\xi=0}=0$
is trivially ensured. Thus, for total mass conservation, we simply
require $\left.\dot{J}_{r^{3}}\rho\right|_{\xi=1}=\left.J_{S}\rho u_{||}\right|_{\xi=1}$
at the outer radius. By expanding the relationship and semi-discretizing,
we obtain the following discrete constraint:

\begin{equation}
\frac{\delta_{t}r_{\xi=1}^{3}}{3}\left\langle 1,\left.\widetilde{f}_{\alpha}\right|_{\xi=1}\right\rangle _{\delta\vec{v}}=J_{S,\xi=1}\left\langle 1,\left.v_{||}\widetilde{f}_{\alpha}\right|_{\xi=1}\right\rangle _{\delta\vec{v}}.\label{eq:mass)conservation_bc_kinetic_constraint}
\end{equation}
Since $\delta_{t}r_{\xi=1}^{3}=\frac{\left[c^{(p+1)}\left(r_{\xi=1}^{(p+1)}\right)^{3}+c^{(p)}\left(r_{\xi=1}^{(p)}\right)^{3}+c^{(p-1)}\left(r_{\xi=1}^{(p-1)}\right)^{3}\right]}{3\Delta t^{(p)}}$,
this provides us with a relationship to solve the new-time radius,
$r_{\xi=1}^{(p+1)}$ that ensures total mass conservation in the system.

\subsubsection{Pressure-driven Lagrangian boundary condition\label{subsubsec:pressure_driven_bc}}

Arguably, the more physically consistent approach to couple the hydrodynamic
simulations with the kinetic approach is to drive the implosion solely
with the external pressure provided by rad-hydro simulations and self-consistently
account for internal (back) pressure to evolve the boundary location.
To achieve this goal, we introduce an auxiliary equation that evolves
the boundary velocity,

\begin{equation}
\frac{du_{||,B}}{dt}=-\frac{1}{\rho_{B}}\left[\partial_{r}P_{B}-r_{B}^{-2}\partial_{r}\left(r_{B}^{2}\tau_{||||,B}\right)\right].\label{eq:pressure_lagrangian_bc_velocity}
\end{equation}
Discretizing the above equation, we obtain:

\begin{eqnarray}
\delta_{t}u_{||,B}^{(p+1)} & =-\frac{1}{\rho_{B}^{(p+1)}}\left[\frac{P_{B+1/2}^{(p+1)}-P_{B-1/2}^{(p+1)}}{\Delta r_{B}^{(p+1)}}-\frac{1}{\left(r_{B}^{(p+1)}\right)^{2}}\frac{\left(r_{B,i+1/2}^{(p+1)}\right)^{2}\tau_{||||,B+1/2}^{(p+1)}-\left(r_{B,i-1/2}^{(p+1)}\right)^{2}\tau_{||||,B-1/2}^{(p+1)}}{\Delta r_{B}^{(p+1)}}\right],\label{eq:pressure_lagrangian_bc_velocity_discrete}
\end{eqnarray}
where the boundary total-mass density, $\rho_{B}$, is obtained as
an average of the inner and ghost-cell boundary values, 
\begin{equation}
\rho_{B}^{(p+1)}=\frac{1}{2}\left[\rho_{B+1/2}^{(p+1)}+\rho_{B-1/2}^{(p+1)}\right],\label{eq:boundary_density_interpolated}
\end{equation}
with

\begin{equation}
\rho^{(p+1)}=\sum_{\alpha}^{N_{s}}\left\langle m_{\alpha},\widetilde{f}_{\alpha}^{(p+1)}\right\rangle _{\delta\vec{v}}.\label{eq:total_mass_density}
\end{equation}
The total hydrodynamic pressure, $P$, is computed as the sum of the
ion and electron pressures,

\begin{equation}
P^{(p+1)}=P_{e}^{(p+1)}+\sum_{\alpha}^{N_{s}}\frac{m_{\alpha}}{3}\left\langle \left(\vec{v}-\vec{u}_{\alpha}^{(p+1)}\right)^{2},\widetilde{f}_{\alpha}^{(p+1)}\right\rangle _{\delta\vec{v}}\label{eq:total_hydro_pressure}
\end{equation}
and the total radial viscous stress, $\tau_{||||}$, is computed as:

\begin{equation}
\tau_{||||}=\frac{2}{3}\sum_{\alpha}^{N_{s}}(P_{\perp,\alpha}-P_{||,\alpha}),\label{eq:viscous_stress}
\end{equation}
where the parallel and perpendicular pressures are defined, respectively,
as: 
\begin{equation}
P_{||,\alpha}=m_{\alpha}\left\langle \left(v_{||}-u_{||,\alpha}\right)^{2},\widetilde{f}_{\alpha}\right\rangle _{\delta\vec{v}},\label{eq:parallel_pressure}
\end{equation}
and 
\begin{equation}
P_{\perp,\alpha}=\frac{m_{\alpha}}{2}\left\langle v_{\perp}^{2},\widetilde{f}_{\alpha}\right\rangle _{\delta\vec{v}}.\label{eq:perpendicular_pressure}
\end{equation}
Once $u_{||,B}$ is evaluated, we source a Maxwellian distribution
at the boundary ghost cell with the prescribed number density, temperature,
and $u_{||,B}$. To ensure total mass conservation, we follow the
procedure in Eqs. (\ref{eq:reduced_continuity_for_bc_mass_conservation})-(\ref{eq:mass)conservation_bc_kinetic_constraint})
to evolve the boundary grid such that the total mass flux at the boundary
vanishes.

\subsection{Adaptive time-stepping strategy\label{subsec:adaptive_Time_stepping}}

Time-scales in ICF implosions range from the stiff collision time-scale,
particle advection and acceleration time-scale, dynamical (hydrodynamic)
time-scale (in which slow physical features evolve), all the way to
the system-time scale (e.g., implosion time). In order to efficiently
and accurately integrate through the various time-scales, we employ
an adaptive time-stepping strategy where the step size for a given
time step, $p$, is computed as:

\begin{equation}
\Delta t^{(p)}=\textnormal{min}\left\{ \Delta t^{(*)},\Delta t_{max},\tau_{v}^{(p)},\tau_{\dot{r}}^{(p)},\tau_{vol}^{(p)},\tau_{{\cal M}}^{(p)}\right\} ,\label{eq:time_stepsize_chooser}
\end{equation}
where 
\begin{equation}
\Delta t^{(*)}=\gamma\Delta t^{(p-1)}\label{eq:time_step_ramping}
\end{equation}
is the predictor time-step size based on the previous time-step size,
with $\gamma=1.15$, and $\Delta t_{max}$ is the maximum allowable
time-step size\textcolor{black}{. The advection-time scale (typically,
on the order of the dynamical time scale during the implosion and
explosion phases of ICF-capsule evolution) is given as }
\begin{equation}
\tau_{v}^{(p)}=0.05\textnormal{min}\left\{ \tau_{v,n_{\alpha}}^{(p)},\tau_{v,u_{\alpha}}^{(p)},\tau_{v,T_{\alpha}}^{(p)}\right\} \;\;\textnormal{for}\;\;\alpha=\left\{ 1,\cdots,N_{s}\right\} ,\label{eq:time_step_advection_timescale}
\end{equation}
with

\begin{equation}
\tau_{v,{\cal M}_{\alpha}}^{(p)}=\frac{L_{{\cal M}_{\alpha}}^{(p)}}{\left|u_{\alpha}^{(p)}\right|+\phi v_{th,\alpha}^{(p)}},\label{eq:time_step_advection_of_moment}
\end{equation}
and

\begin{equation}
L_{{\cal M_{\alpha}}}^{(p)}=\left|\partial_{r}\ln{\cal M}_{\alpha}^{(p)}\right|^{-1},\label{eq:time_step_advection_length_moment}
\end{equation}
the gradient length-scale for moment quantity, ${\cal M}$, and $\phi=3.5$
is an empirically chosen constant that accounts for the possibility
of a high-energy tail population of the distribution function, and
is chosen such that the nonlinear solver (to be discussed in Sec.
\ref{subsec:solvers}) performs robustly for a wide range of problems
when modeling traveling shocks. The grid-advection time scale is given
as

\begin{equation}
\tau_{\dot{r}}^{(p)}=0.05\textnormal{min}\left\{ \tau_{\dot{r},n_{\alpha}}^{(p)},\tau_{\dot{r},u_{\alpha}}^{(p)},\tau_{\dot{r},T_{\alpha}}^{(p)}\right\} \;\;\textnormal{for}\;\;\alpha=\left\{ 1,\cdots,N_{s}\right\} ,\label{eq:time_step_grid_advection_timescale}
\end{equation}
where

\begin{equation}
\tau_{\dot{r},{\cal M}_{\alpha}}^{(p)}=\frac{L_{{\cal M}_{\alpha}}^{(p)}}{\left|\dot{r}^{(p)}\right|+10^{-10}}.\label{eq:time_step_grid_advection_of_moment}
\end{equation}
We remark that, even with the configuration-space grid nonlinear stabilization
strategy employed here, there are certain instances in realistic implosions
where the local volume rate of change of the grid remains large. Time-step-size
limiting is used to address these situations. The time scale in which
the cell volume varies is given as,

\begin{equation}
\tau_{vol}^{(p)}=0.05\left|\partial_{t}\ln J_{r^{2}\xi}^{(p)}\right|^{-1}.\label{eq:time_step_volumetric_change}
\end{equation}
Finally, the exponentiation time-scale for mome\textcolor{black}{nts
(the dynamical time scale for shock collapse) is given a}s, 
\begin{equation}
\tau_{{\cal M}}^{(p)}=0.05\textnormal{min}\left\{ \tau_{n_{\alpha}}^{(p)},\tau_{u_{\alpha}}^{(p)},\tau_{T_{\alpha}}^{(p)}\right\} \;\;\textnormal{for}\;\;\alpha=\left\{ 1,\cdots,N_{s}\right\} ,\label{eq:time_step_moment_exponentiation}
\end{equation}
\begin{equation}
\tau_{{\cal M_{\alpha}}}^{(p)}=\left|\partial_{t}\ln{\cal M}_{\alpha}^{(p)}\right|^{-1}.\label{eq:time_step_exponentiation_of_moment}
\end{equation}
In this manner, the time-step size will adapt to the dynamically varying
features of the physics and the grid.

\subsection{VFP-Ion and Fluid Electron Solver with Implosion Boundary Conditions,
and the Integrated Algorithm\label{subsec:solvers}}

The discretized set of ion-VFP and fluid-electron equations lead to
a system of nonlinear equations, which we solve implicitly using an
Anderson acceleration scheme \citep{anderson_aa_JACM_1965,toth_siam_jna_2015_anderson_proof}.
Consider a fixed point map, 
\begin{equation}
\vec{x}^{l+1}=G\left(\vec{x}^{l}\right),\label{eq:fixed_point_map}
\end{equation}
which maps the solution from one iteration to another. Here 
\begin{equation}
\vec{x}^{l}=\left[\vec{\widetilde{f}}^{l},T_{e}^{l}\right]^{T}\label{eq:nonlinear_solution_vector}
\end{equation}
is the solution vector at iteration $l$ and and $\vec{\widetilde{f}}^{l-1}=\left[\widetilde{f}_{1}^{l-1},\cdots,\widetilde{f}_{N_{s}}^{l-1}\right]$
. The Anderson acceleration algorithm accelerates the iterative convergence
of the solution by employing the history of previous iterations as:
\begin{equation}
\vec{x}^{l+1}=\sum_{i=0}^{m_{l}}\theta_{i}^{l}G\left(\vec{x}^{l-m_{l}+i}\right),\label{eq:anderson_acceleration}
\end{equation}
where $m_{l}=5$ is the length of the residual history we use for
this study, and the coefficients $\theta_{i}^{l}$ are found minimizing
$\left\Vert \sum_{i=0}^{m_{l}}\theta_{i}^{l}\left(G\left(\vec{x}^{l-m_{l}+i}\right)-\vec{x}^{l-m_{l}+i}\right)\right\Vert ,$
subject to the constraint $\sum_{i=0}^{m_{l}}\theta_{i}^{l}=1$. For
$G\left(\vec{x}^{l}\right)$, we consider a Picard linearized solver
where we: 1) solve the fluid electron equation with given ion distribution
functions; and 2) solve the ion VFP equations with given fluid electron
temperature and the collisional transport coefficients from the $l-1$
iteration. The fluid electron equation inner solve is itself solved
with an Anderson acceleration with the fixed point map being a Quasi-Newton
system, i.e., 
\begin{equation}
T_{e}^{l,s+1}=G_{T_{e}}\left(T_{e}^{l,s}\right)=T_{e}^{l,s}-\left(\mathbb{P}_{T_{e}}^{l,s}\right)^{-1}R_{T_{e}}^{l,s}.\label{eq:electron_temperature_fixed_point_map}
\end{equation}
Here, 
\begin{eqnarray}
\mathbb{P}_{T_{e}}^{l,s}\circ=\frac{3}{2}\partial_{t}\left(J_{S\xi}^{l-1}n_{e}^{l-1}\circ\right)+\partial_{\xi}\left[\left(\frac{5}{2}J_{S}^{l-1}u_{||,e}^{l-1}-\frac{3}{2}\dot{J}_{r^{3}}\right)n_{e}^{l-1}\circ\right]+\nonumber \\
\partial_{\xi}\left[J_{S}^{l-1}\mathbb{Q}_{e}^{l-1}\circ\right]-J_{S\xi}^{l-1}q_{e}n_{e}^{l-1}u_{||,e}^{l-1}\partial_{\xi}\left(n_{e}^{l-1}\circ\right)-J_{S\xi}^{l-1}\sum_{\alpha=1}^{N_{s}}\mathbb{W}_{e\alpha}^{l-1}\circ,\label{eq:Te_preconditioner}
\end{eqnarray}
is the preconditioning operator (where $\circ$ simply denotes the
quantity that is being operated on) for the electron temperature equation
subsystem, and 
\begin{eqnarray}
R_{T_{e}}^{l,s}(\vec{\widetilde{f}}^{l-1},T_{e}^{l,s})=\frac{3}{2}\partial_{t}\left(J_{S\xi}^{l-1}n_{e}^{l-1}T_{e}^{l,s}\right)+\partial_{\xi}\left[\left(\frac{5}{2}J_{S}^{l-2}u_{||,e}^{l-1}-\frac{3}{2}\dot{J}_{r^{3}}\right)n_{e}^{l-1}T_{e}^{l,s}\right]+\nonumber \\
\partial_{\xi}\left[J_{S}^{l-1}Q_{||,e}\left(\vec{\widetilde{f}}^{l-1},T_{e}^{l,s}\right)\right]-J_{S\xi}^{l-1}q_{e}n_{e}^{l-1}u_{||,e}^{l-1}E_{||}\left(\vec{\widetilde{f}}^{l-1},T_{e}^{l,s}\right)-J_{S\xi}^{l-1}\sum_{\alpha=1}^{N_{s}}W_{e\alpha}\left(\vec{\widetilde{f}}^{l-1},T_{e}^{l,s}\right)\label{eq:residual_Te}
\end{eqnarray}
is the nonlinear residual we seek the root of. We remind the readers
that $n_{e}^{l-1}=-\sum_{\alpha}^{N_{s}}\left\langle 1,\widetilde{f}_{\alpha}^{l-1}\right\rangle _{\delta\vec{v}}/q_{e}$
and $u_{||,e}^{l-1}=-\sum_{\alpha}^{N_{s}}q_{\alpha}\left\langle v_{||},\widetilde{f}_{\alpha}^{l-1}\right\rangle _{\delta\vec{v}}/q_{e}n_{e}^{l-1}$
are the electron number density and parallel drift velocity, respectively,
computed from quasineutrality and ambipolarity conditions; $\mathbb{Q}_{e}^{l-1}$
and $\mathbb{W}_{e\alpha}^{l-1}$ in Eq. (\ref{eq:Te_preconditioner})
are the linearized operators for the electron heat flux and the ion-electron
energy exchange, respectively, and for the fluid electron subsystem
we use a nonlinear history length of $m_{s}=5$. The Quasi-Newton
system in Eq. (\ref{eq:electron_temperature_fixed_point_map}) is
solved using a direct inversion of a resulting tridiagonal system
for $\mathbb{P}_{T_{e}}^{l,s}$. The nonlinear iteration is continued
until $\left\Vert R_{T_{e}}^{l,s}\right\Vert _{2}\le10^{-6}\left\Vert R_{T_{e}}^{l,0}\right\Vert _{2}$.\textcolor{black}{{} }

The ion VFP equation inner solve is similarly performed with an Anderson
acceleration with a Quasi-Newton fixed point map, 
\begin{equation}
\widetilde{f}_{\alpha}^{l,s+1}=G_{\widetilde{f}_{\alpha}}\left(\widetilde{f}_{\alpha}^{l,s}\right)=\widetilde{f}_{\alpha}^{l,s}-\left(\mathbb{P}_{VFP,\alpha}^{l-1}\right)^{-1}R_{\alpha}^{l,s},\label{eq:vfp_ion_fixed_point_map}
\end{equation}
where 
\begin{equation}
\left(\mathbb{P}_{VFP,\alpha}^{l-1}\right)^{-1}\circ=\left(\mathbb{P}_{t}^{l-1}+\mathbb{P}_{\xi,\alpha}^{l-1}\right)^{-1}\left[\mathbb{I}-\mathbb{P}_{\widetilde{v}}^{l-1}\left(\mathbb{P}_{t}^{l-1}+\mathbb{P}_{\widetilde{v}}^{l-1}\right)^{-1}\right]\circ\label{eq:VFP_preconditioner}
\end{equation}
is the operator-split preconditioner, $\mathbb{I}$ is the identity
operator, 
\begin{equation}
\mathbb{P}_{t}^{l-1}\circ=\partial_{t}\left(J_{S\xi}^{l-1}\circ\right)\label{eq:VFP_temporal_derivative}
\end{equation}
is the temporal derivative, 
\begin{equation}
\mathbb{P}_{\xi,\alpha}^{l-1}\circ=\partial_{\xi}\left[\left(J_{S}^{l-1}v_{||}-\dot{J}_{r^{3}}^{l-1}\right)\circ\right]\label{eq:VFP_preconditioner_configuration}
\end{equation}
is the configuration space operator, 
\begin{eqnarray}
\mathbb{P}_{\widetilde{v},\alpha}^{l-1}\circ=-v_{\alpha}^{*^{-1}}\partial_{\vec{\widetilde{v}}}\cdot\left\{ \left[J_{S\xi}^{l-1}\partial_{t}\vec{v}+\partial_{\xi}\left(\left[J_{S}^{l-1}v_{||}-\dot{J}_{r^{3}}^{l-1}\right]\vec{v}\right)\right]\circ\right\} +r^{l-1}v_{\alpha}^{*^{-1}}J_{\xi}^{l-1}\partial_{\vec{\widetilde{v}}}\cdot\left(\vec{\widetilde{a}}\circ\right)\nonumber \\
-\frac{q_{\alpha}}{m_{\alpha}}E_{||}\left(n_{e}^{l-1},T_{e}^{l}\right)J_{S\xi}^{l-1}v_{\alpha}^{*^{-1}}\partial_{\widetilde{v}_{||}}\circ-J_{S\xi}^{l-1}\left[\sum_{\beta}^{N_{s}}\widetilde{\mathbb{C}}_{\alpha\beta}\left(\widetilde{f}_{\beta}^{l-1},\widetilde{f}_{\alpha}^{l-1}\right)+\widetilde{\mathbb{C}}_{\alpha e}\left(T_{e}^{l},n_{e}^{l-1},u_{||,e}^{l-1}\right)\right]\circ\label{eq:VFP_preconditioner_velocity}
\end{eqnarray}
is the velocity-space operator, and

\begin{eqnarray}
R_{\alpha}^{l,s}(\widetilde{f}_{1}^{l,s},T_{e}^{l})=\partial_{t}\left(J_{S\xi}^{l-1}\widetilde{f}_{\alpha}^{l,s}\right)+\partial_{\xi}\left[\left(J_{S}^{l-1}v_{||}-\dot{J}_{r^{3}}^{l-1}\right)\widetilde{f}_{\alpha}^{l,s}\right]-\nonumber \\
v_{\alpha}^{*^{-1}}\partial_{\vec{\widetilde{v}}}\cdot\left\{ \left[\gamma_{t,\alpha}^{l-1}J_{S\xi}^{l-1}\partial_{t}\vec{v}+\gamma_{x,\alpha}^{l-1}\partial_{\xi}\left(\left[J_{S}^{l-1}v_{||}-\dot{J}_{r^{3}}^{l-1}\right]\vec{v}\right)\right]\widetilde{f}_{\alpha}^{l,s}\right\} +\nonumber \\
r^{l-1}J_{\xi}^{l-1}v_{\alpha}^{*^{-1}}\partial_{\vec{\widetilde{v}}}\cdot\left(\gamma_{S,\alpha}^{l-1}\vec{\widetilde{a}}\widetilde{f}_{\alpha}^{l,s}\right)-\frac{q_{\alpha}}{m_{\alpha}}E_{||}^{l,s}J_{S\xi}^{l-1}v_{\alpha}^{*^{-1}}\partial_{\widetilde{v}_{||}}\widetilde{f}_{\alpha}^{l,s}-J_{S\xi}^{l-1}\left[\sum_{\beta}^{N_{s}}\widetilde{C}_{\alpha\beta}^{l,s}+\widetilde{C}_{\alpha e}^{l,s}\right]\label{eq:residual_vfp}
\end{eqnarray}
is the nonlinear-residual of the ion VFP equation for the $\alpha$
species. Here, $\widetilde{\mathbb{C}}_{\alpha\beta}$ and $\widetilde{\mathbb{C}}_{\alpha e}$
are the Picard-linearized (to the previous outer nonlinear iteration)
$\alpha-\beta$ ion, and ion-electron collision operators, respectively,
and we use a nonlinear history length of $m_{s}=2$ for the Anderson
so\textcolor{black}{lve. Note that $m_{s}$ for the VFP solver is
smaller than that for the fluid electron solver for two reasons: 1)
memory considerations, and 2) because the system is linear (except
for the positivity preserving nonlinear discretization for SMART,
which is not stiff). T}he Quasi-Newton system in Eq. (\ref{eq:vfp_ion_fixed_point_map})
is solved by inverting $\mathbb{P}_{t}+\mathbb{P}_{\xi}$ and $\mathbb{P}_{t}+\mathbb{P}_{\widetilde{v}}$,
where we employ one V cycle of classical geometric multigrid, smoothed
with 3 passes of damped Jacobi and employ agglomeration for restriction
and second order prolongation. The nonlinear iteration is continued
until $\left\Vert R_{\alpha}^{l,s}\right\Vert _{2}\le10^{-2}\left\Vert R_{\alpha}^{l,0}\right\Vert _{2}.$ 

We note that the nonlinearity in the domain size --determined by
the implosion boundary conditions in Sec. \ref{subsec:implosion_bc}--
is not stiff and therefore the configuration-space grid locations
and other derived quantities are Picard-linearized with respect to
the electron temperature and the ion distribution functions inside
the evaluation of the outer fixed point map. All advection operators
in the preconditioner, both in the fluid electron and ion VFP systems,
are discretized using a first-order upwind discretization. The evaluation
of the outer fixed-point map in Eq. (\ref{eq:fixed_point_map}) is
summarized in Alg. \ref{alg:outer_fixed_point_map_evaluation}. 
\begin{algorithm}[t]
1. Compute the new boundary location, $r_{\xi=1}^{l-1}$, and grid
locations, $r^{l-1}(\vec{\widetilde{f}}^{l-1},T_{e}^{l-1}$), and
derived quantities ($J_{S}^{l-1}$, $J_{\xi}^{l-1}$, $J_{S\xi}^{l-1}$)

2. Initialize $s=0$, $T_{e}^{l,0}=T_{e}^{l-1}$

3. Invert $T_{e}$ system for $T_{e}^{l}$

4. For $\alpha=1:N_{s}$

$\;\;\;\;\;\;\;\;\;\;\;\;$Invert $\alpha$-ion species' VFP system
for $\widetilde{f}_{\alpha}^{l}$

$\;\;\;\;\;$end

5. Set $\vec{x}^{l}=\left[\vec{\widetilde{f}}^{l},T_{e}^{l}\right]^{T}$

\caption{Evaluation of the outer fixed point map, $G\left(\vec{x}^{l-1}\right)$,
in Eq. (\ref{eq:fixed_point_map}).\label{alg:outer_fixed_point_map_evaluation}}
\end{algorithm}
The outer fixed point iteration is continue\textcolor{black}{d until
$\left\Vert \vec{R}^{l}\right\Vert _{2}\le\epsilon_{r}\left\Vert \vec{R}^{l=0}\right\Vert _{2}$,
where $\vec{R}=\left[R_{1},\cdots,R_{N_{s}},R_{T_{e}}\right]^{T}$
is the residual vector containing both the VFP and fluid electron
residuals, and $\epsilon_{r}$ is the relative nonlinear convergence
tolerance ($\epsilon_{r}=10^{-3}$ unless otherwise specified).}

The full algorithm containing the NS-MMPDE for the configuration-space-grid
adaptivity, velocity-space-grid adaptivity (for completeness, briefly
summarized in \ref{app:velocity_space_grid_adaptivity}), time-step
adaptivity, and nonlinear solvers is described in Algorithm \ref{alg:integrated_algorithm}.

\begin{algorithm}[h]
1. Adapt time step-size according to Eq. (\ref{eq:time_stepsize_chooser}).

2. Adapt velocity-space grid according to Alg. \ref{alg:vel_grid_update}.

3. Adapt configuration space grid according to Alg. \ref{alg:grid_update}.

4. Solve coupled ion-VFP and fluid electron system according to Eq.
(\ref{eq:fixed_point_map}) and Alg. \ref{alg:outer_fixed_point_map_evaluation}.

\caption{Integrated algorithm of adaptive time-stepping, NS-MMPDE, velocity-space
grid adaptivity, and the AA solver for the hybrid ion Vlasov-Fokker-Planck
and fluid electron equation.\label{alg:integrated_algorithm}}
\end{algorithm}
\textcolor{black}{We close the section by noting that, in the rare
instances when the inner fixed point iteration (fluid electrons and
VFP solve) struggles to converge, we terminate the iteration at the
tenth cycle, and offload the task of convergence to the outer iteration. }

\section{Numerical Results\label{sec:numerical_results}}

We test the proposed integrated algorithm on a set of benchmark problems
that demonstrate its advertised capabilities. In this study, the VFP
and fluid electron temperature equations are non-dimensionalized in
terms of the proton mass, $m_{p}=1.673\times10^{-27}$ {[}$kg${]},
proton charge, $q_{p}=1.602\times10^{-19}$ {[}$C${]}, normalization
number density, $n^{*}=4.203\times10^{27}$ {[}$\textnormal{particles}/m^{3}${]},
normalization temperature, $T^{*}=53.654$ {[}$eV${]} $=8.595\times10^{-18}$
{[}$J${]}; and the resulting proton-proton collision time, $\tau^{*}=\frac{3}{10}\frac{\sqrt{2m_{p}}\epsilon_{0}^{2}\left(2\pi T^{*}\right)^{3/2}}{n^{*}q_{p}^{4}}=1.949\times10^{-13}$
{[}$s${]}, speed, $u^{*}=\sqrt{k_{B}T^{*}/m_{p}}=7.174\times10^{4}$
{[}$m/s${]}, and length, $L^{*}=u^{*}\tau^{*}=1.398\times10^{-8}$
{[}$m${]}. The $v^{*}$ normalized distribution function is initialized
based on the Maxwellian distribution function, 
\begin{equation}
\widetilde{f}_{M}=\frac{n_{M}}{\pi^{3/2}}\left(\frac{v^{*}}{v_{th,M}}\right)^{3}\exp\left[-\frac{\left(\vec{v}-\vec{u}_{M}\right)^{2}}{v_{th,M}^{2}}\right],
\end{equation}
where the numerical Maxwellian moments, $n_{M}$, $\vec{u}_{M}$,
and $v_{th,M}=\sqrt{2T_{M}/m}$ are computed to ensure that the numerical
integrals of the Maxwellian distribution function return the prescribed
density, drift velocity, and temperature \citep{Taitano_JCP_2017_FP_equil_disc}.
Unless otherwise mentioned, we use a density- and temperature-dependent
Coulomb logarithm, defined in Ref. \citep{plasma-formulary}. Also,
unless otherwise stated, an initial Poisson grid generation and optimization
(\ref{app:init_grid_optimization}) is employed to avoid initial numerical
pollution of the simulation due to unresolved gradients.

\subsection{Doubly reflective boundary problem with prescribed grid motion\label{subsec:doubly_reflective_bc_problem}}

We consider an initially perturbed plasma in a spherical cavity, confined
in a reflective system (i.e., symmetry boundary conditions from Sec.
\ref{subsubsec:symmetry_elastic_bc} at both $r|_{\xi=0}=0$ and $r|_{\xi=1}=R$)
to test the discretization convergence rate and conservation properties
of the proposed algorithm. We consider a domain with a normalized
radius of $R=100$, and a logical domain of $\xi\times\widetilde{v}_{||}\times\widetilde{v}_{\perp}\in[0,1]\times\left[-6,6\right]\times\left[0,6\right]$.
A deuterium-tritium plasma is used with masses $m_{D}=2$ and $m_{T}=3$;
charges $q_{D}=q_{T}=1$; number densities $n_{D}=n_{T}=1+0.2\cos\left(k_{r}r\right);$
drift velocities $u_{D}=u_{T}=0;$ and temperatures $T_{D}=T_{T}=T_{e}=1+0.2\cos\left(k_{r}r\right)$,
where $k_{r}=\frac{2\pi}{R}$; and the prescribed analytical grid
is given as:\textcolor{black}{{} 
\begin{equation}
r\left(\xi_{i},t\right)=r_{0,i}+0.45\Delta r_{0}\sin\left(\omega_{grid}t\right)\cos\left(k_{r}r_{0,i}\right).\label{eq:analytical_grid_motion}
\end{equation}
}Here, $\omega_{grid}=\pi/5$ is the grid oscillation time-scale,
\textcolor{black}{$\Delta r_{0}=R/N_{\xi}$ is the average grid size
in the physical space, and $r_{0}=\xi R$ is the initial grid. We
also use a Coulomb log of }10 for all interactions.

\begin{figure}[h]
\begin{centering}
\includegraphics[scale=0.6]{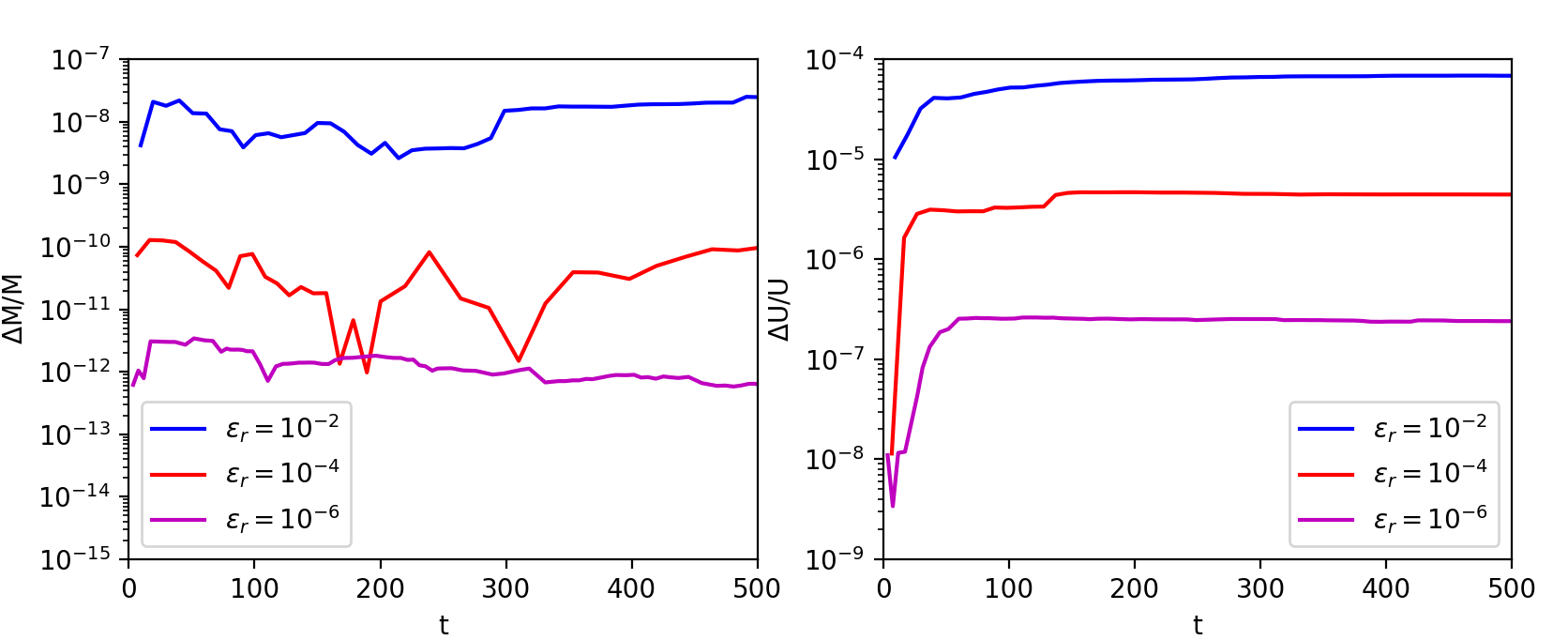} 
\par\end{centering}
\caption{Doubly reflective boundary problem with prescribed grid motion: Discrete
conservation of mass and energy for varying nonlinear convergence
tolerance.\label{fig:Doubly-reflective-boundary-conservation}}
\end{figure}

For this section, the time-step ramping discussed in Sec. \ref{subsec:adaptive_Time_stepping}
is not used to allow for a time-convergence study, and the first time-step
size is taken to be $\frac{\Delta t_{max}}{10}$. We demonstrate the
discrete conservation theorem for mass and energy with a grid of $N_{\xi}\times N_{v_{||}}\times N_{v_{\perp}}=96\times64\times32$,
a time-step size of $\Delta t_{max}=10^{-1}$, and varying nonlinear
convergence tolerance of $\epsilon_{rel}=10^{-2}$, $10^{-4}$, and
$10^{-6}$; refer to Fig. \ref{fig:Doubly-reflective-boundary-conservation}.

As can be seen, discrete conservation error in mass, $\frac{\Delta M^{(p)}}{M^{(0)}}$,
and energy, $\frac{\Delta U^{(p)}}{U^{(0)}}$, reduces commensurately
with the nonlinear convergence tolerance. Here, 
\begin{equation}
M^{(p)}=\sum_{\alpha}^{N_{s}}m_{\alpha}\sum_{i=1}^{N_{\xi}}\Delta\xi J_{S\xi,i}^{(0)}\left\langle 1,\widetilde{f}_{\alpha,i}^{(p)}\right\rangle _{\delta\vec{v}},\label{eq:total_discrete_mass_at_p}
\end{equation}
\begin{equation}
\Delta M^{(p)}=\left|M^{(p)}-M^{(0)}\right|,\label{eq:discrete_mass_difference_at_p}
\end{equation}
\begin{equation}
U^{(p)}=\sum_{\alpha}^{N_{s}}m_{\alpha}\sum_{i=1}^{N_{\xi}}\Delta\xi J_{S\xi,i}^{(p)}\left\langle \frac{v^{2}}{2},\widetilde{f}_{\alpha,i}^{(p)}\right\rangle _{\delta\vec{v}}+\frac{3}{2}\sum_{i=1}^{N_{\xi}}\Delta\xi J_{S\xi,i}^{(p)}n_{e,i}^{(p)}T_{e,i}^{(p)},\label{eq:total+discrete_energy_at_p}
\end{equation}
and 
\begin{equation}
\Delta U^{(p)}=\left|U^{(p)}-U^{(0)}\right|.\label{eq:discrete_energy_difference_at_p}
\end{equation}

Next, we demonstrate the discretization order of convergence of the
proposed algorithm. The temporal, configuration-space, and velocity-space
convergence study is performed by measuring the $L_{2}$-norm of the
relative difference of the temperature, 
\begin{equation}
{\cal E}_{T}^{\Delta t}=\sqrt{\sum_{i=1}^{N_{\xi}}\Delta\xi\left[\sum_{\alpha=1}^{N_{sp}}\left(T_{\alpha,i}^{\Delta t}-T_{\alpha,i}^{\Delta t_{ref}}\right)^{2}+\left(T_{e,i}^{\Delta t}-T_{e,i}^{\Delta t_{ref}}\right)^{2}\right]},
\end{equation}

\begin{equation}
{\cal E}_{T}^{\Delta\xi}=\sqrt{\sum_{i=1}^{N_{\xi}}\Delta\xi\left[\sum_{\alpha=1}^{N_{sp}}\left(T_{\alpha,i}^{\Delta\xi}-T_{\alpha,i}^{\Delta\xi_{ref}}\right)^{2}+\left(T_{e,i}^{\Delta\xi}-T_{e,i}^{\Delta\xi_{ref}}\right)^{2}\right]},
\end{equation}
\begin{equation}
{\cal E}_{T}^{\Delta\widetilde{v}}=\sqrt{\sum_{i=1}^{N_{\xi}}\Delta\xi\left[\sum_{\alpha=1}^{N_{sp}}\left(T_{\alpha,i}^{\Delta\widetilde{v}}-T_{\alpha,i}^{\Delta\widetilde{v}_{ref}}\right)^{2}+\left(T_{e,i}^{\Delta\widetilde{v}}-T_{e,i}^{\Delta\widetilde{v}_{ref}}\right)^{2}\right]}.
\end{equation}
Here, the superscript $\Delta t$ denotes the solution obtained with
a larger time-step size while $\Delta t_{ref}$ denotes the reference
solution, $\Delta\xi$ denotes the solution obtained with a coarse
logical space grid size and $\Delta\xi_{ref}$ denotes the reference
solution, and $\Delta\widetilde{v}$ denotes the solution obtained
with a coarse velocity-space grid and $\Delta\widetilde{v}_{ref}$
denotes the reference grid.

\begin{figure}[h]
\begin{centering}
\includegraphics[width=5.5cm]{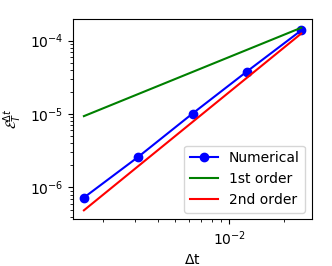}\includegraphics[width=5.5cm]{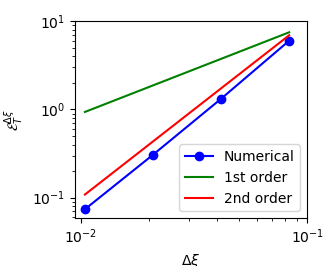}\includegraphics[width=5.5cm]{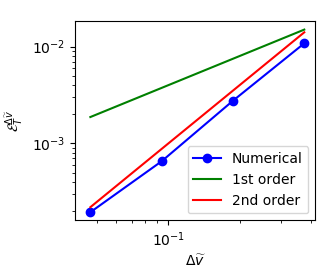} 
\par\end{centering}
\caption{Doubly reflective boundary problem with prescribed grid motion: Temporal
(left), configuration-space (center), and velocity-space (right) convergence
study. As can be seen, second order convergence rate is observed in
all discretization parameters. \label{fig:Doubly-reflective-boundary-convergence-study}}
\end{figure}

The temporal discretization relies on a BDF2 scheme and is second
order accurate (refer to Fig. \ref{fig:Doubly-reflective-boundary-convergence-study}-left).
We compute the $L_{2}$-norm of the relative difference of the temperatures.
The reference solution uses a grid of $N_{\xi}=24$, $N_{v_{||}}=32$,
$N_{v_{\perp}}=16$, and a time-step size of $\Delta t_{ref}=10^{-3}$,
and is time integrated to $t_{max}=0.25.$ The convergence study is
performed by fixing the grids while varying the time-step size.

The configuration-space discretization relies on a SMART discretization
for the flux interpolation, which asymptotically is third order accurate.
However, we see in Fig. \ref{fig:Doubly-reflective-boundary-convergence-study}-center
that we recover only second-order accuracy. This is due to the second-order
accurate operations employed elsewhere (e.g., linear interpolation
employed for the configuration-space cell-face quantities {[}e.g.,
$J_{r^{2}},$ $J_{\xi}$, $v^{*}$, $u_{||}^{*}${]} as well as cell-face
evaluation of the gradients {[}second order centered differencing{]}
of velocity-space metrics {[}e.g., $\partial_{\xi}v^{*}$ and $\partial_{\xi}u_{||}^{*}${]}).
We note that the solution in the \emph{physical} \emph{space} is compared
by interpolating the reference solution to the coarse grid. The reference
solution uses a grid of $N_{\xi}=384$, $N_{v_{||}}=32$, $N_{v_{\perp}}=16$,
and a time-step size of $\Delta t=0.025$ and is time integrated to
$t_{max}=0.25.$ The convergence study is performed by fixing the
time-step size and velocity-space resolution while varying the number
of configuration-space grid points.

Similarly to the configuration space, the velocity-space discretization
also relies on the SMART scheme for the flux interpolation. However,
due to the linear interpolation used elsewhere (e.g., interpolation
of advection coefficients to the cell face), the overall discretization
is second order as is seen in Fig. \ref{fig:Doubly-reflective-boundary-convergence-study}-right.
The reference solution uses a grid of $N_{\xi}=24$, $N_{v_{||}}=512$,
$N_{v_{\perp}}=256$, a time-step size of $\Delta t=0.025$, and time
integrated to $t_{max}=0.25.$ The convergence study is performed
by fixing the time-step size and configuration-space grid while uniformly
refining the transformed velocity-space grid in both $\widetilde{v}_{||}$
and $\widetilde{v}_{\perp}$. 

\subsection{Guderley problem (time-dependent Maxwellian Dirichlet boundary conditions)
\label{subsec:gudurley}}

\begin{figure}
\begin{centering}
\includegraphics[scale=0.62]{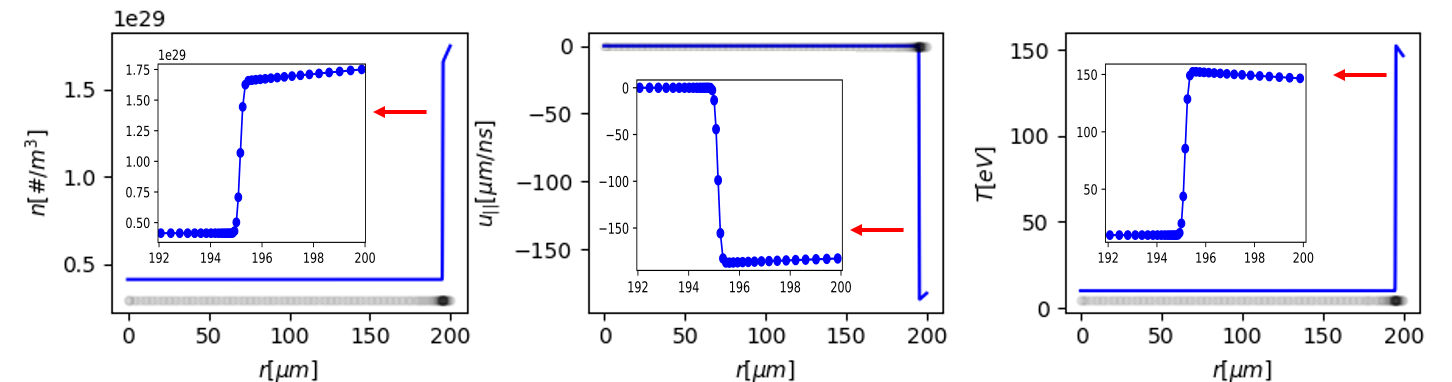} 
\par\end{centering}
\begin{centering}
\includegraphics[scale=0.66]{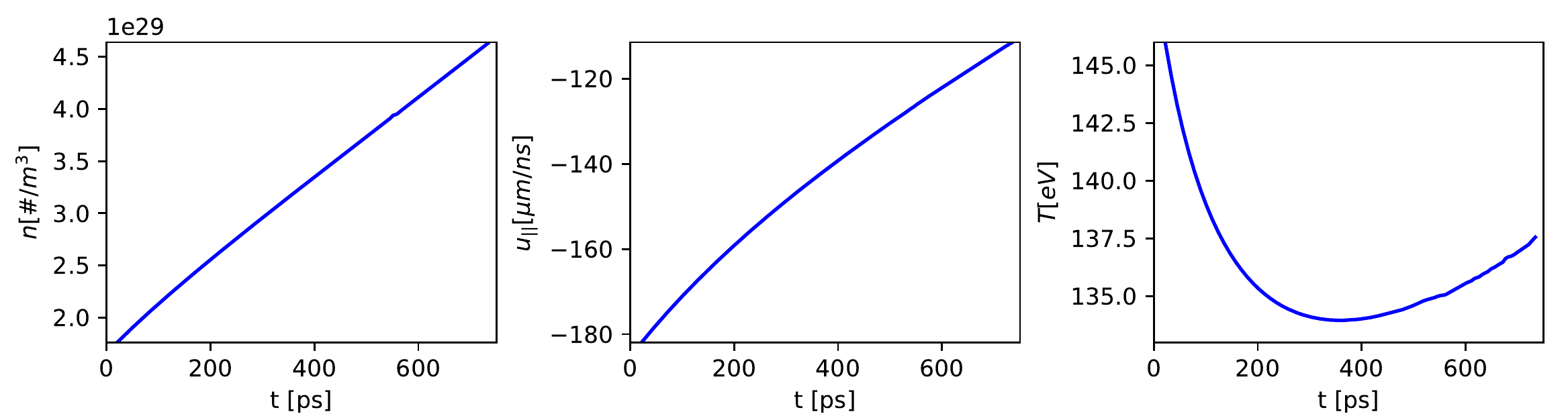} 
\par\end{centering}
\caption{Guderley problem: The initial condition (top) and the time-dependent
Maxwellian boundary conditions (bottom) for the number density (left),
radial drift velocity (center), and temperature (right), obtained
from semi-analytically solving the Guderley problem in Ref. \citep{lazarus_siam_jna_guderley}.
For the initial condition, the black markers denotes the grid density
(i.e., coarse grid regions are lighter in color than finer regions).
The boundary temperature is assumed to be equilibrated across ions
and electrons at all time. \label{fig:guderley_ic_and_bc}}
\end{figure}

We consider the Guderley problem \citep{guderley_luftfahrtf_1942_guderley_problem,lazarus_siam_jna_guderley,vallet_pop_2013_finite_mach_number_guderley,ramsey_2017_JVVUQ_guderley}
of a converging/diverging shock to test the proposed time-dependent
boundary conditions in spherical geometry and to demonstrate the ability
of the algorithm to capture the hydrodynamic asymptotic limit for
short mean-free-paths. Semi-analytic solutions to the Euler equations
exist for the Guderley problem (see \ref{app:gudurley_problem} for
details), which we will appeal to for verification. Suppose that a
strong shock is spherically converging into a uniform, Deuterium-Tritium,
$\langle\text{DT}\rangle$, plasma with constant mass density of $\rho_{0}=m_{\left\langle DT\right\rangle }n_{\left\langle DT\right\rangle }=0.173$
{[}$g/cc${]} and temperature $T_{0}=10$ {[}$eV${]} (i.e., conditions
relevant to double-shell and revolver ICF implosions\citep{kirkpatrick_ppcf_1975,kirkpatrick_ppcf_1981,colgate_1988_laur,colgate_1992_laur,lackner_aip_conf_proc_1994,amendt_pop_2002,amendt_pop_2007,montgomery_pop_2018,molvig_2016_prl_revolver,molvig_2018_pop_revolver}).
We model the $\langle\text{DT}\rangle$ plasma as an equimolar, averaged
single ion, fully ionized, and initialize the distribution function
with density, drift velocity, and pressure/temperature profiles obtained
from a Guderley solution. The Guderley profiles are obtained 20 {[}$ps${]}
after the shock was initialized from a radius of $R_{0}=200$ {[}$\mu m${]},
and the Mach number there is $M_{0}=7.23$. Similarly, the hydrodynamic
boundary conditions for the spherical domain are sourced from the
same Guderley solution. The logical grid sizes for these calculations
are chosen as $N_{\xi}=$192 and 384, $N_{v_{\perp}}=64$ and $N_{v_{||}}=128$
with domain limits $\xi\times\widetilde{v}_{||}\times\widetilde{v}_{\perp}\in\left[0,1\right]\times\left[-8,8\right]\times\left[0,8\right]$.
The profiles for the initial and the boundary conditions for the moments
are shown in Fig. \ref{fig:guderley_ic_and_bc}. 

\begin{figure}[h]
\centering{}\includegraphics[scale=0.4]{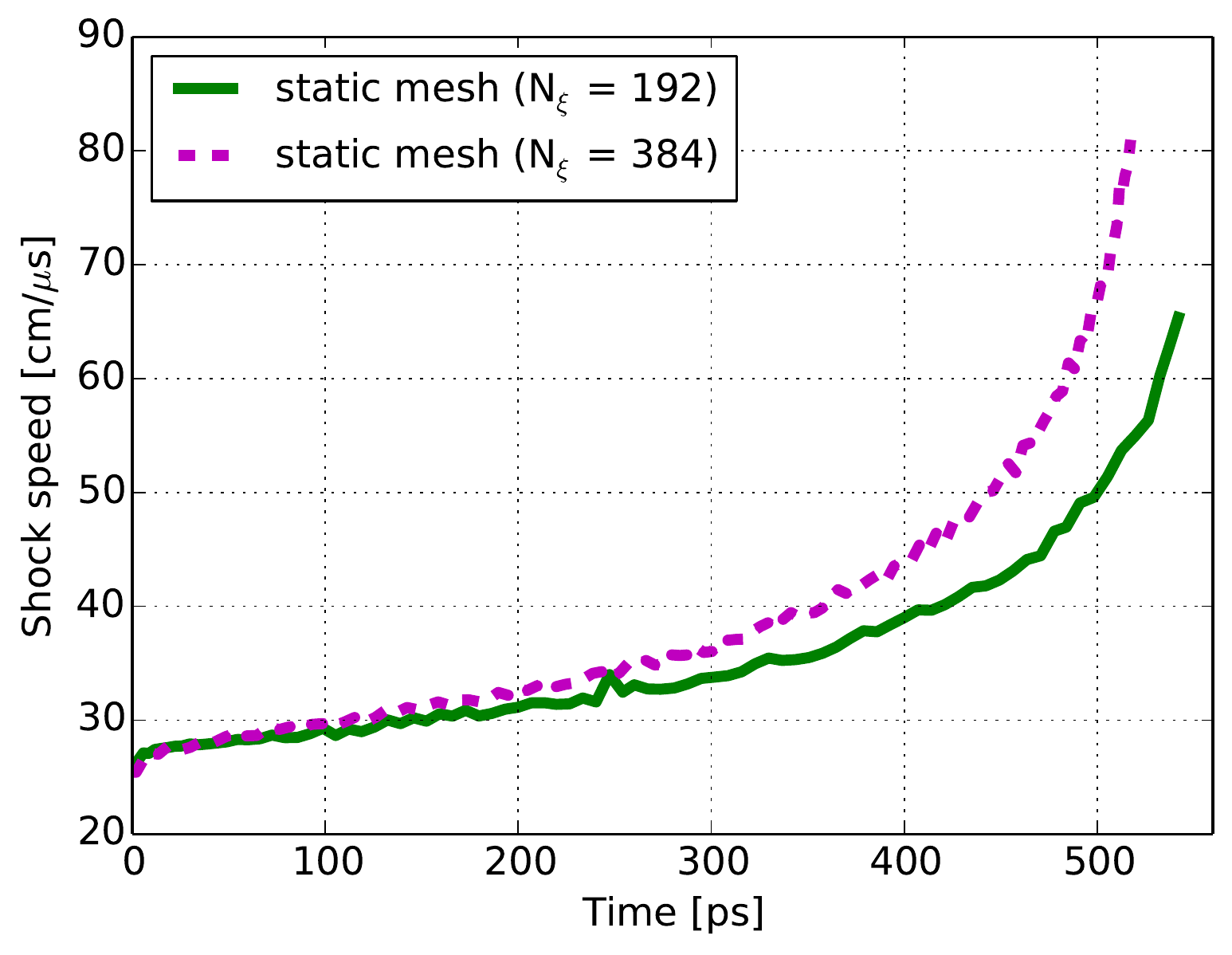}\includegraphics[scale=0.4]{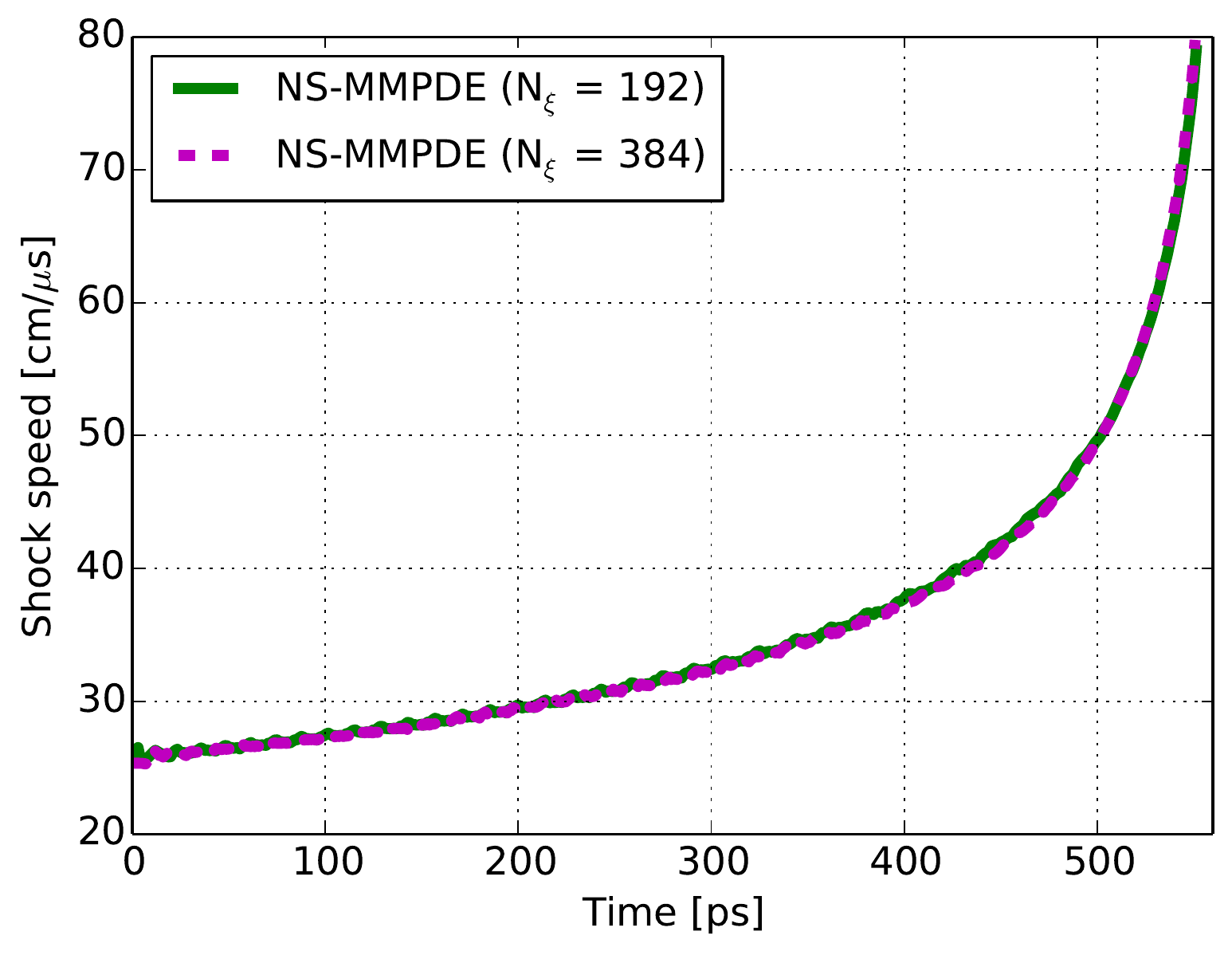}\caption{Guderley problem: The shock speed versus time \textcolor{black}{with
a static and uniform configuration-space grid} (left) and a moving
mesh with an optimized initial mesh generation (right).\label{fig:stat_nx_comp_time_speed}}
\end{figure}

\begin{figure}[h]
\centering{}\includegraphics[scale=0.4]{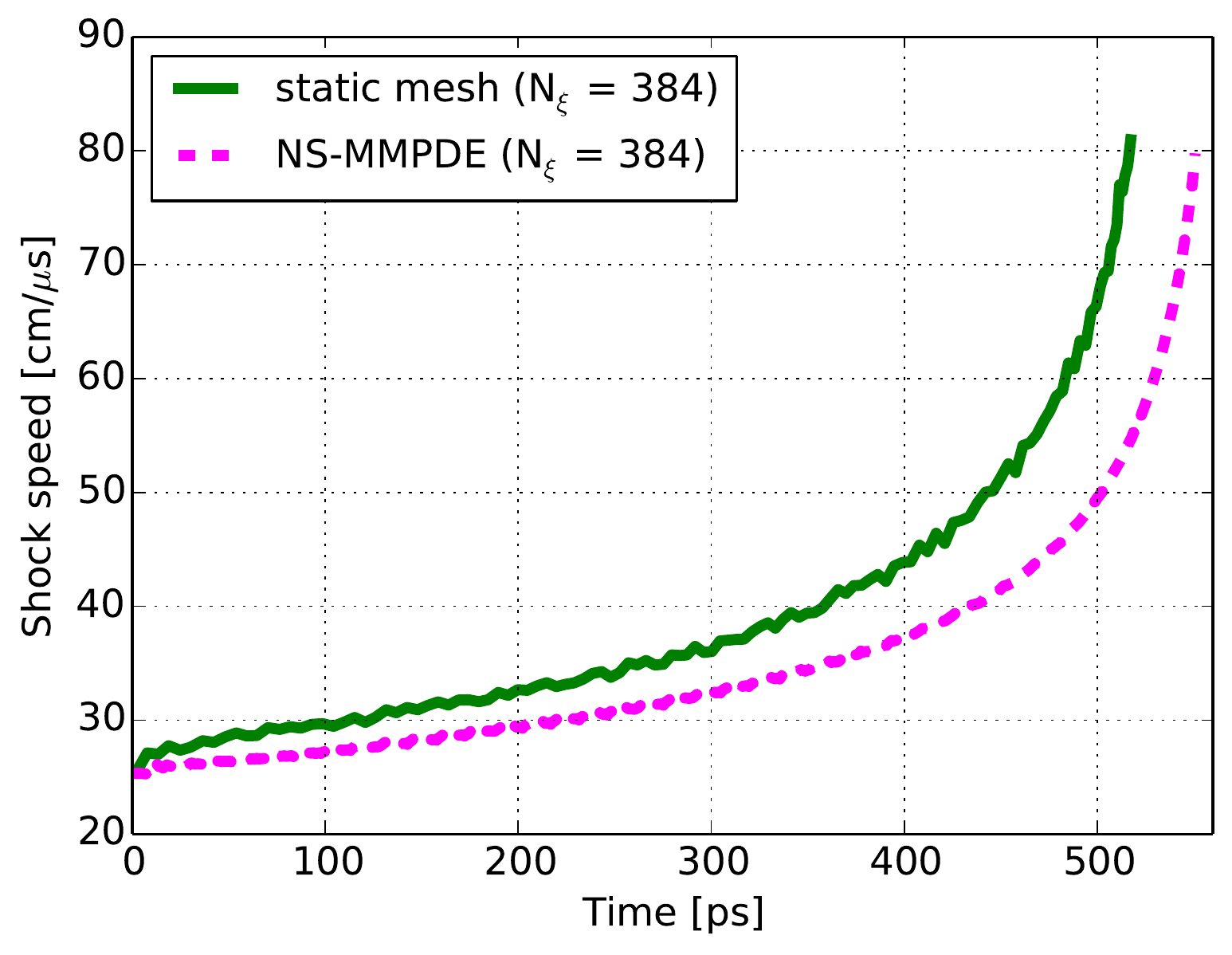} \caption{Guderley problem: The shock speed versus time comparing the two simulations
for $N_{\xi}=384$. Green (solid) uses\textcolor{black}{{} a static
and uniform configuration-space mesh and magenta (dashed) uses NS-MMPDE.\label{fig:384_comp_time_speed}}}
\end{figure}

We start our verification study b\textcolor{black}{y employing a static
and uniform mesh in the configuration space (while keeping the the
velocity-space mesh adaptive). The purpose of this case is to serve
as the baseline to highlight the computational savings and the accuracy
gains afforded by the fully adaptive phase-space mesh. For the verification
metric, we choose the shock speed, }
\begin{equation}
u_{s}=\left|\frac{dr_{s}}{dt}\right|,\label{eq:shock_speed}
\end{equation}
where $r_{s}$ is the shock position. The simulated shock trajectory
is defined either as: 1) the location of the maximum ion temperature,
or 2) the location of the maximum ion viscous (energy) dissipation
rate. The first definition is the most reliable (i.e., it has less
dependence on small-scale fluctuations in the moment quantities),
but only unambiguously identifies the shock location in the converging
phase. For our simulations, we choose definition 1 for the converging
shock phase, and 2 for the diverging phase. We define the viscous
dissipation rate as: 
\begin{equation}
\left(\frac{dT_{i}}{dt}\right)_{\text{visc}}=-\frac{2}{3}\left(\frac{\tau_{||||}}{n_{i}}\right)\frac{\partial_{r}\left(r^{2}u_{||,i}\right)}{r^{2}},\label{eq:shock_location}
\end{equation}
where, $T_{i}$ is the ion temperature, $u_{||,i}$ is the radial
component of the ion drift velocity, $n_{i}$ is the total ion number
density, and $\tau_{||||}$ is the radial component of the ion viscosity
tensor {[}Eq. (\ref{eq:viscous_stress}){]}. Fig. \ref{fig:stat_nx_comp_time_speed}-left
shows a comparison of the shock speed versus time for the two different
spatial-grid resolutions. The two curves show strong deviations from
each other near the shock collapse time of approximately 500 {[}ps{]}.
The noticeable fluctuations in the curves suggest that the simulations
are poorly resolving the shock structure, where the spatial gradients
can change dramatically in the course of the shock's collapse. 
We see a dramatic improvement in the quality of the solution when
we employ an initial grid optimization in conjunction with the NS-MMPDE
strategy with $\delta_{min}=5\times10^{-3}$ and\textbf{ $\lambda_{\omega}=10^{-3}$}.
Indeed, as shown in Fig.\ \ref{fig:stat_nx_comp_time_speed}-right,
the differences between $N_{\xi}=192$ and $N_{\xi}=384$ disappear,
and the fluctuations seen in Fig.\ \ref{fig:stat_nx_comp_time_speed}-left
are nearly eliminated for both cases. Moreover, comparing the static
and the moving mesh cases for $N_{\xi}=384$ reveals the origin of
the deviations in the shock trajectory near the collapse time: As
seen in Fig.\ \ref{fig:384_comp_time_speed}, the shock trajectory
with the uniform mesh veers away from the correct trajectory immediately
after the simulation begins. 
The initial shock structure is not well resolved in the static-mesh
simulation, and this seems to be the cause of the strong deviation.
The drifting Maxwellian for the temperature, density, and drift velocity
are not the true distribution functions for the shock, and this leads
to the production of transients in the solution which, if not properly
resolved, can cause the solution to deviate strongly from initial
condition.

\begin{figure}[h]
\centering{}\includegraphics[scale=0.4]{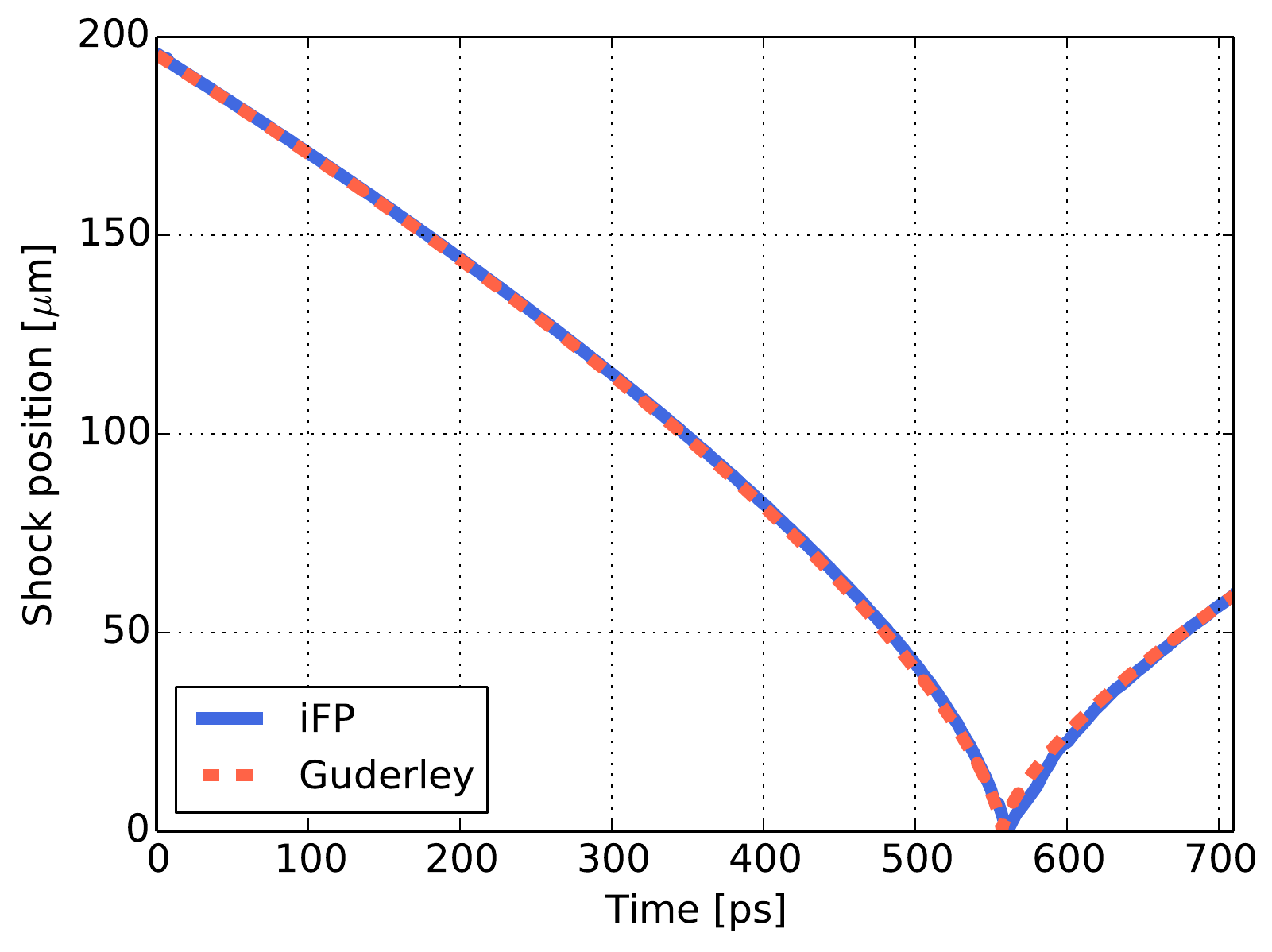}\includegraphics[scale=0.4]{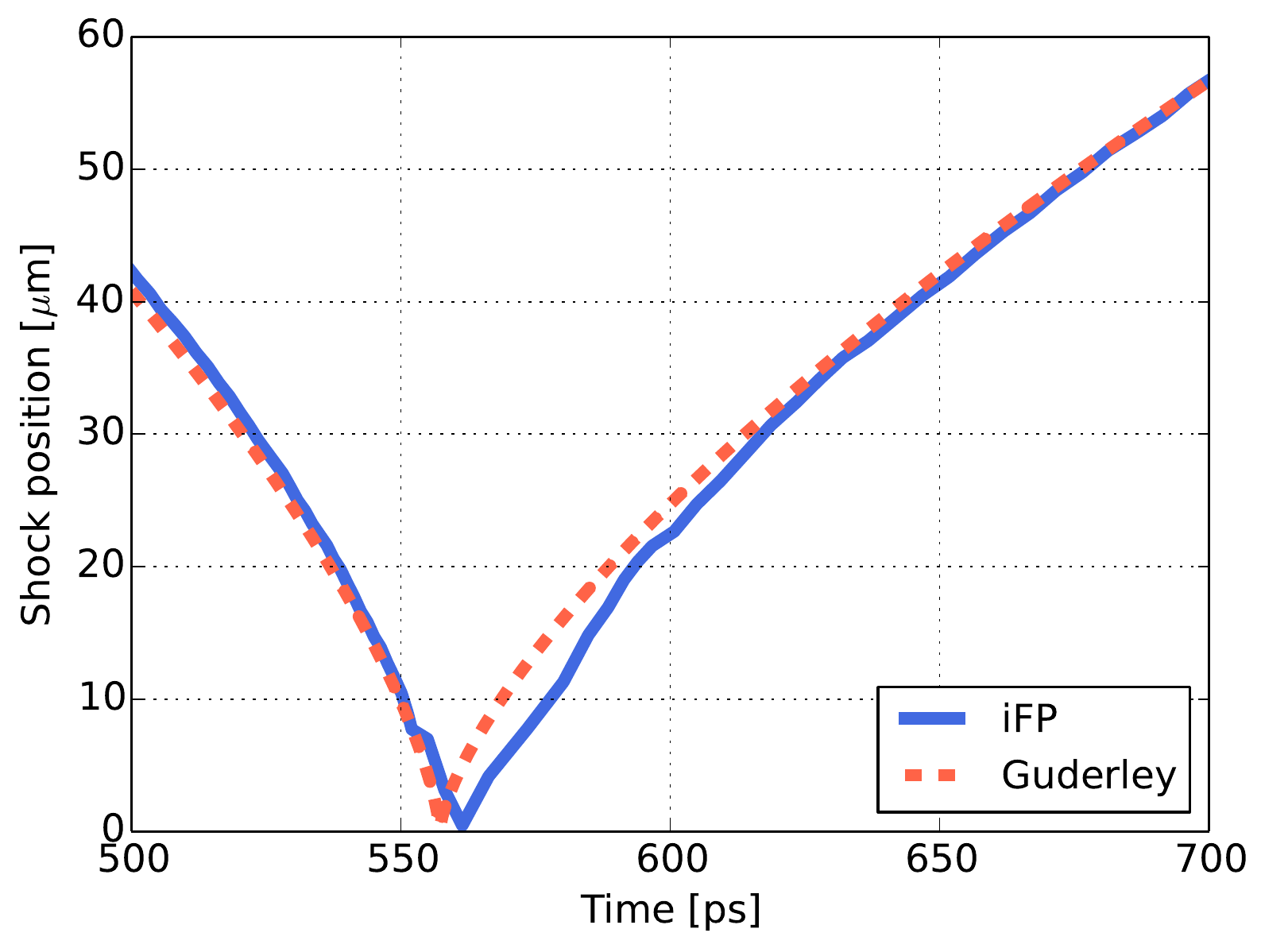}\caption{Guderley problem: The shock trajectory vs.\ time. The blue (``solid'')
curve is obtained from our algorithm for $N_{\xi}=192$ with the proposed
NS-MMPDE and initial grid optimization algorithms, and the red (``dashed'')
curve is the semi-analytical Guderley prediction over the entire simulation
(left) and the isolated time in the vicinity of the shock collapse
(right).\label{fig:gud_comp_shock_traj}}
\end{figure}

\begin{figure}[h]
\begin{centering}
\includegraphics[scale=0.3]{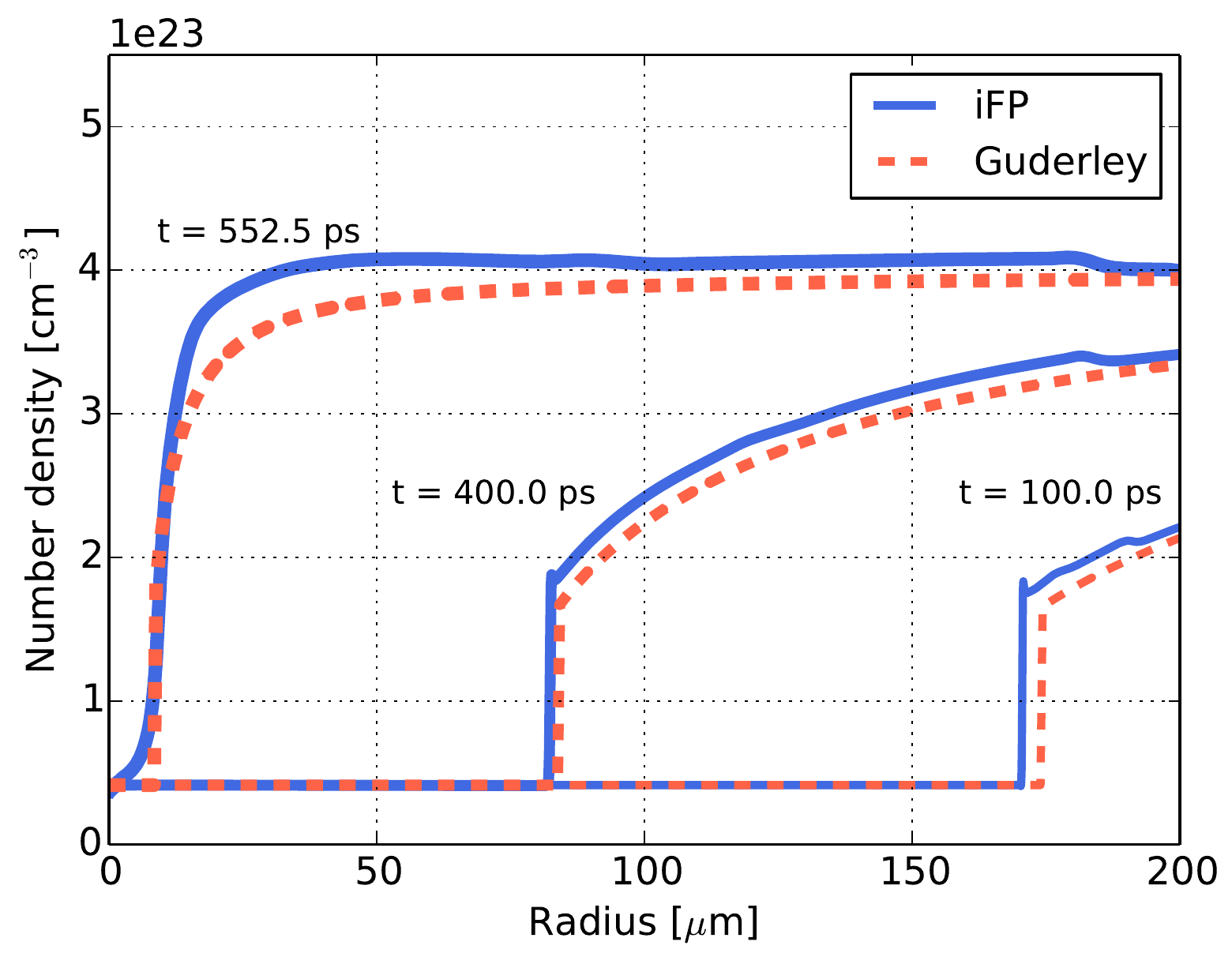}\includegraphics[scale=0.3]{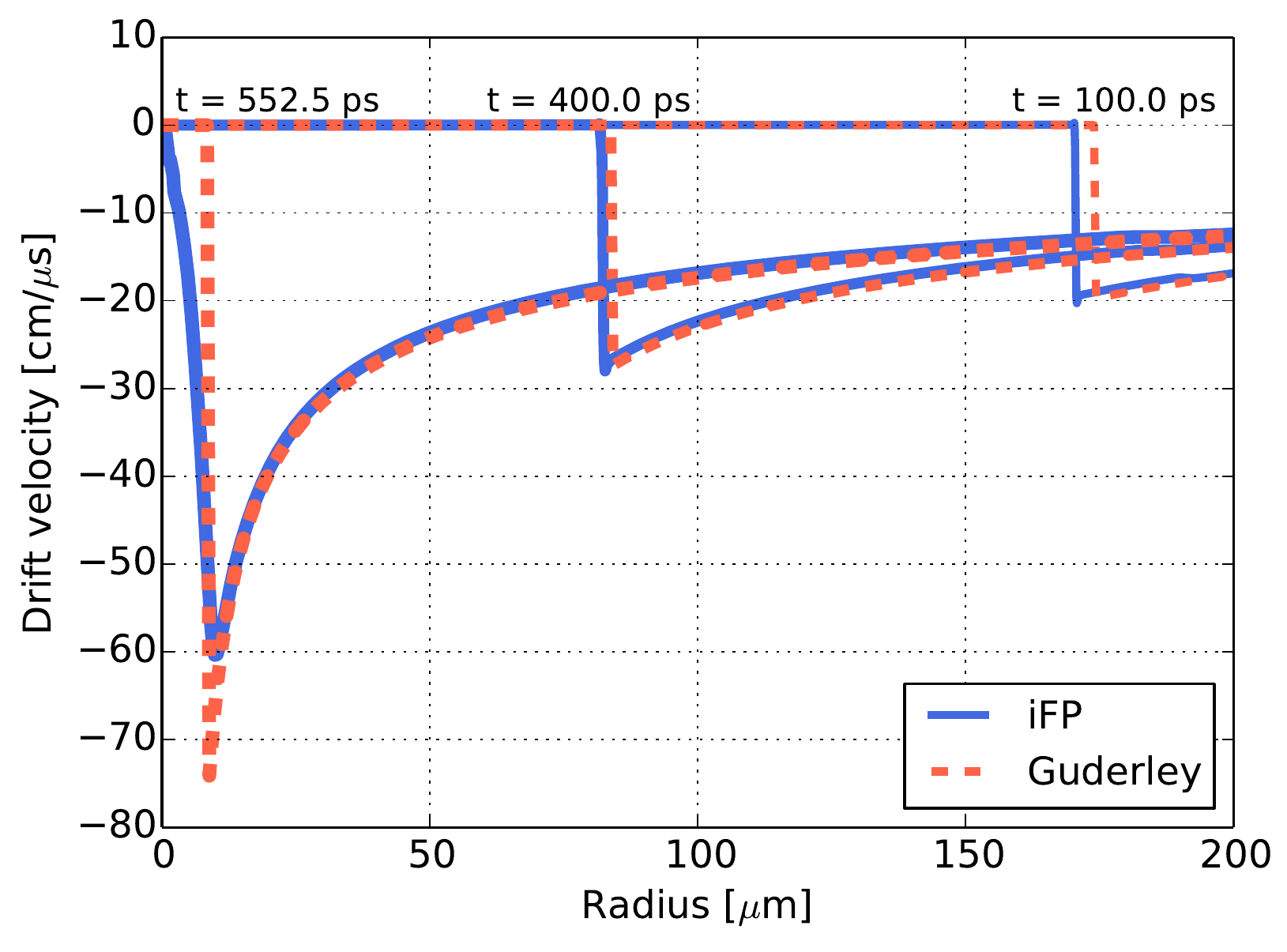}\includegraphics[scale=0.3]{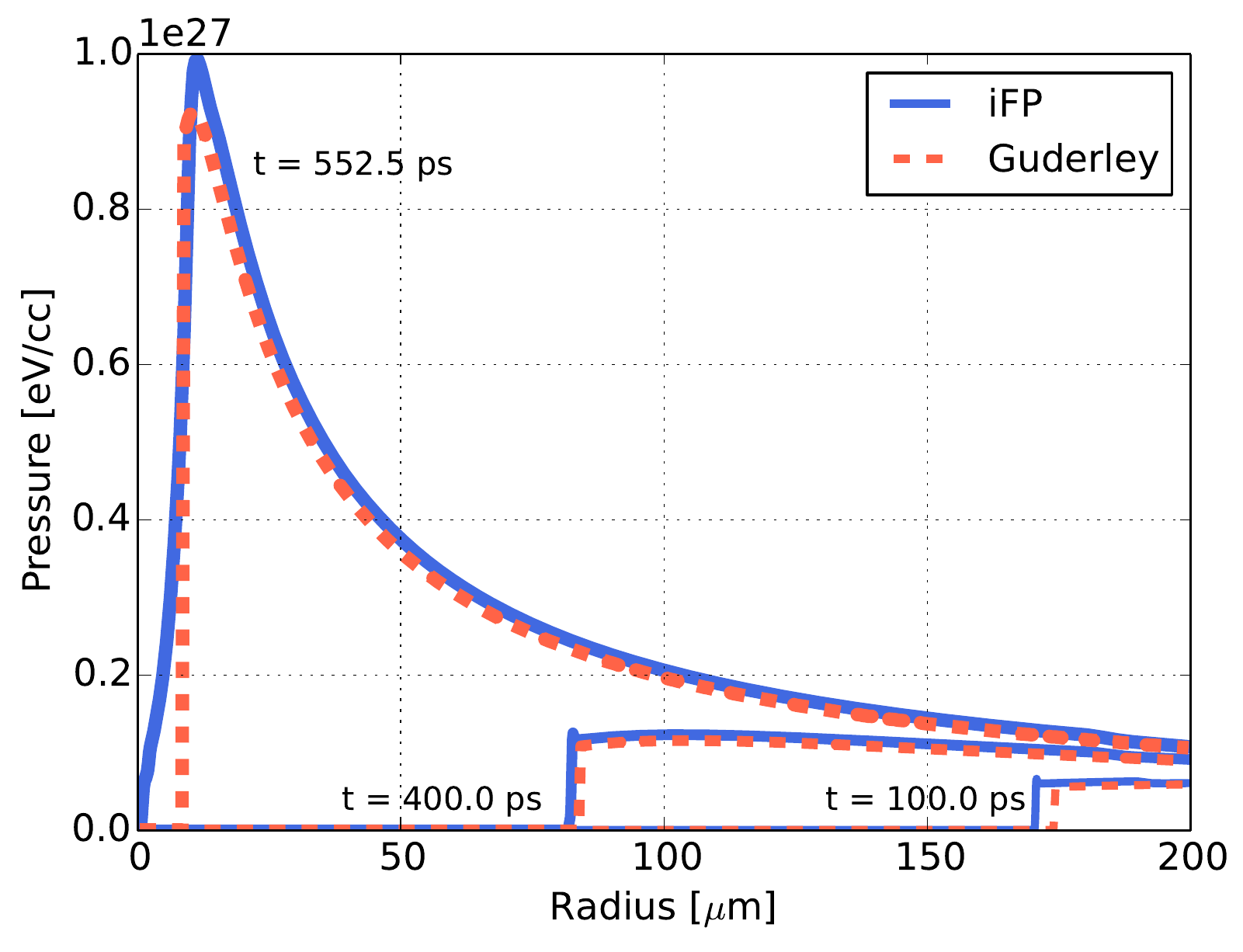} 
\par\end{centering}
\begin{centering}
\includegraphics[scale=0.3]{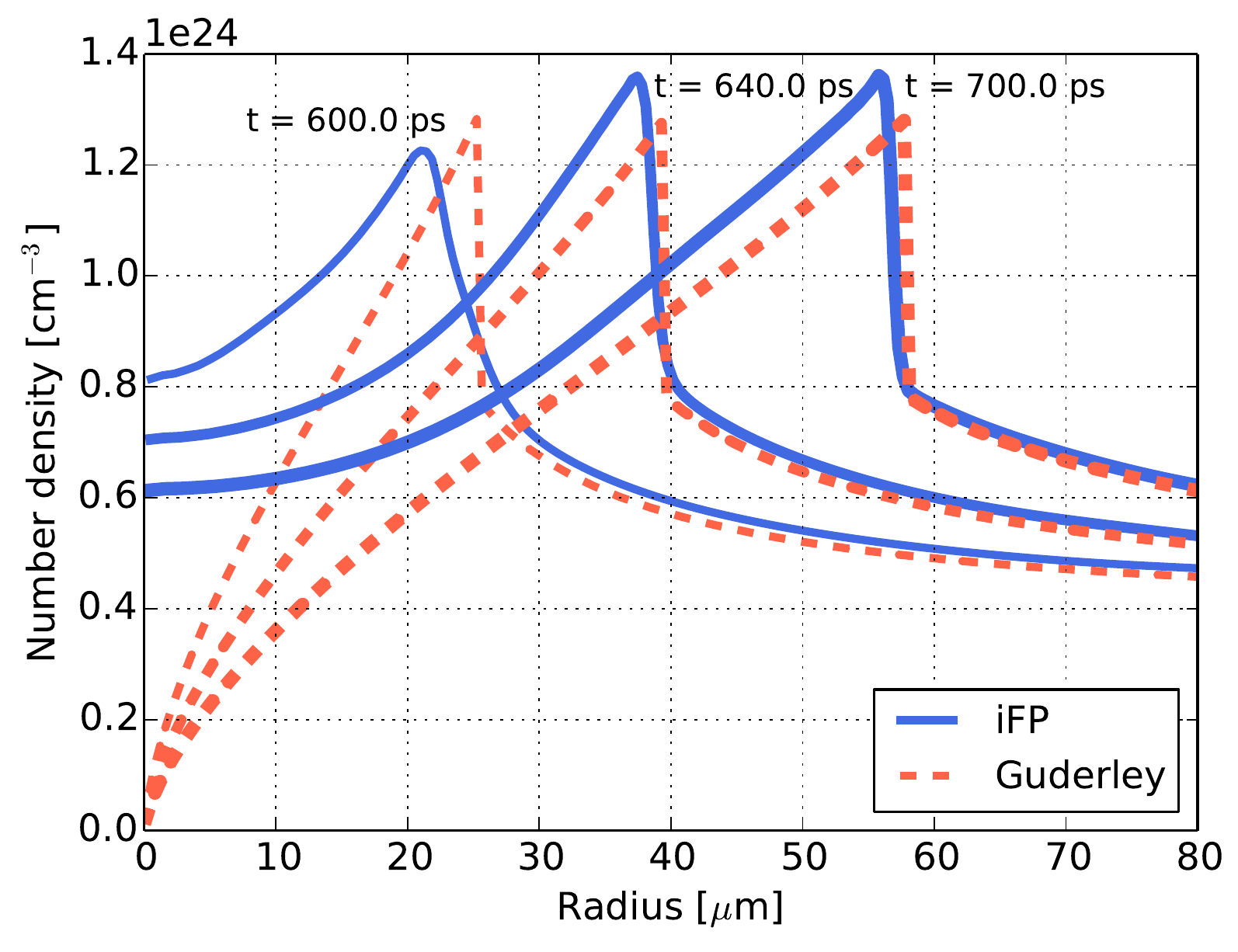}\includegraphics[scale=0.3]{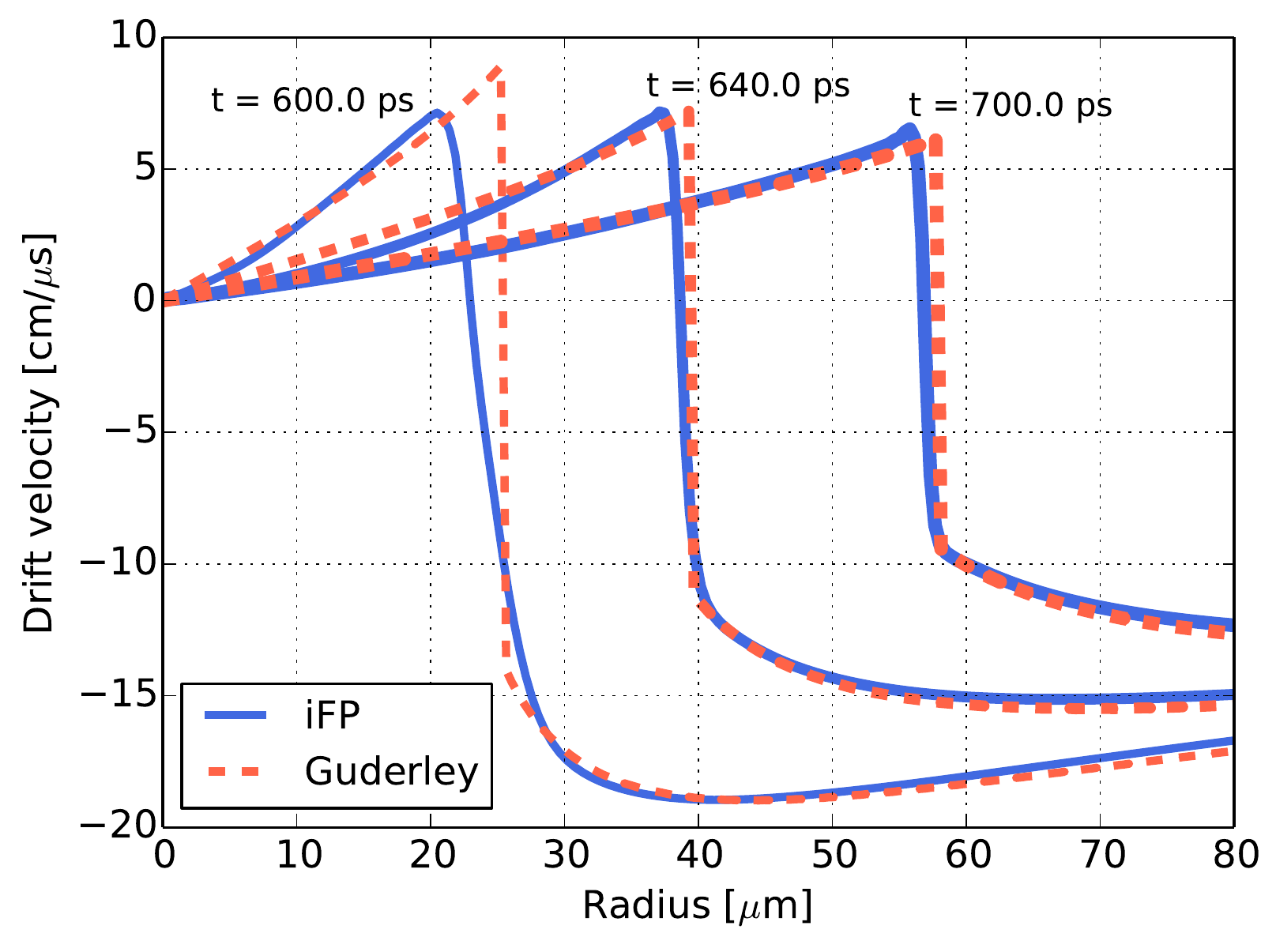}\includegraphics[scale=0.3]{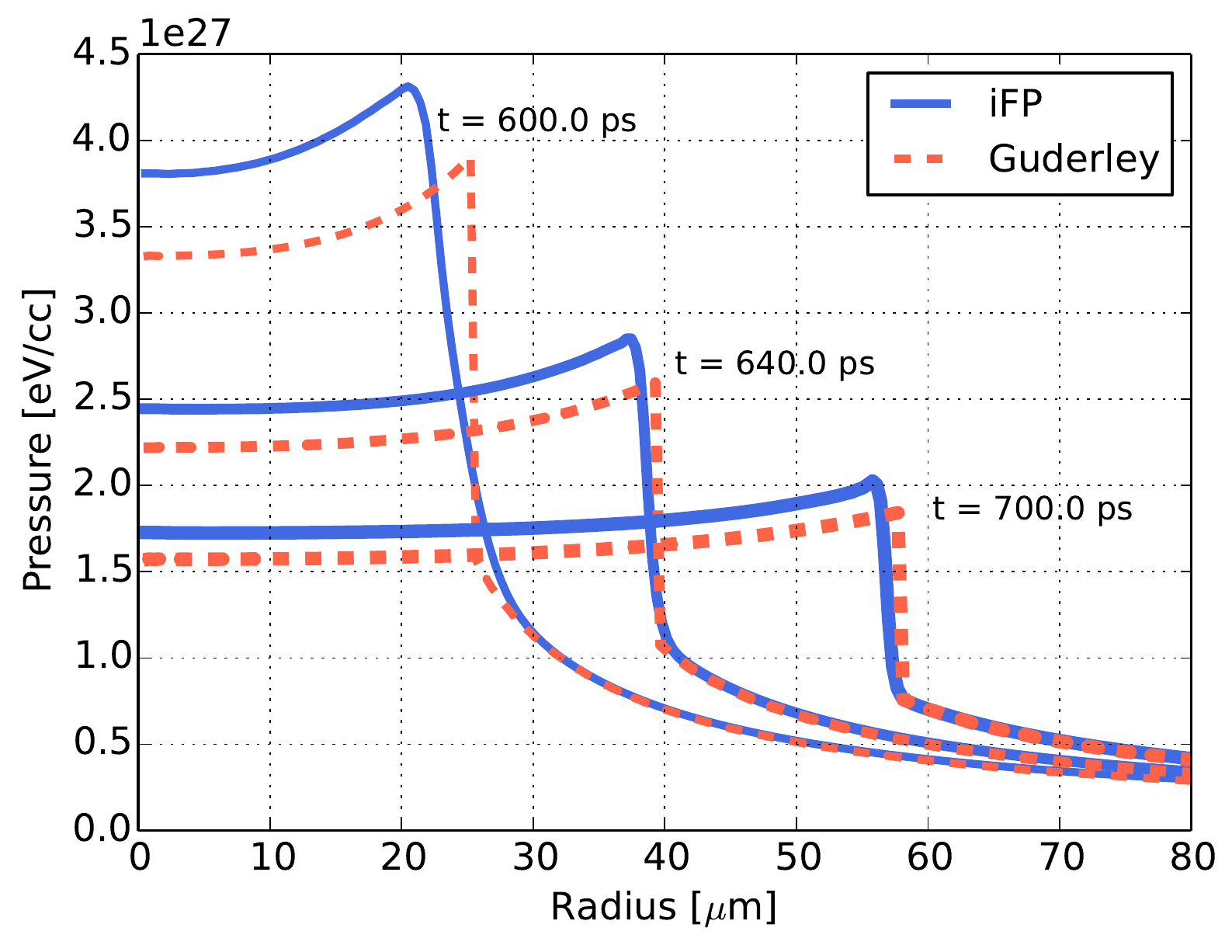} 
\par\end{centering}
\caption{Guderley problem: Pre-collapse (top) and post-collapse (bottom) hydro
profiles for number density (left), drift velocity (center), and total
hydrodynamic pressure (right) from simulation (``solid'') and Guderley
(``dashed'').\label{fig:gud_pre}}
\end{figure}

We expect that such a high $\langle\text{DT}\rangle$ density ($0.173$
{[}$g/cc${]}) justifies the use of boundary conditions sourced from
a dissipation-less fluid theory (i.e., Euler equations), and that
a meaningful comparison to this theory is actually possible. To this
end, we compare our $N_{\xi}=192$ NS-MMPDE results to the corresponding
Guderley solution. Fig. \ref{fig:gud_comp_shock_traj}-left shows
the shock trajectory as a function of time from the simulation and
the semi-analytic Guderley prediction. We see that the shock collapse
times agree very well. Remarkably, the Guderley shock collapse time
and the simulation agree to within $0.75\%$. 
However, if we take a closer look in the neighborhood of the collapse,
as depicted in the ``zoomed in'' trajectory plot shown in Fig. \ref{fig:gud_comp_shock_traj}-right,
we see that the Guderley and simulated shock trajectories differ in
the early post-collapse phase. Insight into this discrepancy may be
gained by directly comparing the simulation and Guderley hydrodynamic
profiles. Fig. \ref{fig:gud_pre}-top shows pre-collapse profiles
for density (left), drift velocity (center), and total hydrodynamic
pressure, $P=n_{i}T_{i}+n_{e}T_{e}$ (right). We see that there is
some separation between the simulation and Guderley densities at an
early time (t = 100 {[}$ps${]} after the simulation is initialized).
Moreover, the transients created in the simulation initialization
have yet to be fully eliminated. By t = 400 {[}$ps${]}, however,
the simulation and Guderley profiles largely agree, although we see
some remaining differences near the shock collapse time (t = 552.5
{[}$ps${]}), which persist into the diverging phase -- as depicted
in Fig. \ref{fig:gud_pre}-bottom. Nevertheless, the qualitative features
are nearly identical between simulation and Guderley. 

The strong deviations in the hydro profiles seen in Fig. \ref{fig:gud_pre}-bottom
hint at the origin of the trajectory discrepancy in the post-collapse
phase -- as shown in Fig.\ \ref{fig:gud_comp_shock_traj}-right.
The Guderley solution predicts a zero density and (infinite temperature)
at the origin at the shock collapse time. This behavior is unphysical,
and it results from the infinite collisionality assumption in the
Euler equations \citep{vallet_pop_2013_finite_mach_number_guderley}.
The VFP equation models the full dissipation physics, and thus captures
the correct behavior near shock collapse (where ion heat conduction,
viscosity, and kinetic effects become important).

\begin{figure}[h]
\begin{centering}
\includegraphics[scale=0.5]{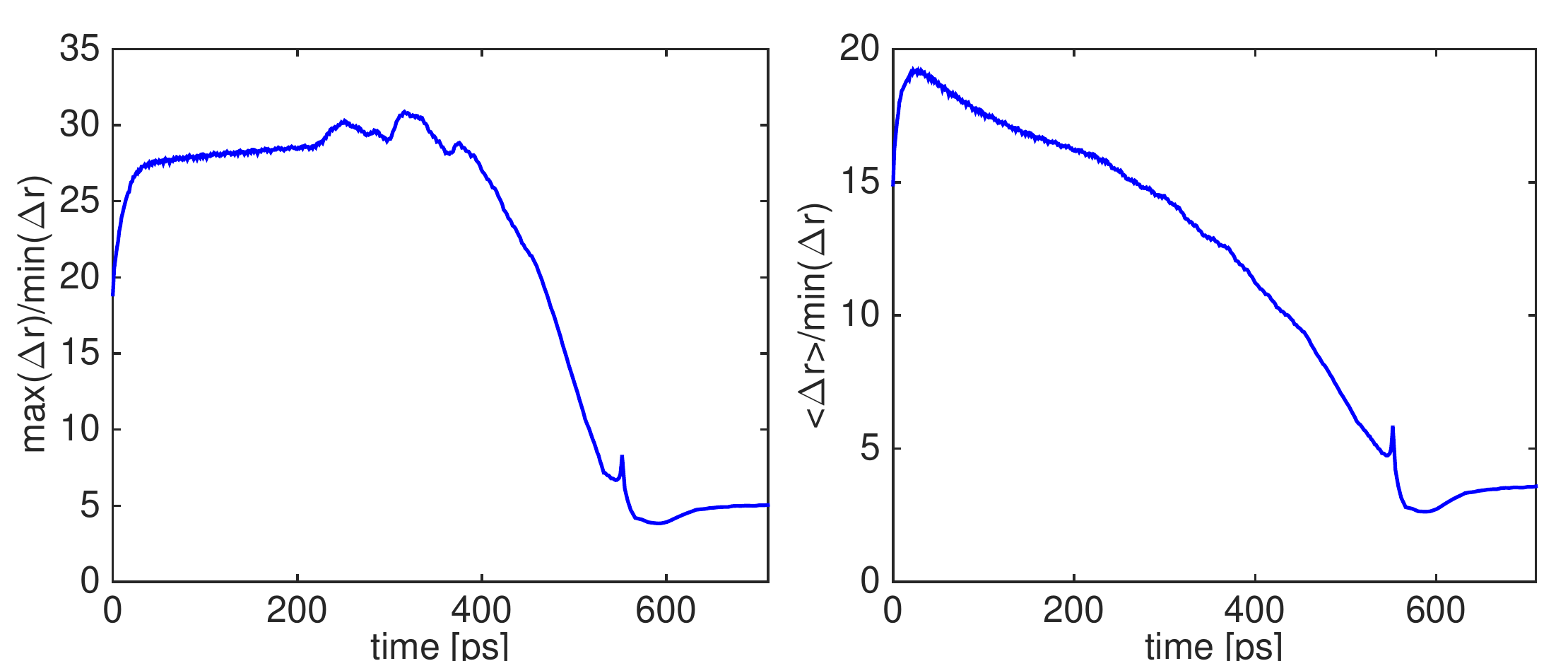} 
\par\end{centering}
\caption{Guderley problem: Ratios of maximum to minimum (left) and average
to minimum (right) grid sizes versus time.\label{fig:guderley_dr_ratios}}
\end{figure}

\begin{figure}[h]
\begin{centering}
\includegraphics[scale=0.5]{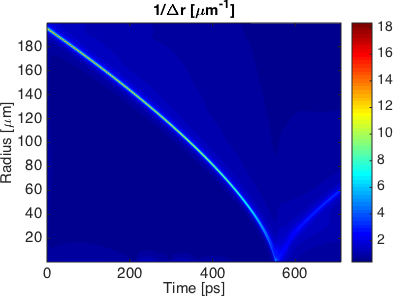} 
\par\end{centering}
\caption{Guderley problem: RT diagram of $1/\Delta r$ for $N_{\xi}=192$.\label{fig:guderley_inv_dr_vs_rt} }
\end{figure}

\begin{figure}[h]
\begin{centering}
\includegraphics[scale=0.7]{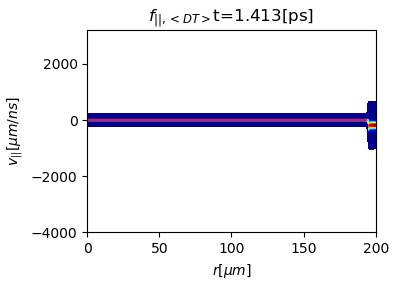}\includegraphics[scale=0.7]{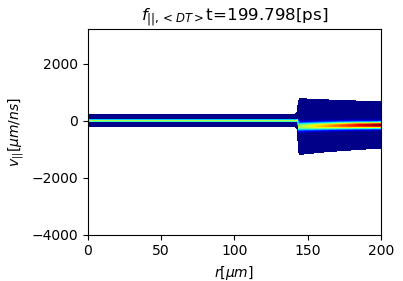} 
\par\end{centering}
\begin{centering}
\includegraphics[scale=0.7]{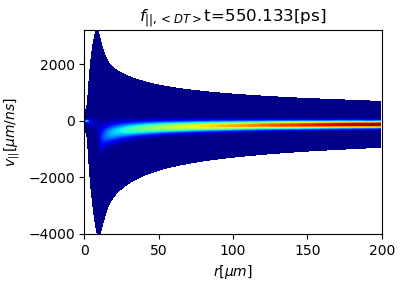}\includegraphics[scale=0.7]{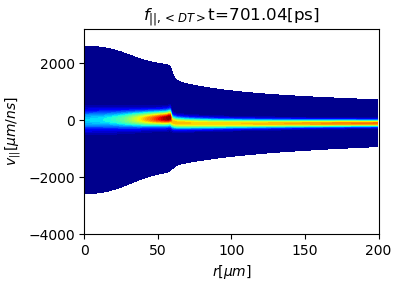} 
\par\end{centering}
\caption{Guderley problem: Parallel velocity-space marginal distribution functions
at $t\approx0,\;200\;,550,\;700$ {[}ps{]}.\label{fig:pdf_perpendicular} }
\end{figure}

In Fig. \ref{fig:guderley_dr_ratios} we show, as a function of time,
the $\max\left(\Delta r\right)/\min\left(\Delta r\right)$ and $\left\langle \Delta r\right\rangle /\min\left(\Delta r\right)$
(where $\left\langle \Delta r\right\rangle =\sum_{i=1}^{N_{\xi}}\Delta r_{i}/N_{\xi}$)
to demonstrate the computational complexity reduction (relative to
a uniform grid) afforded by the proposed NS-MMPDE scheme. 
As can be seen, both ratios exhibit large values and a reduction of
approximately 20 in the computational complexity is achieved by the
mesh motion for the Guderley problem. In Fig. \ref{fig:guderley_inv_dr_vs_rt},
a Lagrangian radius-time (RT) diagram of $1/\Delta r$ is shown to
demonstrate how the grid tracks the shock front as a function of time.
As can be seen, the grid tightly concentrates near the sharp converging
shock front, and diffuses away after shock collapse due to kinetic
effects smearing the gradients. Finally, to demonstrate the computational
complexity reduction afforded by the velocity-space adaptivity scheme,
we present the parallel velocity-space marginal distribution function,
\begin{equation}
f_{||,<DT>}=2\pi\int_{0}^{\infty}dv_{\perp}v_{\perp}f_{\left\langle DT\right\rangle },
\end{equation}
in the $r-v_{||}$ plane in Fig. \ref{fig:pdf_perpendicular}. 

For the calculation, $\max\left(v_{th}\right)/\min\left(v_{th}\right)\approx15,$
thus a total computational complexity reduction of $\max$ $\left(\frac{\left\langle \Delta r\right\rangle }{\min\left(\Delta r\right)}\right)$
$\left(\frac{\max\left(v_{th}\right)}{\min\left(v_{th}\right)}\right)^{2}$
$\approx4500$ is achieved through the moving phase-space mesh stra\textcolor{black}{tegy.
We stress that, due to the large velocity scales encountered in the
problem, it is not possible to simulate the present problem on a static
and uniform velocity-space grid. We }remind the readers that this
problem is intended to be a verification of the proposed numerical
method to recover the hydrodynamic solution and that to our knowledge,
these results are unprecedented for a kinetic code. Because of the
nature of the problem, kinetic effects are not significant and the
true capability of the proposed algorithm is highlighted later in
a multi-scale ICF capsule implosion simulation (Sec. \ref{subsec:fuel_only_compression_yield}),
where kinetic physics manifest in a non-trivial manner. 

\subsection{Van Dyke problem (elastic moving wall boundary condition)\label{subsec:gudurley_moving_wall}}

Consider a spherical piston surrounding a gas at constant pressure
and density. At $t=0$, the piston begins moving at a constant velocity
inward. A shock is created in the gas, and converges to the center.
We simulate this setup directly, using the moving elastic wall boundary
condition discussed in Sec. \ref{subsubsec:symmetry_elastic_bc}.
Once more, we consider a fully ionized $\langle\text{DT}\rangle$
gas with an initial mass density of $\rho_{\left\langle DT\right\rangle }=m_{\left\langle DT\right\rangle }n_{\left\langle DT\right\rangle }=0.173$
{[}$g/cc${]} and temperature of $T_{\left\langle DT\right\rangle }=10$
{[}$eV${]}. The piston velocity was chosen to be $u_{W}=-19$ {[}$cm/\mu s${]}
inwards, and the initial radius was $R_{0}=326$ {[}$\mu m${]}. To
avoid any violent, abrupt changes in the ion distribution function
at the wall, we accelerate the wall to the desired piston velocity
(-19 {[}$cm/\mu$s{]}) at a constant rate. Considering two different
acceleration times (2.5 and 5 {[}$ps${]}), we found negligible differences
in the final solution. Our convergence study begins with a static
mesh ($N_{\xi}=192$; $N_{v_{\perp}}=64$; $N_{v_{||}}=128$).

\begin{figure}[h]
\centering{}\includegraphics[scale=0.4]{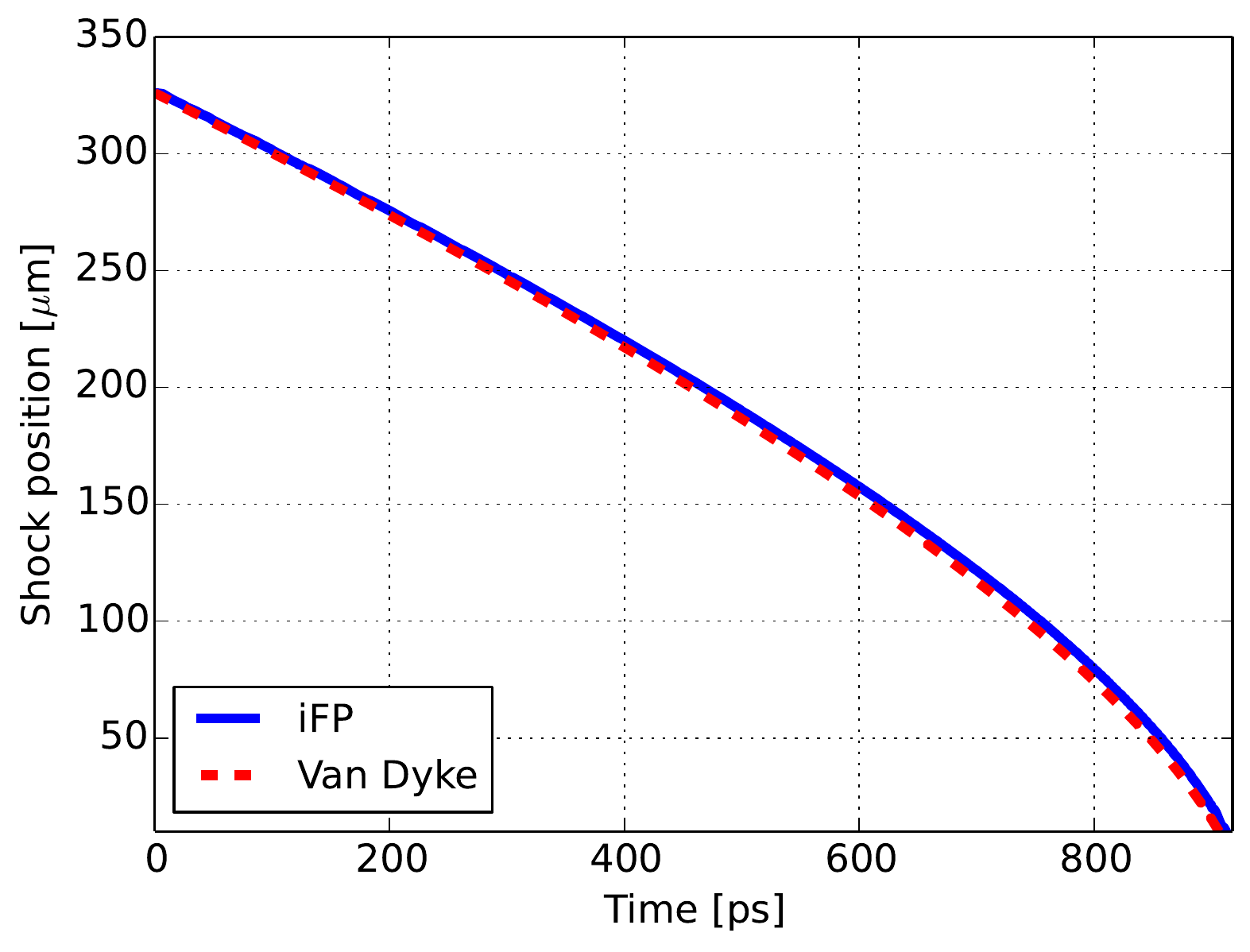}\includegraphics[scale=0.4]{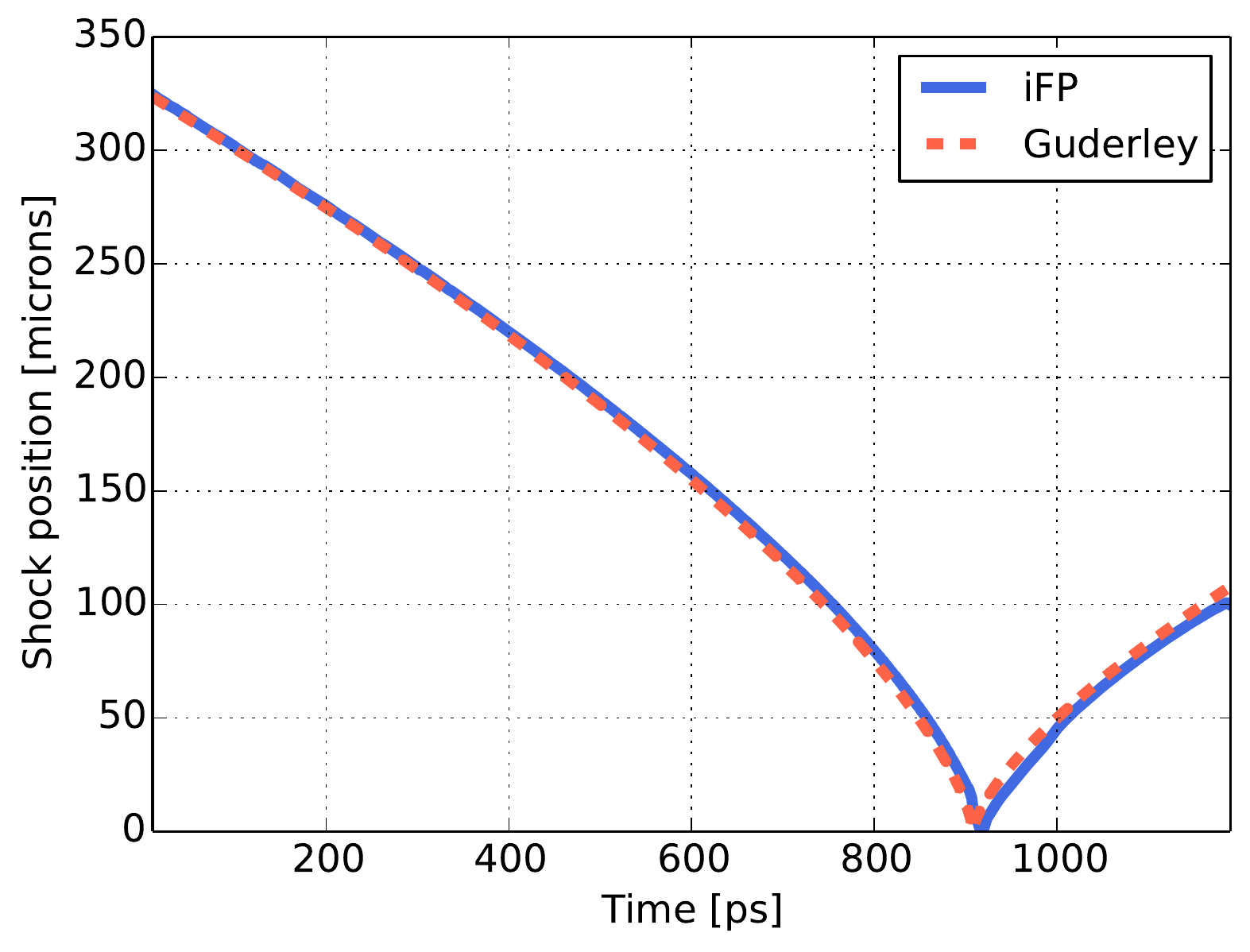}
\caption{Van Dyke problem: The shock trajectory versus time. The blue (``solid'')
curve is obtained from simulation using the elastic moving wall boundary
condition, and the red (``dashed'') curve is the Van Dyke (left)
and Guderley (right) predictions. \label{fig:moving_wall_shock_traj}}
\end{figure}

This simulation predicted a shock collapse time of 938 {[}ps{]}, which
is 30 {[}$ps${]} later than the Van Dyke prediction \citep{van_dyke_JFM_1982_van_dyke_problem}
of 908 {[}$ps${]}. Nonetheless, this is only about a $3\%$ difference.
The collapse time further improves when we use the NS-MMPDE algorithm,
going from $938\rightarrow928$ {[}$ps${]}. This figure is largely
insensitive to the grid resolution, and the non-linear convergence
tolerance; changing only to 927 {[}$ps${]} for $N_{\xi}=384$ (keeping
all other parameters constant), and 930 {[}$ps${]} when $N_{\xi}=192$
and the relative tolerance is changed from $\epsilon_{rel}=10^{-2}$
to $\text{\ensuremath{\epsilon_{rel}=}}10^{-3}$. The largest gain
was attained by enhancing the grid resolution near the piston using
an initial grid optimization. We once more used $N_{\xi}=192$. In
this case, the simulated collapse time was 909 {[}$ps${]}, putting
the agreement with the Van-Dyke solution within $\approx0.1\%$. Fig.
\ref{fig:moving_wall_shock_traj}-left shows that this is not just
a coincidence; the simulated shock trajectory precisely matches that
of Van Dyke. 

\begin{figure}[h]
\begin{centering}
\includegraphics[scale=0.3]{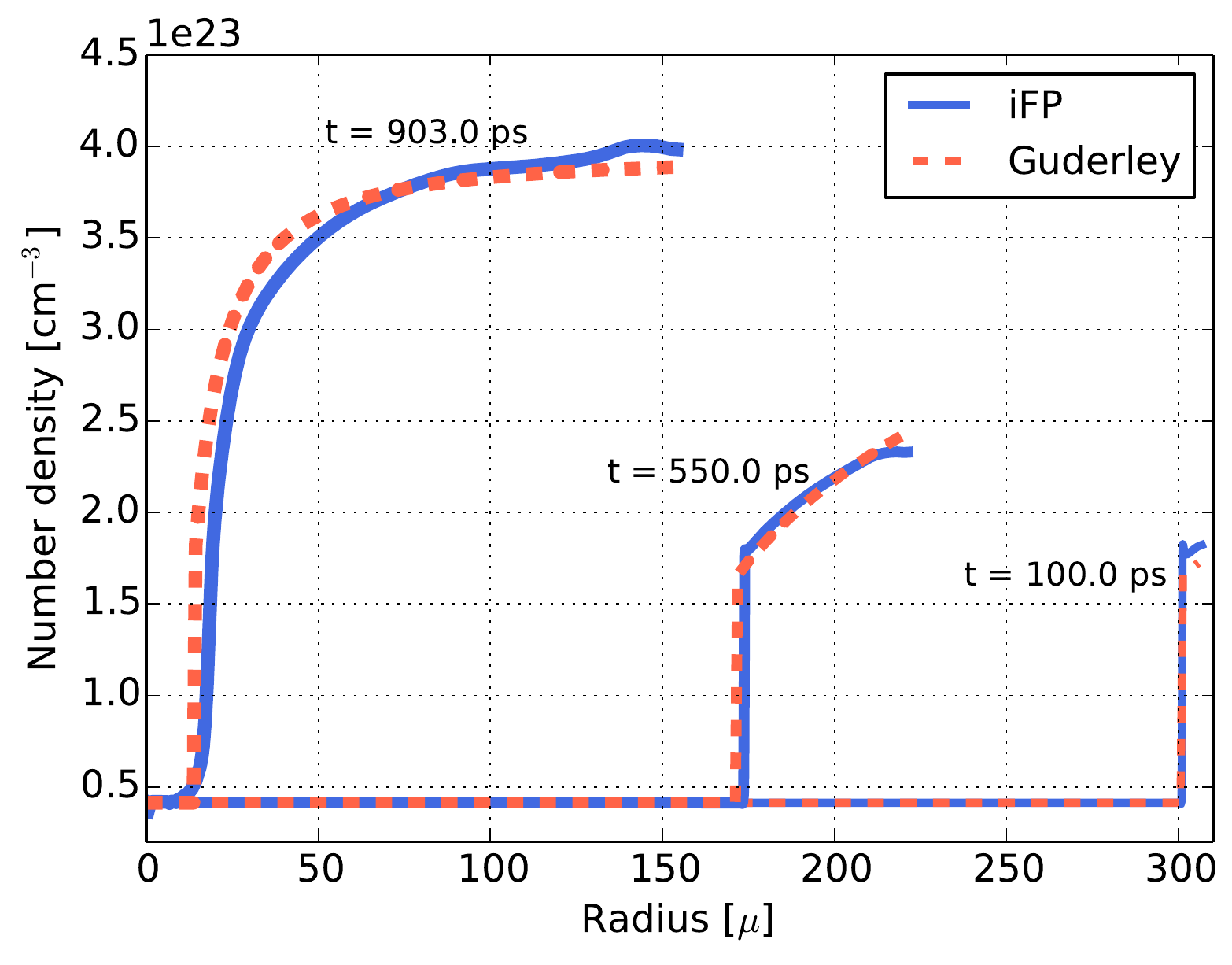}\includegraphics[scale=0.3]{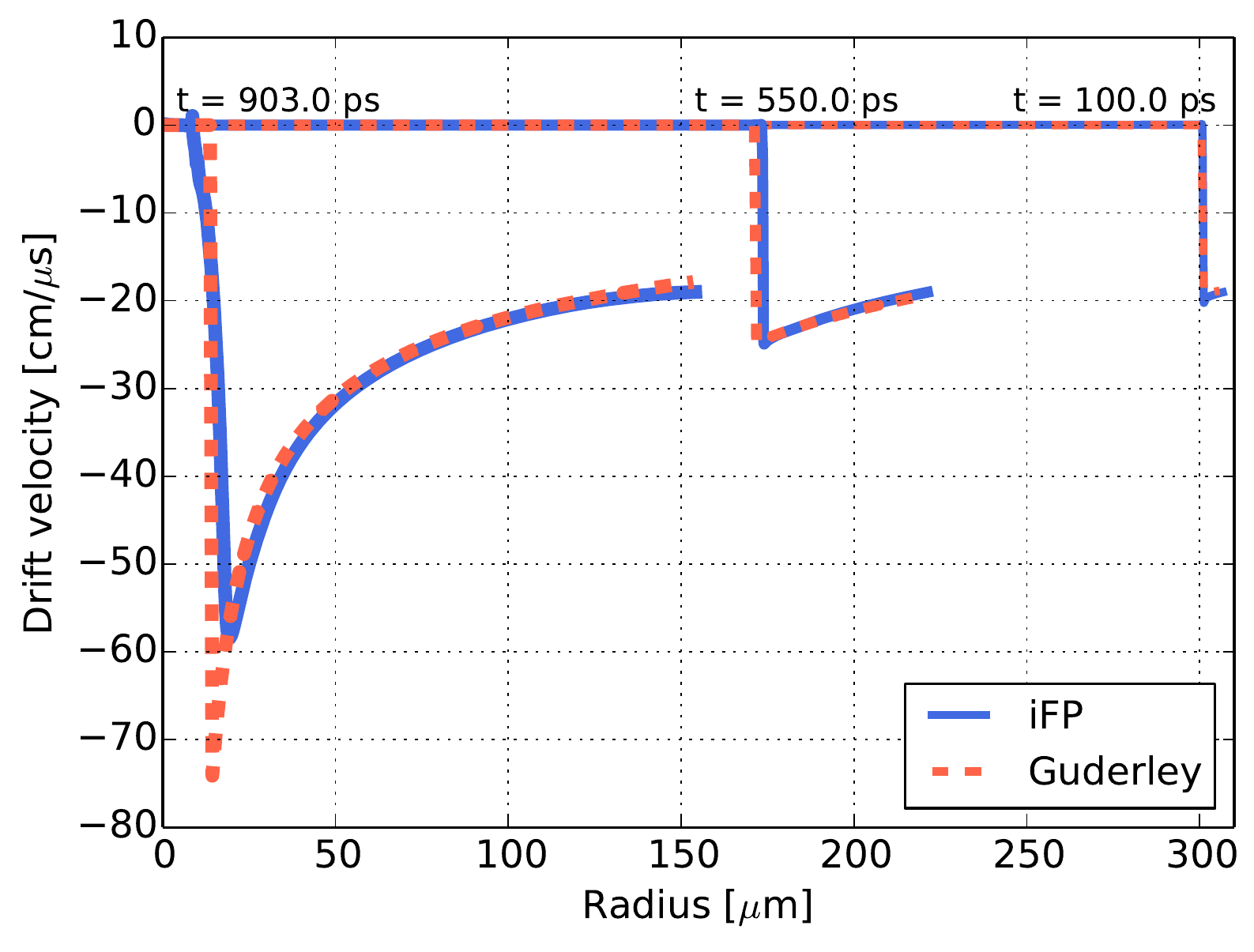}\includegraphics[scale=0.3]{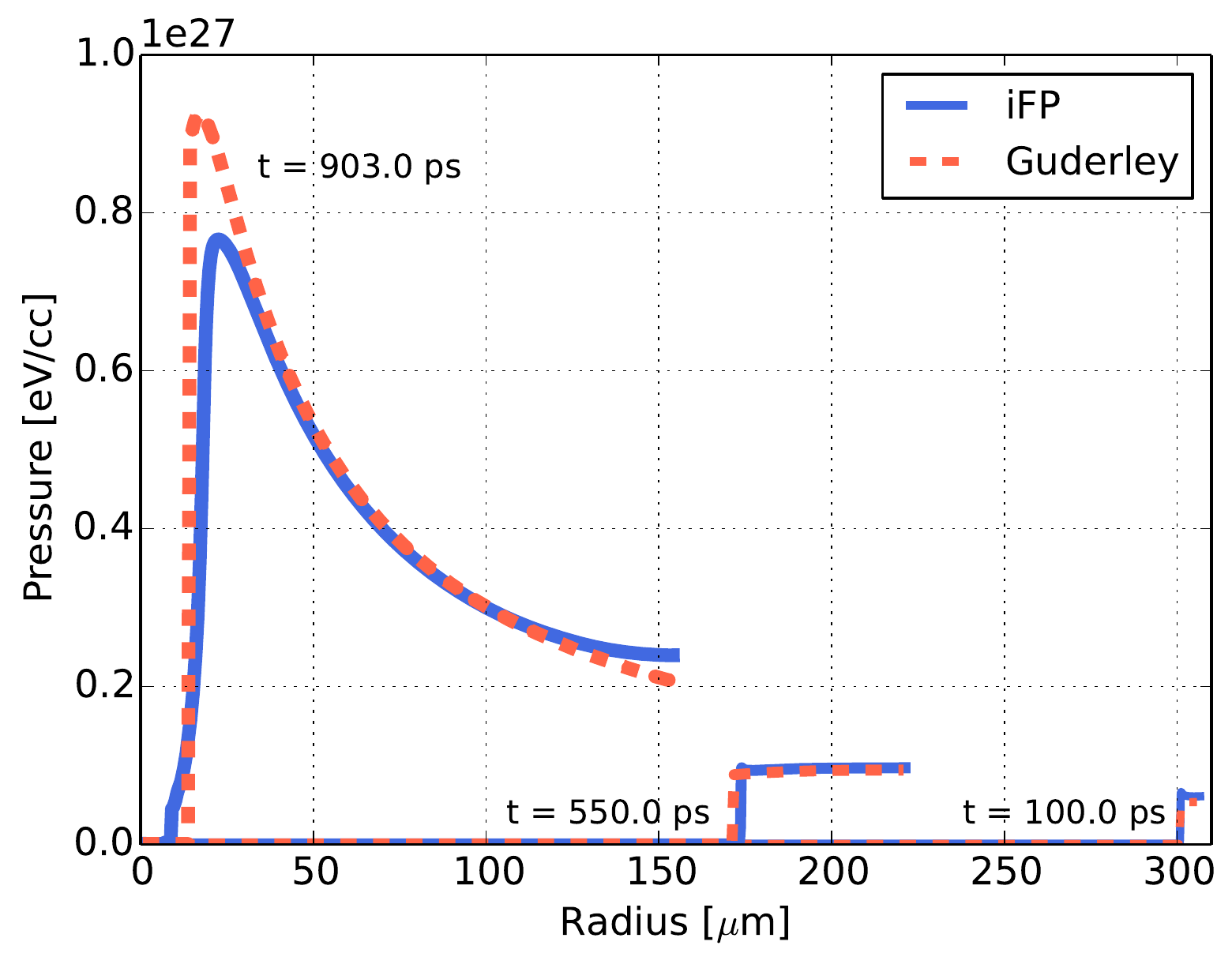} 
\par\end{centering}
\begin{centering}
\includegraphics[scale=0.3]{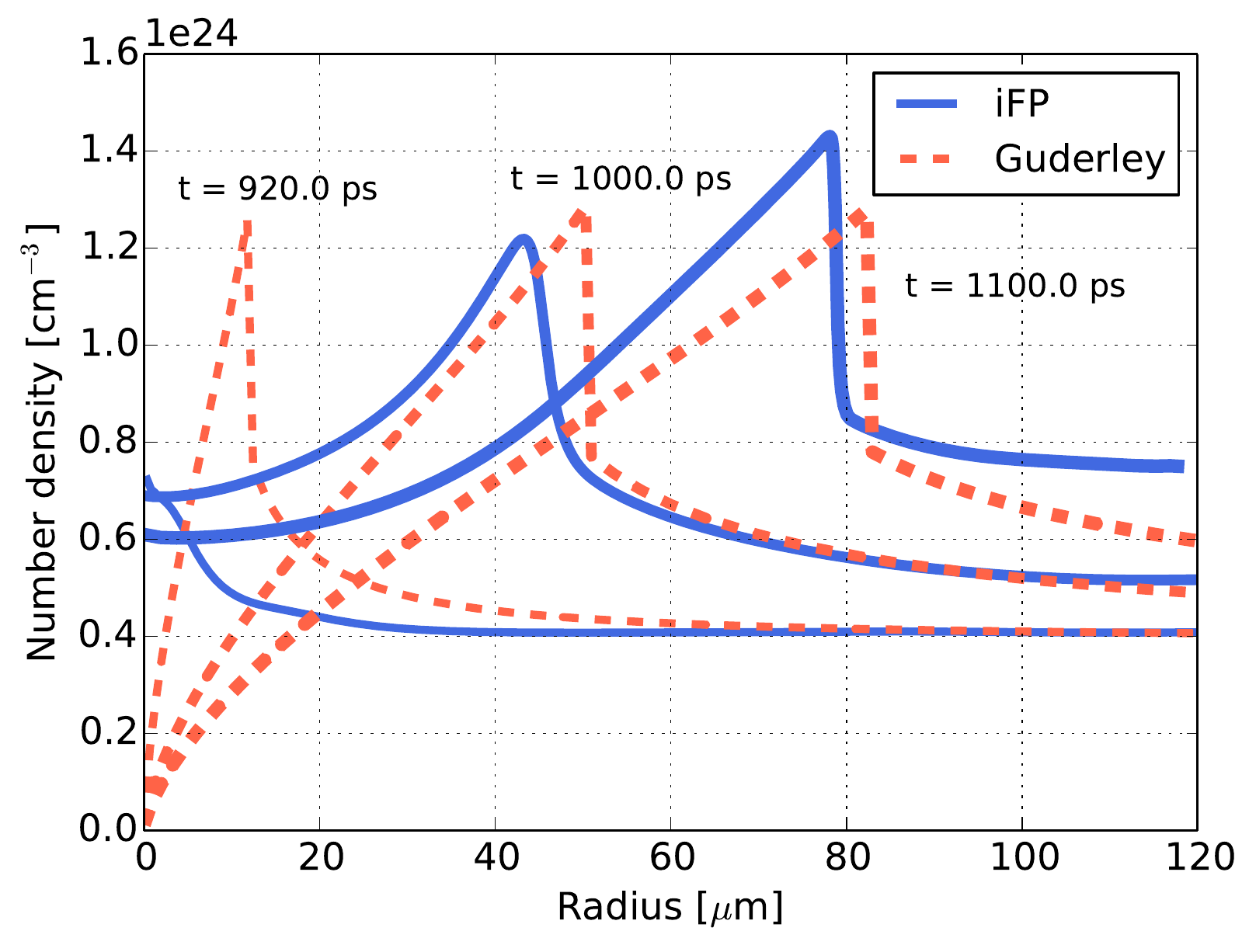}\includegraphics[scale=0.3]{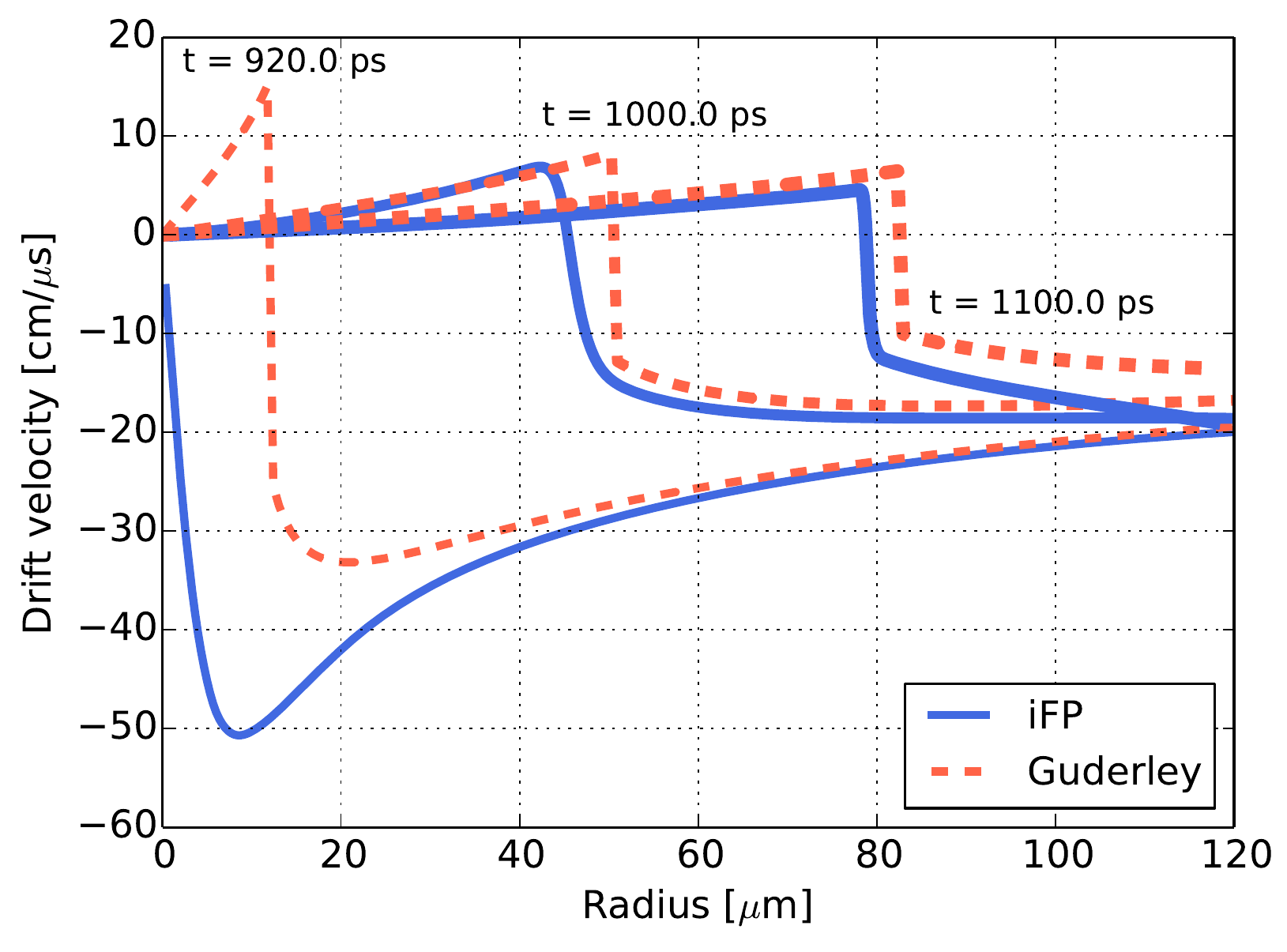}\includegraphics[scale=0.3]{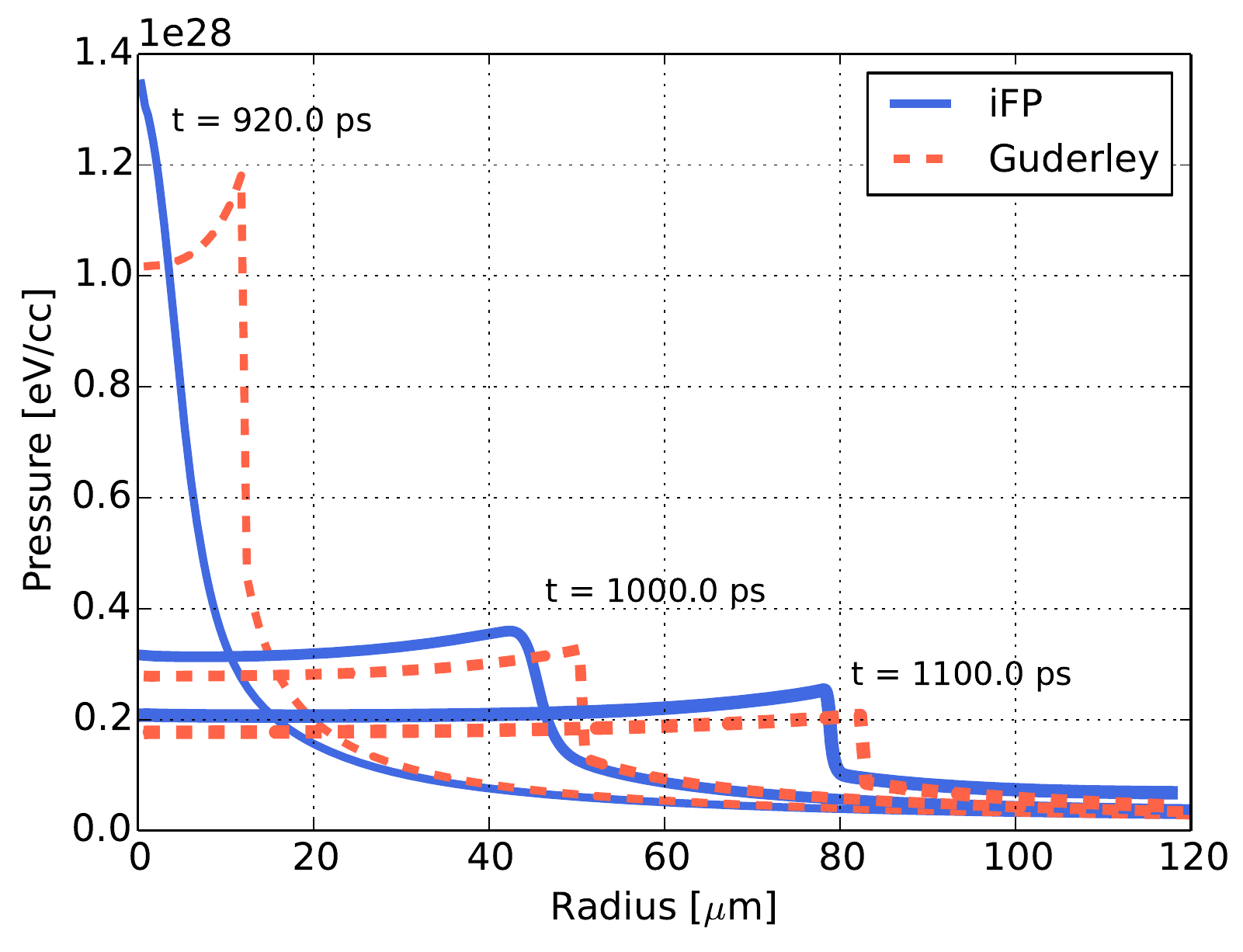} 
\par\end{centering}
\caption{Van Dyke problem: Pre-collapse (top) and post-collapse (bottom) hydro
profiles for density (left), drift velocity (center), and total hydrodynamic
pressure (right) -- given the moving wall boundary condition --
from simulation (``solid'') and Guderley (``dashed'').\label{fig:wall_gud_pre}}
\end{figure}

Finally, the Van Dyke solution asymptotically approaches the Guderley
solution as the shock collapses. Fig. \ref{fig:moving_wall_shock_traj}-right
shows decent agreement between simulation and the equivalent Guderley-shock
trajectories. As before, we can also compare the numerical and Guderley
hydro profiles. Figs. \ref{fig:wall_gud_pre}-top shows pre-collapse
profiles for density (left), drift velocity (center), and total hydrodynamic
pressure (right). Similarly to the time-dependent boundary condition
simulations in Section \ref{subsec:gudurley}, the simulation and
Guderley profiles agree exceedingly well. The shock collapse times
agree within $0.02\%$. Nevertheless, Figs. \ref{fig:wall_gud_pre}-bottom
show that significant differences remain in the post-collapse profiles.
This is believed to be due to the finite kinetic effects near shock
collapse time inducing wall heating of the plasma and altering the
dynamics of the reflected shock. Further, since the Guderley problem
is strictly only applicable for an \emph{infinite} domain, the solutions
continue to deviate as the wall comes closer to the diverging shock
front.

\begin{figure}[h]
\begin{centering}
\includegraphics[scale=0.5]{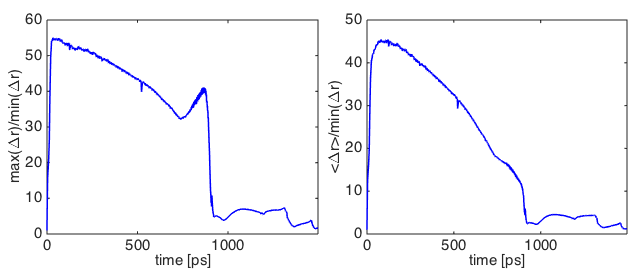} 
\par\end{centering}
\caption{Van Dyke problem: Ratios of maximum to minimum (left) and average
to minimum (right) grid sizes versus time.\label{fig:van_dyke_dr_ratios}}
\end{figure}

\begin{figure}[h]
\begin{centering}
\includegraphics[scale=0.5]{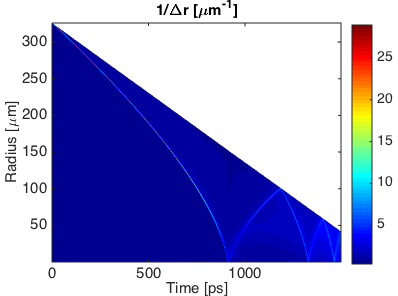} 
\par\end{centering}
\caption{Van Dyke problem: RT diagram of $1/\Delta r$ for $N_{\xi}=192$.\label{fig:van_dyke_inv_dr_vs_rt} }
\end{figure}

\begin{figure}[t]
\begin{centering}
\includegraphics[scale=0.7]{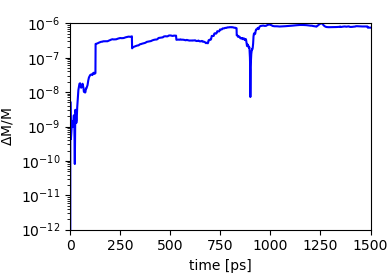}
\par\end{centering}
\textcolor{black}{\caption{\textcolor{black}{Van Dyke problem: Relative mass difference as a
function of time. }\textcolor{black}{\label{fig:van_dyke_mass_conservation}}}
}
\end{figure}

Similarly to the Guderley problem, we demonstrate the grid savings
in the configuration space by showing the maximum- and average-to-minimum
grid size ratios in Fig. \ref{fig:van_dyke_dr_ratios}. 
As can be seen, a complexity reduction in configuration space of roughly
50 is achieved for the Van Dyke problem. Further, since $\max\left(v_{th}\right)/\min\left(v_{th}\right)\approx15,$
a total computational complexity reduction of $\max\left(\frac{\left\langle \Delta r\right\rangle }{\min\left(\Delta r\right)}\right)\left(\frac{\max\left(v_{th}\right)}{\min\left(v_{th}\right)}\right)^{2}\approx11250$
is achieved through the moving phase-space grid strategy. In Fig.
\ref{fig:van_dyke_inv_dr_vs_rt}, a Lagrangian RT diagram of $1/\Delta r$
to demonstrate how the grid tracks the shock front as a function of
time. 
As can be seen, the grid tracks several shock reflections from the
center and the wall, until gradients broaden due to finite transport
effects and the grid \textcolor{black}{relaxes. Finally, in Fig. \ref{fig:van_dyke_mass_conservation}
we demonstrate the discrete mass conservation {[}from conservation
theorems in Eqs. \eqref{eq:moving_wall_bc_dist_func} - \eqref{eq:j_prime_for_elastic_wall}{]}
for the elastic moving wall boundary condition.}\textcolor{black}{}
\textcolor{black}{As can be seen, the total mass variation is well
below the relative nonlinear convergence tolerance of $\epsilon_{r}=10^{-3}$
as designed.}

\subsection{Fuel-only compression yield simulation of shot 22862 at the Omega
Facility (pressure-driven Lagrangian boundary condition)\label{subsec:fuel_only_compression_yield}}

We reproduce the results reported in Ref. \citep{taitano_pop_2018}
in which a fuel-only simulation was performed for a compression yield
target to study the so-called \emph{Rygg effect }\citep{rygg_pop_2006,hermann_pop_2009,taitano_pop_2018}.
We consider this problem for three reasons: 1) to test the implementation
of the pressure-driven Lagrangian boundary conditions on a realistic
implosion experiment; 2) to demonstrate the performance of our nonlinear
solver strategy applied to a realistic implosion scenario; and 3)
to demonstrate the algorithmic advantage of our proposed NS-MMPDE
scheme over the classic MMPDE scheme.

\begin{figure}[h]
\begin{centering}
\includegraphics[scale=0.65]{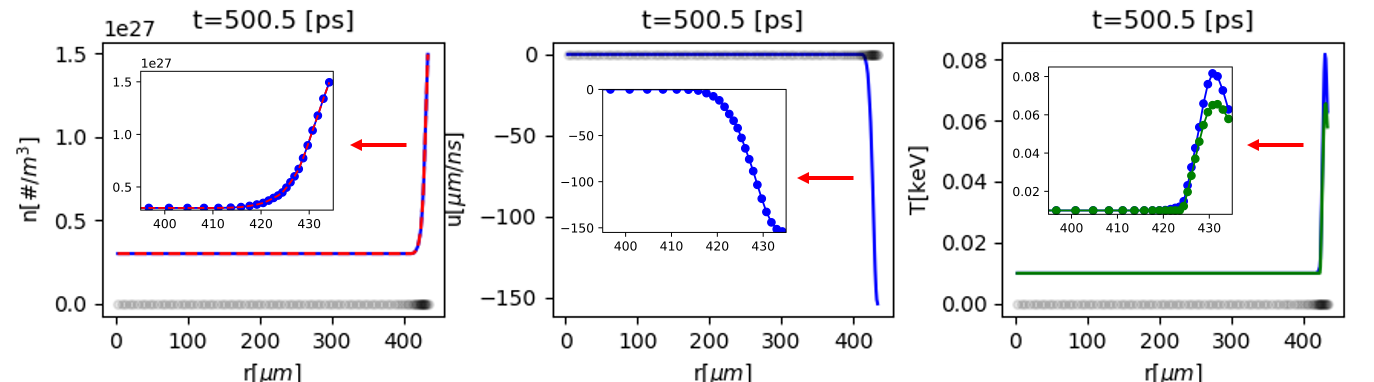} 
\par\end{centering}
\begin{centering}
\includegraphics[scale=0.6]{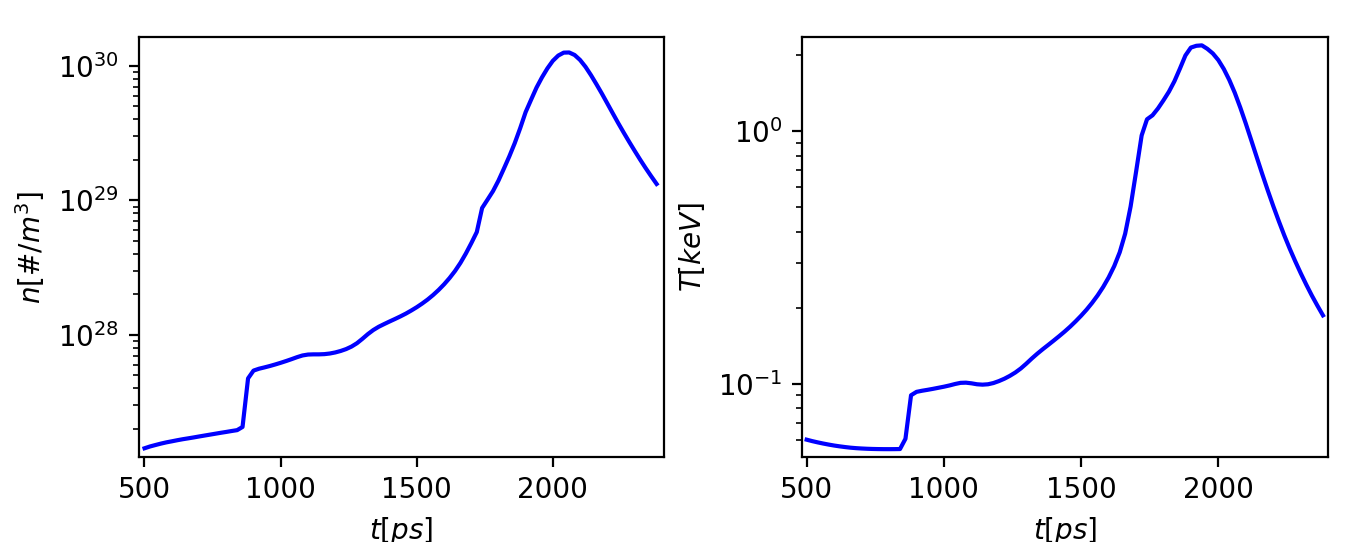} 
\par\end{centering}
\caption{Fuel-only compression yield simulation: The initial condition (top)
and the time-dependent Maxwellian boundary conditions (bottom) for
the number density, $n$, radial drift velocity, $u$, and temperature,
$T$, obtained from the LILAC rad-hydro calculation for the last Lagrangian
fuel zone at the time in which the first shock breaks out of the fuel-pusher
interface (\textasciitilde 500.5 {[}$ps${]}). For the initial condition
for number densities, the blue (red) curve denotes Deuterium (Helium-3);
for the temperature, blue (green) denotes ions (electrons); and the
black markers denotes the grid density (i.e., coarse grid regions
are lighter in color than finer regions). We remind the readers that
the boundary position and velocity are evolved from the mass conservation
constraint and the Lagrangian equation of motion {[}Eqs. (\ref{eq:mass)conservation_bc_kinetic_constraint})
and \ref{eq:pressure_lagrangian_bc_velocity_discrete}{]}, respectively,
and are not explicitly provided to drive the simulation. Further the
boundary temperatures and drift velocities are assumed to be equilibrated
across species at all times. \label{fig:fuel_only_simulation_bc_and_ic}}
\end{figure}

The Rygg effect refers to the anomalous (non-hydrodynamic) scaling
of experimental yield observed by Ref. \citep{rygg_pop_2006}. In
the reference, experiments were performed to test the validity of
the hydrodynamic assumptions in a plastic (CH) capsule with D-$^{3}$He
filled capsules at the Omega laser ICF facility. In the experiments,
the fuel species concentrations were varied from shot to shot such
that the initial mass density, $\rho=\sum_{\alpha}^{N_{s}}m_{\alpha}n_{\alpha}$,
and total hydrodynamic pressure, $P=\sum_{\alpha}^{N_{s}}\left(1+Z_{\alpha}\right)n_{\alpha}T_{\alpha}$,
was kept fixed, thus ensuring \emph{hydro-equivalence}. Here, $m_{D}=2m_{p}$,
$m_{^{3}He}=3m_{p}$ and $Z_{\alpha}$ is the ionization state of
ion-species $\alpha$. Accordingly, if hydrodynamic predictions are
valid, the nuclear yield of D-D reaction, $Y_{DD}\propto n_{D}^{2}$,
should follow a simple $f_{D}^{2}$ scaling law, where $f_{D}$ is
the deuterium number fraction, defined as:

\begin{equation}
f_{D}=\frac{n_{D}}{n_{D}+n_{He3}}.\label{eq:number_fraction_relationship}
\end{equation}
The hydrodynamic variables were obtained from the LILAC Lagrangian
rad-hydro simulation for Omega facility experimental shot 22862 at
the last fuel region. The simulation was initialized at $t_{0}=500.5$
{[}$ps${]} at the moment in which the shock breaks out of the fuel-pusher
interface; refer to Fig. \ref{fig:fuel_only_simulation_bc_and_ic}.
We note that at this point the fuel species', Deuterium and Helium-3,
are assumed to be fully ionized ($Z_{D}=1$ and $Z_{^{3}He}=2$).

\begin{figure}[h]
\begin{centering}
\includegraphics[scale=0.8]{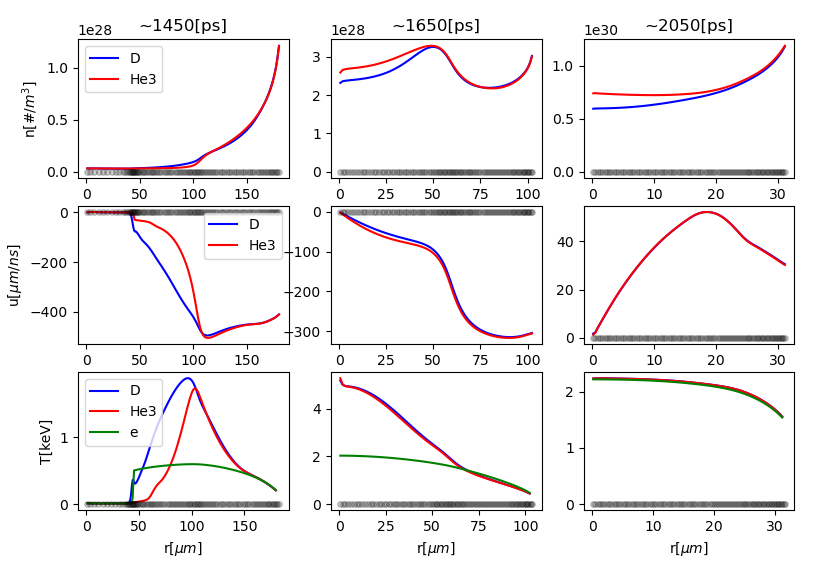} 
\par\end{centering}
\caption{Fuel-only compression yield simulation: The density (top), drift velocity
(center), and temperature (bottom) for the $f_{D}=0.5$ case prior
to shock convergence (\textasciitilde 1450 {[}$ps${]}), upon shock
reflection (\textasciitilde 1650 {[}$ps${]}) and at near peak compression
and stagnation (\textasciitilde 2050 {[}$ps${]}). The black markers
denotes the grid density. \label{fig:fuel_only_simulation_moments_pressure_bc}}
\end{figure}

We consider a grid of $N_{\xi}=96$, $N_{v_{||}}=128$, $N_{v_{\perp}}=64$,
and a transformed velocity-domain size of $\widetilde{v}_{||}\in\left[-6,6\right]$
and $\widetilde{v}_{\perp}\in\left[0,6\right]$ with grid parameters,
$\lambda_{v^{*}}=\lambda_{u^{*}}=10^{-3}$, $\delta_{min}=0.025$,
$\lambda_{\omega}=10^{-3}$, $\tau_{r}=0.1$ {[}$ps${]}, and $\Delta t_{max}=0.2$
{[}$ps${]}. In Fig. \ref{fig:fuel_only_simulation_moments_pressure_bc},
the simulation results are shown for the case of $f_{D}=0.5$ at the
beginning of the simulation, upon shock collapse, and at peak compression.

\begin{figure}[h]
\begin{centering}
\includegraphics[scale=0.7]{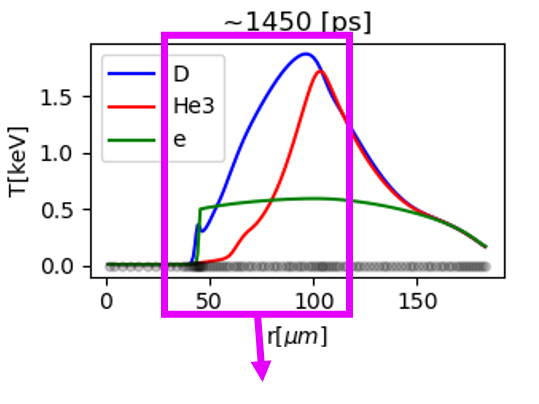} 
\par\end{centering}
\begin{centering}
\includegraphics[scale=0.7]{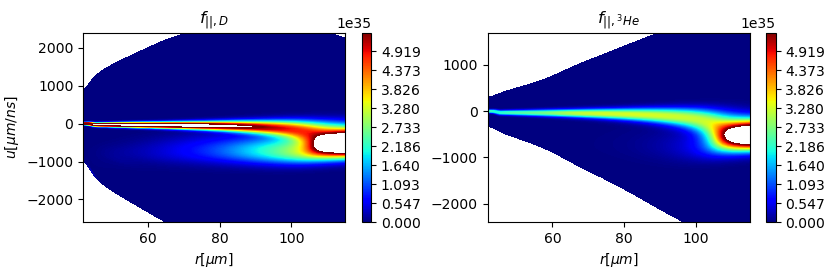} 
\par\end{centering}
\caption{Fuel-only compression yield simulation: The temperature (top) of the
various plasma species, and the corresponding ion marginal parallel-velocity
distribution function (bottom) for Deuterium (left) and Helium-3 (right).
\label{fig:fuel_only_simulation_pdfs}}
\end{figure}

As can be seen, an appreciable fuel separation is seen in all quantities
at shock collapse time (\textasciitilde 1450{[}$ps${]}). The separations
are supported due to the difference in $v_{th}=\sqrt{2T/m}$ amongst
the ion species and the long D-$^{3}$He mean-free-paths --which
scales as $\lambda_{D-^{3}He}\propto T^{2}/\left(Z_{D}^{2}Z_{^{3}He}^{2}\right)$--
supported within the shock; refer to Fig. \ref{fig:fuel_only_simulation_pdfs}.

\begin{figure}[h]
\begin{centering}
\includegraphics[scale=0.8]{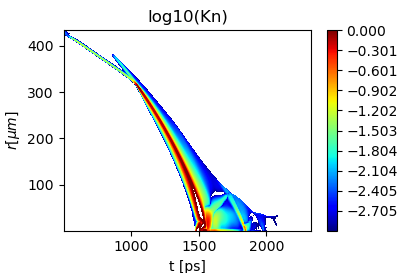} 
\par\end{centering}
\caption{Fuel-only compression yield simulation: The Deuterium Knudsen number
RT diagram is shown. Here, the Knudsen number is defined as $Kn=v_{th,D}\tau_{eff}\partial_{r}\ln n_{ion}$,
where $n_{ion}=n_{D}+n_{^{3}he}$, $v_{th,D}=\sqrt{2T_{D}/m_{D}}$,
$\tau_{eff}=\left(\tau_{DD}^{-1}+\tau_{D^{3}He}^{-1}+\tau_{De}^{-1}\right)^{-1}$,
and $\tau_{D\beta}\propto T_{\beta}^{3/2}/\left(Z_{D}^{2}Z_{\beta}^{2}n_{\beta}m_{\beta}^{1/2}\right)$.
As can be seen, the Knudsen number is relatively large near the shock
trajectory while at peak compression, it reduces to below $10^{-3}$.
\label{fig:knudsen_rt_compression_yield}}
\end{figure}

As can be seen, a distinct beam feature forms in the Deuterium species,
allowing a significant population to run ahead of the Helium-3. These
structures are inherently kinetic and far from a linear perturbation
in pitch angle, necessitating a general description for the distribution
function. In fact, one observes that the plasma near the shock front
is appreciably kinetic throughout the entire course of the implosion
process as can be seen in the Knudsen number RT diagram in Fig. \ref{fig:knudsen_rt_compression_yield}.

As seen, although the Knudsen number does reduce to $Kn\le10^{-3}$
at peak-compression time, hence equilibrating the distribution function
roughly to a Maxwellian, the early separation in the species densities
persists due to the early \emph{kinetic imprinting} in the solution.
As a consequence, Deuterium species are depleted from the hot core
and transported to the periphery of the domain, leading to a degradation
in the D-D yield.

\begin{figure}
\begin{centering}
\includegraphics[scale=0.8]{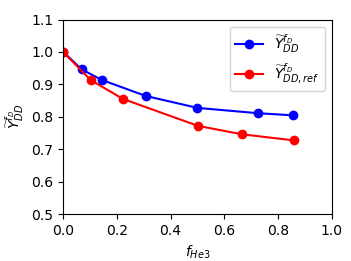} 
\par\end{centering}
\caption{Fuel-only compression yield simulation: $\widetilde{Y}_{DD}^{f_{D}}$
for varying $f_{D}^{0}$. The blue markers denote values obtained
in this study, while the reference values obtained from Ref. \citep{taitano_pop_2018}
are shown in red. We note that the observed disagreement with the
reference result is expected due to the significant algorithmic improvements
we have made to iFP (e.g., improved treatment for the phase-space
grid adaptivity, dynamic time-step adaptivity, etc.) \label{fig:fuel_only_Y_DD_pressure_bc}}
\end{figure}

In Fig. \ref{fig:fuel_only_Y_DD_pressure_bc} the normalized (to the
$f_{D}=1$ case) $DD$-neutron nuclear yield, 
\begin{equation}
\widetilde{Y}_{DD}^{f_{D}}=\frac{Y_{DD}^{f_{D}}}{Y_{DD}^{f_{D}=1}}\left(\frac{n_{D}^{f_{D}=1}}{n_{D}^{f_{D}}}\right)^{2},\label{eq:scaled_yield}
\end{equation}
is shown, where

\begin{equation}
Y_{DD}=2\pi\int_{t_{0}}^{t_{max}}dt\int r^{2}dr\int d^{3}vf_{D}\left(r,\vec{v}\right)\int d^{3}v'f_{D}\left(r,\vec{v}'\right)\left|\vec{v}-\vec{v}'\right|g\left(\left|\vec{v}-\vec{v}'\right|\right),\label{eq:raw_yield}
\end{equation}
$g\left(\left|\vec{v}-\vec{v}'\right|\right)$ is the $DD-n$ cross
section \citep{bosch-hale}, and $n_{D}^{f_{D}}$ is the initial (un-shocked)
Deuterium number density for a given $f_{D}$. 

As can be seen, with decreasing $f_{D}$ (increasing $f_{He3}$),
the scaled relative yield decreases. In a single-fluid hydrodynamic
formalism, the total mass and pressure would be independent of species
concentrations, leading to no variation in $\widetilde{Y}_{DD}^{fD}$.
In kinetic theory however, differential-ion motion, viscous heating,
and other kinetic physics break hydro-equivalence and lead to dependence
of $\widetilde{Y}_{DD}^{f_{D}}$ with Deuterium concentration. For
further insights and discussion into the yield variation physics,
we refer the reader to Ref. \citep{taitano_pop_2018}.

\begin{figure}[h]
\begin{centering}
\includegraphics[scale=0.7]{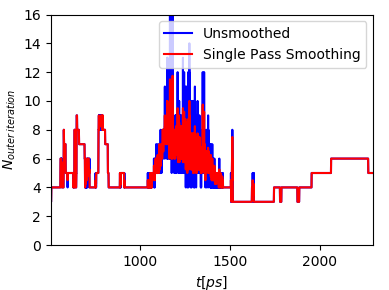} 
\par\end{centering}
\caption{Fuel-only compression yield simulation: Instantaneous number of outer
fixed point iteration, $N$, versus time (blue) and the same quantity
with a single pass of binomial smoothing (red), $N^{*,(p)}=\frac{N^{(p+1)}+2N^{(p)}+N^{(p-1)}}{4}$,
where the superscript ({*}) denotes the smoothed value, and (p) denotes
the time-index. \label{fig:iteration_vs_t_compression_yield}}
\end{figure}

In Fig. \ref{fig:iteration_vs_t_compression_yield}, we demonstrate
the performance of our solver strategy by showing, for the $f_{D}=0.5$
case, the instantaneous number of outer fixed-point/nonlinear iteration
(i.e., number of evaluation of the fixed point map summarized in Alg.
\ref{alg:outer_fixed_point_map_evaluation}) as a function of time.

As can be seen, the majority of the time-steps feature a reasonable
number of iterations with an average over the entire simulation at
approximately 5. A noticeable exception is in the $1000\le t\le1400$
{[}$ps${]} time interval, where nontrivial structures are evolving
in the simulation (e.g., second shock launched into the system, interaction
and merging of the two shocks, and the subsequent fast evolution of
physical structures), leading to occasional spikes in the number of
nonlinear iterations. However, as can be seen from a single pass of
binomial smoothing, the times in which the nonlinear iteration exceeds
10 is rare.

\begin{figure}[h]
\begin{centering}
\includegraphics[scale=0.45]{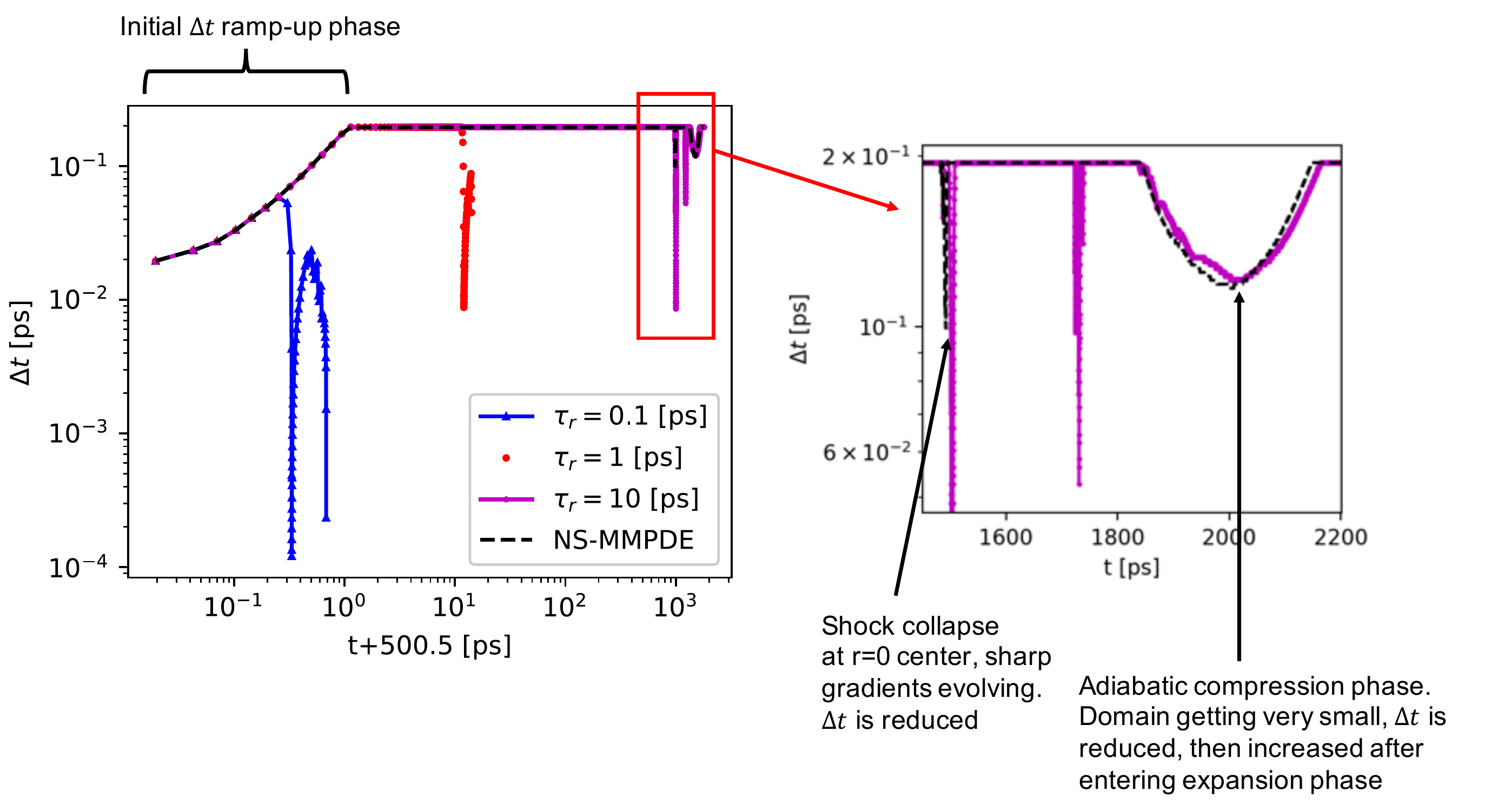} 
\par\end{centering}
\caption{Fuel-only compression yield s\textcolor{black}{imulation: Instantaneous
$\Delta t$ versus time for NS-MMPDE and MMPDE ($\tau_{r}=0.1,\;1,\;10$
{[}ps{]}) for various grid relaxation time-scales. For the standard
MMPDE, $\tau_{r}=0.1$ (blue) and $1$ {[}ps{]} (red) cases lead to
unstable time-integration as can be seen with the algorithm producing
impractically small $\Delta t$. }\label{fig:dt_vs_t_compression_yield}}
\end{figure}

Further, once again for the $f_{D}=0.5$ case, we show the larger
$\Delta t$ afforded by the NS-MMPDE scheme (relative to the standard
MMPDE) by comparing the instantaneous $\Delta t$ as a function of
time; refer to Fig. \ref{fig:dt_vs_t_compression_yield}. 

\begin{figure}[t]
\begin{centering}
\includegraphics[scale=0.7]{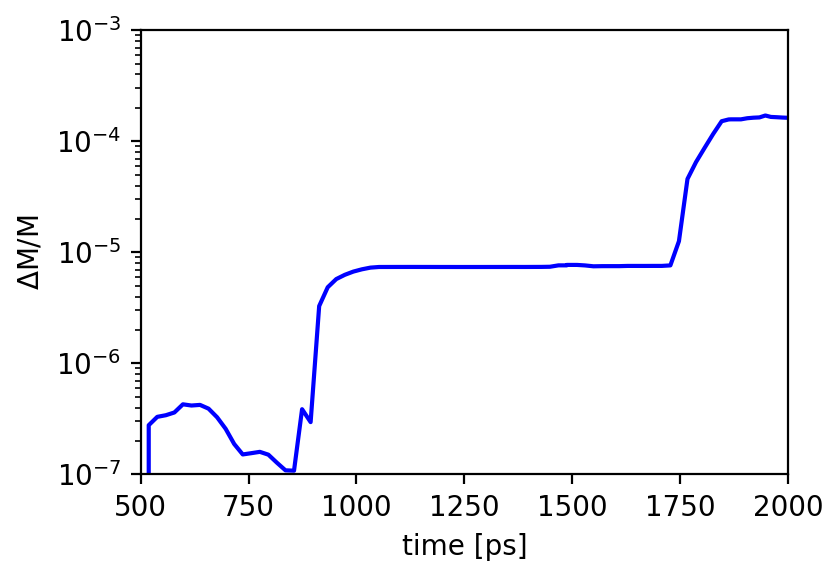}
\par\end{centering}
\textcolor{black}{\caption{\textcolor{black}{Fuel-only compression yield simulation: Relative
total mass variation as a function of time. }\textcolor{black}{\label{fig:pressure_driven_lagrangian_mass_cons_bc}}}
}
\end{figure}

As seen, the NS-MMPDE scheme permits a generally larger $\Delta t$
with less erratic behavior than the standard MMPDE approach. The erratic
behavior of MMPDE is attributed to the periodic \emph{jitter} in the
grid caused by the traveling shock front, which occasionally causes
the grid near the shock front to evolve rapidly {[}and consequently
a smaller $\Delta t$ is taken according to Eq. (\ref{eq:time_stepsize_chooser}){]}.
In fact, the erratic grid evolution eventually leads to very small
$\Delta t$, rendering the simulation impractical for engineering-scale
simulations. The only stable integration for the classic MMPDE scheme
is obtained for a $\tau_{r}=10$ {[}$ps${]} case, where the grid
evolution time-scale is set to be longer than the dynamical time-scales
in the system. Even then, occasional erratic behaviors are seen at
around 1725 {[}$ps${]}. In\textcolor{black}{{} contrast, the proposed
NS-MMPDE scheme detects regions where the grid is evolving rapidly
in the predictor stage and attempts to minimize it through the optimization
procedure, leading to a more robust behavior in the time-integration
throughout the simulation. Thus, the proposed NS-MMPDE scheme greatly
relaxes the process of having to empirically }\textcolor{black}{\emph{tune}}\textcolor{black}{{}
the $\tau_{r}$. Finally, we demonstrate the discrete mass conservation
property for the pressure-driven Lagrangian boundary condition {[}from
conservation theorems shown in Eqs. (\ref{eq:reduced_continuity_for_bc_mass_conservation})-(\ref{eq:mass)conservation_bc_kinetic_constraint}){]},
as shown in Fig. \ref{fig:pressure_driven_lagrangian_mass_cons_bc}.}\textcolor{black}{}
\textcolor{black}{As can be seen, the mass is conserved to within
the nonlinear convergence tolerance of $10^{-3}$. }

\section{Conclusions and Future Work\label{sec:conclusions_and_future_work}}

We have demonstrated a conservative (mass, energy) phase-space moving-grid
strategy in a spherically imploding system with a variety of boundary
conditions - relevant to modeling ICF capsule implosions. The phase-space
grid-adaptivity algorithm leverages the work from Ref. \citep{taitano_cpc_2020_1d2v_cartesian_phase_space}
by adapting the velocity-space grid based on the instantaneous thermal
speed and bulk flow of the individual plasma species (i.e., multiple
velocity-space grid). In the configuration space, the grid is evolved
based on the MMPDE formalism. To deal with strong shocks and complex
features encountered in implosion problems, we have developed a nonlinear
stabilization strategy for the classic MMPDE algorithm (NS-MMPDE).
The approach nonlinearly stabilizes MMPDE against numerical instabilities
that can arise based on a fixed grid relaxation time-scale, $\tau_{r}$.
The strategy is based on splitting the grid evolution process in two
stages: 1) predict the grid based on MMPDE; and 2) pass the predicted
grid through a nonlinear constrained optimization stage which optimizes
the grid against the predicted value and the volumetric rate change.
Our fully implicit solver is based on a nested Anderson accelerated
fixed point iteration strategy that efficiently deals with the nonlinear
coupling between the kinetic ion and fluid electrons. We have demonstrated
the capability of the NS-MMPDE strategy (coupled with the previous
velocity-space adaptivity capability and the adaptive time-stepping
strategy) to robustly simulate ICF implosion problems including the
Guderley/Van-Dyke problem --as far as we are aware, the first to
do so in the literature with a Vlasov-Fokker-Planck approach. We have
also demonstrated with a realistic Omega implosion experiment the
integrated capability of the proposed algorithm in effectively dealing
with multi-length and -velocity scale problems. Future work will include
the extension of the proposed capability to: 1) kinetic electron treatment
by advancing the work performed by Ref. \citep{anderson_jcp_2020_vlasov_ampere_adaptive}
to spherical geometry (e.g., kinetic electron and Amp{è}re equations);
2) to multiple spatial dimensions (e.g., 2D3V) in a general curvilinear
coordinate system; 3) couple with radiation transport physics; 4)
couple with a high-order low-order (HOLO) algorithm \citep{kord_ANS_2002,trt_siam_park_2013,aristova_mmcs_2013_quasi_diffusion,Taitano_2015_jcp_cmec_va,chacon_jcp_2017_holo_review}
to step over the stiff collision time-scales when dense and cold ablators
are included in the simulation\textcolor{black}{.} We note that all
extensions have already been performed and will be documented in follow-on
publications. 

\section*{Acknowledgments}

We greatly appreciate Dr. Brian J. Albright from LANL for the many
helpful discussions and particularly suggestions regarding the physics
applications for the proposed capability. We appreciate Dr. Olivier
Larroche from CEA for providing the LILAC rad-hydro initial and boundary
conditions that were used for the Rygg problem and the many helpful
discussions regarding clarifications on the FPION and FUSE capabilities.
We thank Dr. Hong Sio from the Lawrence Livermore National Laboratory
for discussions in simulating exploding pusher capsules, which ultimately
lead to the need for improving our mesh motion strategy (and the eventual
development of the NS-MMPDE scheme). \textcolor{black}{We also thank
the anonymous referees, whose suggestions helped to make the manuscript
easier to understand. This }work was sponsored by the Metropolis Postdoctoral
Fellowship for W.T.T. between the years 2015-2017, the Thermonuclear
Burn Initiative of the Advanced Simulation and Computing Program between
years 2018-2020, the revolver LDRD project for B.D.K for year 2019,
and the Institutional Computing program at the Los Alamos National
Laboratory. This work was performed under the auspices of the National
Nuclear Security Administration of the U.S. Department of Energy at
Los Alamos National Laboratory, managed by Triad National Security,
LLC under contract 89233218CNA000001.

 \bibliographystyle{ieeetr}
\bibliography{./mybib,./kinetic,./fokker-planck,./transport,./numerics,./general,./icf_physics,./mmpde,./phase_space_adaptivity}

\appendix

\section{Details on the Fluid Electron Model\label{app:fluid_electron_model}}

The friction between the $\alpha$-ion species and electrons is modeled
as: 
\begin{equation}
\vec{F}_{\alpha e}=-m_{e}n_{e}\nu_{e\alpha}\left(\vec{u}_{\alpha}-\left\langle \vec{u}_{\alpha}\right\rangle \right)+\alpha_{0}m_{e}n_{e}\nu_{e\alpha}\left(\vec{u}_{e}-\left\langle \vec{u}_{\alpha}\right\rangle \right)+\beta_{0}\frac{n_{e}\nu_{e\alpha}\nabla_{x}T_{e}}{\sum_{\alpha}^{N_{s}}\nu_{e\alpha}},
\end{equation}
where

\begin{equation}
\nu_{e\alpha}=\frac{2n_{e}e^{4}\Lambda_{e\alpha}}{3\epsilon_{0}^{2}m_{e}^{1/2}\left(2\pi T_{e}\right)^{3/2}},
\end{equation}
is the collision frequency between electrons and the $\alpha$-ion
species

\begin{equation}
\left\langle \vec{u}_{\alpha}\right\rangle =\frac{\sum_{\alpha}^{N_{s}}\nu_{e\alpha}\vec{u}_{\alpha}}{\sum_{\alpha}^{N_{s}}\nu_{e\alpha}},
\end{equation}
is the collision frequency averaged drift velocity, 
\begin{equation}
\alpha_{0}=\frac{4\left(16Z_{eff}^{2}+61\sqrt{2}Z_{eff}+72\right)}{217Z_{eff}^{2}+604\sqrt{2}Z_{eff}+288},
\end{equation}
\begin{equation}
\beta_{0}=\frac{30Z_{eff}\left(11Z_{eff}+15\sqrt{2}\right)}{217Z_{eff}^{2}+604\sqrt{2}Z_{eff}+288},
\end{equation}
and the effective charge is defined as: 
\begin{equation}
Z_{eff}=-\frac{\sum_{\alpha}^{N_{s}}q_{\alpha}^{2}n_{\alpha}}{q_{e}n_{e}}.
\end{equation}
The electron heat flux is given as:

\begin{equation}
\vec{Q}_{e}=\beta_{0}n_{e}T_{e}\left(\vec{u}_{e}-\left\langle \vec{u}_{\alpha}\right\rangle \right)-\kappa_{e}\nabla_{x}T_{e},
\end{equation}
where the electron-thermal conductivity is given as: 
\begin{equation}
\kappa_{e}=\frac{\gamma_{0}n_{e}T_{e}}{m_{e}\sum_{\alpha}^{N_{s}}\nu_{e\alpha}},
\end{equation}
with 
\begin{equation}
\gamma_{0}=\frac{25Z_{eff}\left(433Z_{eff}+180\sqrt{2}\right)}{4\left(217Z_{eff}^{2}+604\sqrt{2}Z_{eff}+288\right)}.
\end{equation}
The generalized Ohm's law for an electrostatic plasma is given as:
\begin{equation}
\vec{E}=\frac{\sum_{\alpha}^{N_{s}}\vec{F}_{\alpha e}+\nabla_{x}P_{e}}{q_{e}n_{e}},
\end{equation}
and the electron-ion energy exchange as:

\begin{equation}
W_{e\alpha}=-\vec{F}_{\alpha e}\cdot\vec{u}_{\alpha}+3\nu_{e\alpha}\frac{m_{e}}{m_{\alpha}}n_{e}\left(T_{e}-T_{\alpha}\right).
\end{equation}
\textcolor{black}{For a detailed derivation and discussions of the
model, refer to \citep{simakov_PoP_2014_e_transp_wh_multi_ion}.}

\section{Cartesian-Cartesian to Spherical-Cylindrical Phase-Space Coordinate
Transformation\label{app:CC_2_SC_transformation}}

\begin{figure}[h]
\begin{centering}
\includegraphics[scale=0.5]{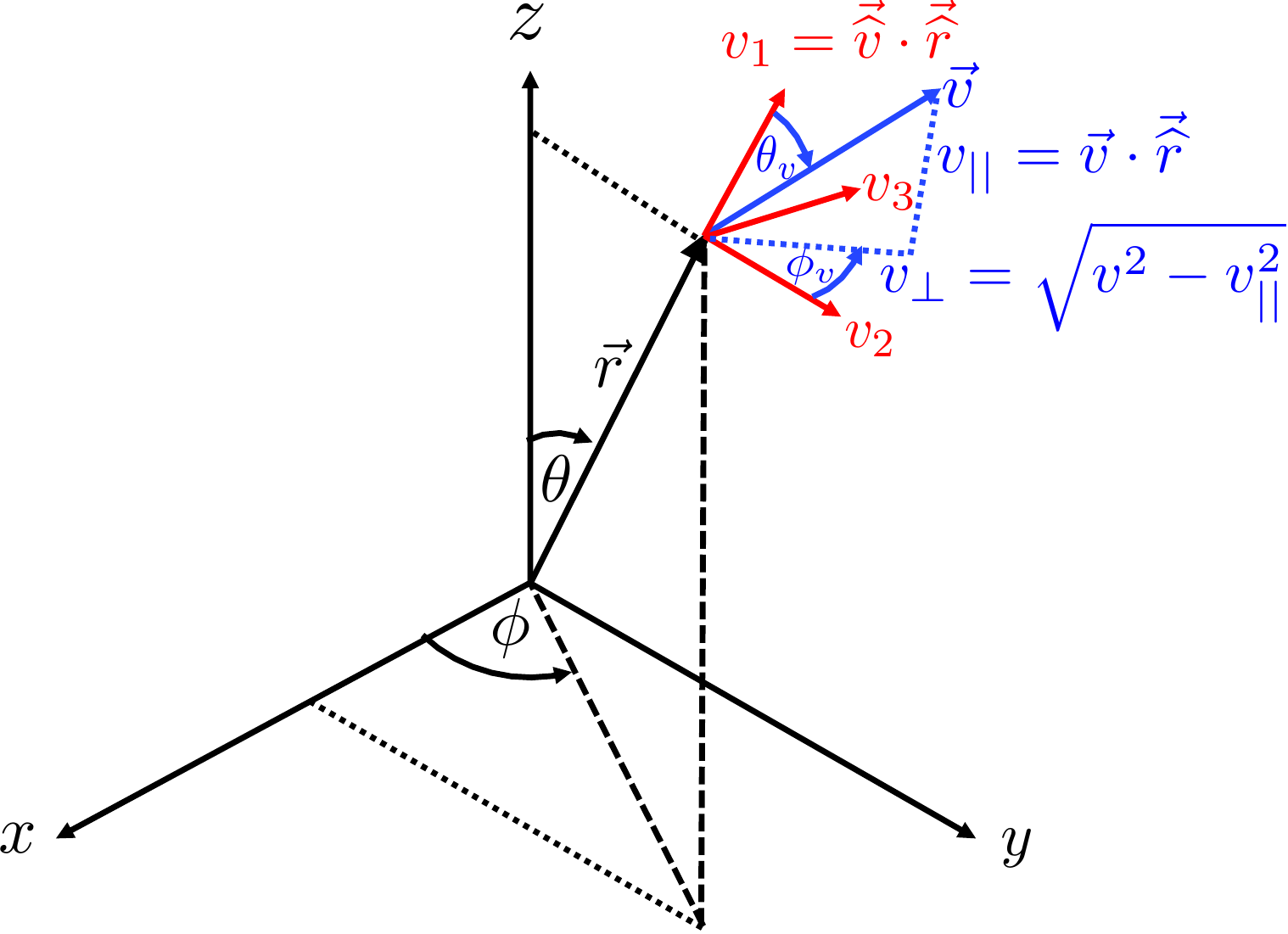} 
\par\end{centering}
\caption{Illustration of spherical-cylindrical hybrid phase-space coordinate
system. \label{fig:CC2SCy}}
\end{figure}

We derive the Vlasov equation in a 1D spherically symmetric configuration
space, $\left(r\right)$, and a 2V cylindrically symmetric velocity
space, $\left(v_{||},v_{\perp}\right)$, from the 3D3V Cartesian coordinate
system. An illustration of the spherical-cylindrical hybrid coordinate
system is illustrated in Fig. \ref{fig:CC2SCy}. 

We define a map to transform from a Cartesian to spherical coordinate
system in the configuration space as:

\begin{equation}
r=\sqrt{x^{2}+y^{2}+z^{2}},
\end{equation}
\begin{equation}
\theta=\textnormal{cos}^{-1}\left(\frac{z}{\sqrt{x^{2}+y^{2}+z^{2}}}\right),
\end{equation}
\begin{equation}
\phi=\textnormal{tan}^{-1}\left(\frac{y}{x}\right).
\end{equation}
Thus, we also obtain, 
\[
\vec{r}=x\widehat{\vec{x}}+y\widehat{\vec{y}}+z\widehat{\vec{z}}=r\left(\sin\theta\cos\phi\widehat{\vec{x}}+\sin\theta\sin\phi\widehat{\vec{y}}+\cos\theta\widehat{\vec{z}}\right).
\]

Similarly for the velocity, we first define a map from Cartesian to
spherical coordinate system as: 
\begin{equation}
v=\sqrt{v_{x}^{2}+v_{y}^{2}+v_{z}^{2}},
\end{equation}

\begin{equation}
\phi_{v}=\textnormal{tan}^{-1}\left(\frac{v_{3}}{v_{2}}\right),
\end{equation}
\begin{equation}
\theta_{v}=\cos^{-1}\left(\frac{v_{1}}{v}\right),
\end{equation}
where 
\begin{equation}
v_{1}=\vec{v}\cdot\vec{\widehat{r}},
\end{equation}
\begin{equation}
v_{2}=\vec{v}\cdot\widehat{\bm{\theta}},
\end{equation}
\begin{equation}
v_{3}=\vec{v}\cdot\widehat{\bm{\phi}}.
\end{equation}
Thus we obtain, 
\begin{equation}
\vec{v}=v_{1}\widehat{\vec{r}}+v_{2}\widehat{\bm{\theta}}+v_{3}\widehat{\bm{\phi}}=v\left(\cos\theta_{v}\widehat{\vec{r}}+\sin\theta_{v}\cos\phi_{v}\widehat{\vec{\theta}}+\sin\theta_{v}\sin\phi_{v}\widehat{\vec{\phi}}\right).
\end{equation}
The map to cylindrical velocity coordinate is then, 
\begin{equation}
v_{||}=\vec{v}\cdot\vec{\widehat{r}},
\end{equation}
\begin{equation}
v_{\perp}=\sqrt{v^{2}-v_{||}^{2}}=v\sin\theta_{v},
\end{equation}
where $v_{||}$ represents the velocity component parallel to the
radial (configuration space) axis and $v_{\perp}$ represents the
perpendicular velocity component. Note, the hat here denotes a unit
basis vector (i.e., $\vec{\widehat{r}}=\partial_{r}\vec{r}/\left|\partial_{r}\vec{r}\right|\equiv\vec{r}/|\vec{r}|$,
$\hat{\bm{\theta}}=\partial_{\vec{r}}\theta/\left|\partial_{\vec{r}}\theta\right|$,
and $\hat{\bm{\phi}}=\partial_{\vec{r}}\phi/\left|\partial_{\vec{r}}\phi\right|$)
and $\left\{ v_{1},v_{2},v_{3}\right\} $ are components of the orthogonal
velocity basis vector aligned with the orthogonal unit vectors of
the spherical coordinate system (i.e., $\vec{\widehat{v}}_{1}\times\vec{\widehat{r}}=\vec{\widehat{v}}_{2}\times\widehat{\bm{\theta}}=\vec{\widehat{v}}_{3}\times\widehat{\bm{\phi}}=0$).
With this, the $v_{||}$ component of the velocity acts as the traditional
$z$ (axial) component in cylindrical coordinate system, $v_{\perp}$
acts as the $\rho$ (radial) component, and $\phi_{v}$ as the azimuthal
angle in the plane of $\vec{v}_{2}$ and $\vec{v}_{3}$.

The Vlasov equation is given as:

\begin{equation}
\partial_{t}f+\nabla_{x}\cdot\left(\vec{v}f\right)+\nabla_{v}\cdot\left(\vec{a}f\right)=0,\label{eq:vlasov_cc2sc}
\end{equation}
which can be rewritten as:

\begin{equation}
\nabla_{T}\cdot\left(\vec{\dot{T}}f\right)=0.\label{eq:vlasov_hamiltonian}
\end{equation}
Here, $\vec{T}=\left\{ t,x,y,z,v_{x},v_{y},v_{z}\right\} $ is the
coordinate vector, $\vec{\dot{T}}\equiv\left\{ 1,v_{x},v_{y},v_{z},a_{x},a_{y},a_{z}\right\} $
is the advection velocity in a Cartesian system, and $\nabla_{T}=\left\{ \partial_{t},\partial_{x},\partial_{y},\partial_{z},\partial_{v_{x}},\partial_{v_{y}},\partial_{v_{z}}\right\} $.
We define the transformed coordinate vector, $\vec{\bm{\tau}}=\left\{ t,r,\theta,\phi,v_{||},v_{\perp},\phi_{v}\right\} $,
the respective advection velocity, $\dot{\bm{\tau}}=\left\{ 1,\dot{r},\dot{\theta},\dot{\phi},\dot{v}_{||},\dot{v}_{\perp},\dot{\phi}_{v}\right\} ,$
and the basis vectors, $\vec{g}_{i}=\frac{\partial\vec{r}}{\partial\tau^{i}}$
and $\vec{g}^{i}=\frac{\partial\tau^{i}}{\partial\vec{r}}$. Here
the superscript (subscript) denotes contravariant (covariant) components
and $i$ denotes the components of the new coordinate system, $\left\{ t,r,\theta,\phi,v_{||},v_{\perp},\phi_{v}\right\} $.

We begin by rewriting $\nabla_{T}\cdot\left(\vec{\dot{T}}f\right)$
as:

\begin{equation}
\nabla_{T}\cdot\left(\vec{\dot{T}}f\right)=\vec{g}^{i}\cdot\frac{\partial}{\partial\tau^{i}}\left(\dot{\tau}^{j}\vec{g}_{j}f\right),\label{eq:vlasov_cc2scy_step1}
\end{equation}
where

\begin{equation}
\vec{\dot{T}}=\dot{\tau}^{j}\vec{g}_{j}.\label{vlasov_cc2scy_step2}
\end{equation}
We use the chain rule to obtain:

\begin{equation}
\vec{g}^{i}\cdot\frac{\partial}{\partial\tau^{i}}\left(\dot{\tau}^{j}\vec{g}_{j}f\right)=\vec{g}^{i}\cdot\left[\frac{\partial\left(\dot{\tau}^{j}f\right)}{\partial\tau^{i}}\vec{g}_{j}+\dot{\tau}^{j}f\frac{\partial\vec{g}_{j}}{\partial\tau^{i}}\right].\label{vlasov_cc2scy_step3}
\end{equation}
We introduce the Cristoffel symbol of second kind,

\begin{equation}
\Gamma_{ji}^{k}=\frac{\partial\vec{g}_{j}}{\partial\tau^{i}}\cdot\vec{g}^{k}\label{vlasov_cc2scy_step4}
\end{equation}
to rewrite Eq. \ref{vlasov_cc2scy_step3} as:

\begin{equation}
\vec{g}^{i}\cdot\left[\frac{\partial\left(\dot{\tau}^{j}f\right)}{\partial\tau^{i}}\vec{g}_{j}+\dot{\tau}^{j}f\frac{\partial\vec{g}_{j}}{\partial\tau^{i}}\right]=\vec{g}^{i}\cdot\left[\frac{\partial\left(\dot{\tau}^{j}f\right)}{\partial\tau^{i}}\vec{g}_{j}+\dot{\tau}^{j}f\Gamma_{ji}^{k}\vec{g}_{k}\right].\label{eq:vlasov_cc2scy_step5}
\end{equation}
Finally, by using the identities, 
\begin{equation}
\vec{g}^{i}\cdot\vec{g}_{j}=\delta_{j}^{i},
\end{equation}
\begin{equation}
\Gamma_{ji}^{k}\delta_{k}^{i}=\Gamma_{jk}^{k}=\frac{1}{J}\frac{\partial J}{\partial\tau^{j}},
\end{equation}
and the chain rule, one obtains: 
\begin{equation}
\nabla_{T}\cdot\left(\vec{\dot{T}}f\right)=\frac{1}{J}\frac{\partial}{\partial\tau^{i}}\left(J\dot{\tau}^{i}f\right)=0.\label{eq:vlasov_cc2scy_step_6}
\end{equation}
Here, $\delta_{j}^{i}$ is the Kronecker delta function ($\delta_{j}^{i}=1$
if $i=j$ and $0$ otherwise), and $J$ is the Jacobian of the transformation,
defined as:

\begin{equation}
J=\left(\textnormal{det}\tensor g^{ij}\right)^{-1/2},
\end{equation}
where $\tensor g^{ij}$ is the contravariant metric tensor, with the
components given as $g^{ij}=\vec{g}^{i}\cdot\vec{g}^{j}$. Note that,
for our specific coordinate system, $J=v_{\perp}r^{2}\textnormal{sin}\theta$.

We proceed by expressing the components of advection velocity in the
transformed coordinate as: 
\begin{equation}
\dot{\tau}^{i}=\dot{\vec{T}}\cdot\vec{g}^{j}=\dot{\tau}^{i}\vec{g}_{i}\cdot\vec{g}^{j}=\frac{\partial\tau^{i}}{\partial t}+\vec{v}\cdot\frac{\partial\tau^{i}}{\partial\vec{r}}+\vec{a}\cdot\frac{\partial\tau^{i}}{\partial\vec{v}}.
\end{equation}
The advection velocity (derivation provided in \ref{app:evaluation_of_advection_velocity})
is then given as:

\begin{equation}
\dot{\bm{\tau}}^{i}=\left[\begin{array}{c}
1\\
v_{||}\\
\frac{v_{\perp}}{r}\textnormal{cos}\phi_{v}\\
\frac{v_{\perp}}{r}\frac{\textnormal{sin}\phi_{v}}{\textnormal{sin}\theta}\\
\frac{v_{\perp}^{2}}{r}+\vec{a}\cdot\frac{\partial v_{||}}{\partial\vec{v}}\\
-\frac{v_{||}v_{\perp}}{r}+\vec{a}\cdot\frac{\partial v_{\perp}}{\partial\vec{v}}\\
-\frac{v_{\perp}}{r}\frac{\sin\phi_{v}}{\tan\theta}+\vec{a}\cdot\frac{\partial\phi_{v}}{\partial\vec{v}}
\end{array}\right];
\end{equation}
and from the electrostatic approximation and spherical symmetry, only
the radial component of the electric field will exist, $\vec{E}=E_{||}\widehat{\vec{r}}$.
Thus, 
\begin{equation}
\vec{a}\cdot\frac{\partial v_{||}}{\partial\vec{v}}=\frac{q}{m}E_{||},\;\;\vec{a}\cdot\frac{\partial v_{\perp}}{\partial\vec{v}}=0,\;\;\vec{a}\cdot\frac{\partial\phi_{v}}{\partial\vec{v}}=0,
\end{equation}
yielding:

\begin{eqnarray}
\frac{\partial}{\partial\xi^{i}}\left(J\dot{\tau}^{i}f\right)=\partial_{t}\left(Jf\right)+\partial_{r}\left(Jv_{||}f\right)+\underbrace{\partial_{\theta}\left(J\frac{v_{\perp}}{r}\textnormal{cos}\phi_{v}f\right)}_{\textcircled a}+\underbrace{\partial_{\phi}\left(J\frac{v_{\perp}}{r}\frac{\textnormal{sin}\phi_{v}}{\textnormal{sin}\theta}f\right)}_{\textcircled b}+\nonumber \\
\partial_{v_{||}}\left(J\left[\frac{v_{\perp}^{2}}{r}+\frac{q}{m}E_{||}\right]f\right)+\partial_{v_{\perp}}\left(J\left[-\frac{v_{||}v_{\perp}}{r}\right]f\right)+\underbrace{\partial_{\phi_{v}}\left(J\left[-\frac{v_{\perp}}{r}\frac{\sin\phi_{v}}{\textnormal{tan}\theta}\right]\right)}_{\textcircled c}=0.
\end{eqnarray}
Due to spherical symmetry in configuration space and azimuthal symmetry
in velocity space (e.g., $\frac{\partial f}{\partial\theta}=\frac{\partial f}{\partial\phi}=\frac{\partial f}{\partial\phi_{v}}=0$),
it is easily shown that terms $\textcircled a$, $\textcircled b$,
and $\textcircled c$ vanish together. By re-defining the Jacobian,
$J\leftarrow J_{Sv_{\perp}}=v_{\perp}r^{2}$, one obtains: 
\begin{equation}
\partial_{t}\left(J_{Sv_{\perp}}f\right)+\partial_{r}\left(J_{Sv_{\perp}}v_{||}f\right)+\frac{q}{m}E_{||}\partial_{v_{||}}\left(J_{Sv_{\perp}}f\right)+\frac{1}{r}\partial_{\vec{v}}\cdot\left(J_{Sv_{\perp}}\vec{\widetilde{a}}\right)=0,\label{eq:vlasov_scy_derivation}
\end{equation}
where $\partial_{\vec{v}}\cdot\vec{A}=\left[\partial_{v_{||}}A_{||},\partial_{v_{\perp}}A_{\perp}\right]^{T}$,
is the divergence operator acting on a vector $\vec{A}=\left[A_{||},A_{\perp}\right]^{T}$,
and $\widetilde{\vec{a}}=\left[v_{\perp}^{2},-v_{||}v_{\perp}\right]^{T}$
is the acceleration vector associated with the inertial term due to
the coordinate transformation. Further, by dividing Eq. (\ref{eq:vlasov_scy_derivation})
by $v_{\perp}$ and redefining the Jacobian as $J_{Sv_{\perp}}\leftarrow J_{S}=r^{2}$,
we obtain: 
\begin{equation}
\partial_{t}\left(J_{S}f\right)+\partial_{r}\left(J_{S}v_{||}f\right)+\frac{q}{m}E_{||}\partial_{v_{||}}\left(J_{S}f\right)+\frac{1}{r}\partial_{\vec{v}}\cdot\left(J_{S}\vec{\widetilde{a}}\right)=0,\label{eq:vlasov_scy_derivation-1}
\end{equation}
where $\partial_{\vec{v}}\cdot\vec{A}=\left[\partial_{v_{||}}A_{||},v_{\perp}^{-1}\partial_{v_{\perp}}\left(v_{\perp}A_{\perp}\right)\right]^{T}$,
is the modified divergence operator acting on a vector $\vec{A}=\left[A_{||},A_{\perp}\right]^{T}.$

\subsection{Evaluation of the advection velocity in the transformed coordinate
system\label{app:evaluation_of_advection_velocity}}

The contravariant component of the advection velocity is derived for
individual components, $\left\{ \dot{t},\dot{r},\dot{\theta},\dot{\phi},\dot{v}_{||},\dot{v}_{\perp},\dot{\phi}_{v}\right\} $.
The $\dot{t}$ component is trivially evaluated to unity, as $t$
depends neither on space nor velocity.

The $\dot{r}$ is also trivially evaluated as:

\begin{equation}
\dot{r}=\vec{v}\cdot\frac{\partial r}{\partial\vec{r}}=\vec{v}\cdot\vec{\widehat{r}}=v_{||}.
\end{equation}

The $\dot{\theta}$ is evaluated starting from:

\begin{equation}
\vec{v}\cdot\frac{\partial\theta}{\partial\vec{r}}=\vec{v}\cdot\widehat{\bm{\theta}}\left|\partial_{\vec{r}}\theta\right|.
\end{equation}
Here, 
\begin{equation}
\vec{v}\cdot\widehat{\bm{\theta}}=v\left(\cos\theta_{v}\widehat{\vec{r}}+\sin\theta_{v}\cos\phi_{v}\widehat{\bm{\theta}}+\sin\theta_{v}\sin\phi_{v}\widehat{\bm{\phi}}\right)\cdot\widehat{\bm{\theta}}=v\sin\theta_{v}\cos\phi_{v}=v_{\perp}\cos\phi_{v},
\end{equation}
and 
\begin{equation}
\left|\frac{\partial\theta}{\partial\vec{r}}\right|=\frac{1}{r}.
\end{equation}
Thus, putting the two terms together, we obtain 
\begin{equation}
\dot{\theta}=\vec{v}\cdot\frac{\partial\theta}{\partial\vec{r}}=\frac{v_{\perp}\textnormal{cos}\phi_{v}}{r}.
\end{equation}

The $\dot{\phi}$ is evaluated similarly from, 
\begin{equation}
\dot{\phi}=\vec{v}\cdot\frac{\partial\phi}{\partial\vec{r}}=\vec{v}\cdot\hat{\bm{\phi}}\left|\partial_{\vec{r}}\phi\right|.
\end{equation}
Here, 
\begin{equation}
\vec{v}\cdot\widehat{\bm{\phi}}=v\left(\cos\theta_{v}\widehat{\vec{r}}+\sin\theta_{v}\cos\phi_{v}\widehat{\bm{\theta}}+\sin\theta_{v}\sin\phi_{v}\widehat{\bm{\phi}}\right)\cdot\widehat{\bm{\phi}}=v\sin\theta_{v}\sin\phi_{v}=v_{\perp}\sin\phi_{v},
\end{equation}
and 
\begin{equation}
\left|\frac{\partial\phi}{\partial\vec{r}}\right|=\frac{1}{r\sin\theta}.
\end{equation}
Thus, putting the two terms together, we obtain

\begin{equation}
\dot{\phi}=\frac{v_{\perp}}{r}\frac{\sin\phi_{v}}{\sin\theta}.
\end{equation}

The $\dot{v}_{||}$ is evaluated from, 
\begin{equation}
\dot{v}_{||}=\vec{v}\cdot\frac{\partial v_{||}}{\partial\vec{r}}+\vec{a}\cdot\frac{\partial v_{||}}{\partial\vec{v}}.
\end{equation}
Since $v_{||}=\vec{v}\cdot\widehat{\vec{r}}=\vec{v}\cdot\frac{\vec{r}}{\sqrt{x^{2}+y^{2}+z^{2}}}$,
\begin{equation}
\vec{v}\cdot\frac{\partial v_{||}}{\partial\vec{r}}=\vec{v}\cdot\frac{\partial}{\partial\vec{r}}\left(\vec{v}\cdot\frac{\vec{r}}{r}\right)=\vec{v}\cdot\left\{ \vec{v}\cdot\left[\frac{r^{2}\tensor I-\vec{r}\otimes\vec{r}}{r^{3}}\right]\right\} .
\end{equation}
With $\vec{r}=\left\{ x,y,z\right\} $, $r=\sqrt{x^{2}+y^{2}+z^{2}}$,
and completing the square, we obtain 
\begin{equation}
\vec{v}\cdot\left\{ \vec{v}\cdot\left[\frac{r^{2}\tensor I-\vec{r}\otimes\vec{r}}{r^{3}}\right]\right\} =\frac{v^{2}-\left(\vec{v}\cdot\widehat{\vec{r}}\right)^{2}}{r}=\frac{v^{2}-v_{||}^{2}}{r}.
\end{equation}
Since $v_{\perp}^{2}=v^{2}-v_{||}^{2}$, 
\begin{equation}
\dot{v}_{||}=\frac{v_{\perp}^{2}}{r}+\vec{a}\cdot\frac{\partial v_{||}}{\partial\vec{v}}.
\end{equation}

The $\dot{v}_{\perp}$ is evaluated from, 
\begin{equation}
\dot{v}=\vec{v}\cdot\frac{\partial v_{\perp}}{\partial\vec{r}}+\vec{a}\cdot\frac{\partial v_{\perp}}{\partial\vec{v}}.
\end{equation}
Since $v_{\perp}=\sqrt{v^{2}-v_{||}^{2}}$, 
\begin{equation}
\vec{v}\cdot\frac{\partial v_{\perp}}{\partial\vec{r}}=\vec{v}\cdot\frac{\partial}{\partial\vec{r}}\sqrt{v^{2}-v_{||}^{2}}=\vec{v}\cdot\left[-\frac{v_{||}\partial_{\vec{r}}v_{||}}{\sqrt{v^{2}-v_{||}^{2}}}\right]=-\frac{v_{||}}{v_{\perp}}\vec{v}\cdot\left\{ \vec{v}\cdot\left[\frac{r^{2}\tensor I-\vec{r}\otimes\vec{r}}{r^{3}}\right]\right\} .
\end{equation}
Thus, 
\begin{equation}
-\frac{v_{||}}{v_{\perp}}\vec{v}\cdot\left\{ \vec{v}\cdot\left[\frac{r^{2}\tensor I-\vec{r}\otimes\vec{r}}{r^{3}}\right]\right\} =-\frac{v_{||}v_{\perp}}{r}
\end{equation}
and finally 
\begin{equation}
\dot{v}_{\perp}=-\frac{v_{||}v_{\perp}}{r}+\vec{a}\cdot\frac{\partial v_{\perp}}{\partial\vec{v}}.
\end{equation}

The $\dot{\phi}_{v}$ is obtained from:

\begin{equation}
\dot{\phi}_{v}=\vec{v}\cdot\frac{\partial\phi_{v}}{\partial\vec{r}}+\vec{a}\cdot\frac{\partial\phi_{v}}{\partial\vec{v}},
\end{equation}
where given $\phi_{v}=\tan^{-1}\left(v_{3}/v_{2}\right)$ , we obtain
\begin{equation}
\frac{\partial\phi_{v}}{\partial\vec{r}}=\frac{1}{v_{2}^{2}+v_{3}^{2}}\left(v_{2}\frac{\partial v_{3}}{\partial\vec{r}}-v_{3}\frac{\partial v_{2}}{\partial\vec{r}}\right).
\end{equation}
From, 
\begin{equation}
v_{2}=\vec{v}\cdot\widehat{\bm{\theta}},\;\;\textnormal{and}\;\;v_{3}=\vec{v}\cdot\widehat{\bm{\phi}},
\end{equation}
\begin{equation}
\frac{\partial v_{2}}{\partial\vec{r}}=\vec{v}\cdot\frac{\partial\widehat{\bm{\theta}}}{\partial\vec{r}},\;\;\textnormal{and}\;\;\frac{\partial v_{3}}{\partial\vec{r}}=\vec{v}\cdot\frac{\partial\widehat{\bm{\phi}}}{\partial\vec{r}},
\end{equation}
and 
\begin{eqnarray}
\vec{v}=\left(v_{1}\sin\theta\cos\phi+v_{2}\cos\theta\cos\phi-v_{3}\sin\phi\right)\widehat{\vec{x}}\nonumber \\
+\left(v_{1}\sin\theta\sin\phi+v_{2}\cos\theta\sin\phi+v_{3}\cos\phi\right)\widehat{\vec{y}}\nonumber \\
+\left(v_{1}\cos\theta-v_{2}\sin\theta\right)\widehat{\vec{z}},
\end{eqnarray}
one can show, 
\begin{eqnarray}
\vec{v}\cdot\frac{\partial\widehat{\bm{\theta}}}{\partial\vec{r}}=-\frac{1}{r\sin\theta}\left[\left(v_{2}\cos\phi\sin^{2}\theta+v_{3}\cos\theta\sin\phi\right)\widehat{\vec{x}}+\right.\nonumber \\
\left.+\left(-v_{3}\cos\theta\cos\phi+v_{2}\cos^{2}\theta\sin\phi+v_{2}\left(1+\sin^{2}\theta\sin\phi\right)/2\right)\widehat{\vec{y}}+v_{2}\sin^{2}\theta\cos\theta\widehat{\vec{z}}\right]
\end{eqnarray}
and 
\begin{equation}
\vec{v}\cdot\frac{\partial\widehat{\bm{\phi}}}{\partial\vec{r}}=-\frac{v_{3}}{r\sin\theta}\left(\cos\phi\widehat{\vec{x}}+\sin\phi\widehat{\vec{y}}\right).
\end{equation}
We can then trivially obtain: 
\begin{equation}
\vec{v}\cdot\left[\vec{v}\cdot\frac{\partial\widehat{\bm{\phi}}}{\partial\vec{r}}\right]=-\frac{1}{r}\left[v_{3}v_{1}+\frac{v_{3}v_{2}}{\tan\theta}\right],\;\;\textnormal{and}\;\;\vec{v}\cdot\left[\vec{v}\cdot\frac{\partial\widehat{\bm{\theta}}}{\partial\vec{r}}\right]=-\frac{1}{r}\left[v_{2}v_{1}-\frac{v_{3}^{2}}{\tan\theta}\right].
\end{equation}
Assembling the terms, 
\begin{eqnarray}
\vec{v}\cdot\frac{\partial\phi_{v}}{\partial\vec{r}}=\frac{1}{v_{2}^{2}+v_{3}^{2}}\left[v_{2}\vec{v}\cdot\vec{v}\cdot\frac{\partial\widehat{\bm{\phi}}}{\partial\vec{r}}-v_{3}\vec{v}\cdot\vec{v}\cdot\frac{\partial\widehat{\bm{\theta}}}{\partial\vec{r}}\right]\nonumber \\
=-\frac{1}{r\left(v_{2}^{2}+v_{3}^{2}\right)}\left[v_{2}\left(v_{3}v_{1}+\frac{v_{3}v_{2}}{\tan\theta}\right)-v_{3}\left(v_{2}v_{1}-\frac{v_{3}^{2}}{\tan\theta}\right)\right]=-\frac{v_{3}}{r}\frac{1}{\tan\theta}.
\end{eqnarray}
Finally, since $v_{3}=v_{\perp}\sin\phi_{v}$,

\begin{equation}
\dot{\phi}_{v}=-\frac{v_{\perp}}{r}\frac{\sin\phi_{v}}{\tan\theta}+\vec{a}\cdot\frac{\partial\phi_{v}}{\partial\vec{v}}.
\end{equation}

\section{Physical to Logical Phase-Space Coordinate Transformation\label{app:physical_2_logical_transformation}}

Starting from our 1D-2V spherically symmetric in configuration space
and azimuthally symmetric cylindrical velocity-space coordinate system,
$\vec{T}=\left\{ t,r,v_{||},v_{\perp}\right\} $, we transform our
Vlasov equation, Eq. (\ref{eq:vlasov_scy_derivation}),

\begin{equation}
\partial_{t}\left(J_{Sv_{\perp}}f\right)+\partial_{r}\left(J_{Sv_{\perp}}v_{||}f\right)+\frac{q}{m}E_{||}\partial_{v_{||}}\left(J_{Sv_{\perp}}f\right)+\frac{1}{r}\partial_{\vec{v}}\cdot\left(J_{Sv_{\perp}}\widetilde{\vec{a}}f\right)=0,
\end{equation}
(where $\widetilde{\vec{a}}=\left[v_{\perp}^{2},-v_{||}v_{\perp}\right]^{T}$
is the acceleration vector associated with the inertial term due to
the coordinate transformation) to logical in configuration and normalized
in velocity coordinate system, $\vec{\tau}=\left\{ t,\xi,\widetilde{v}_{||},\widetilde{v}_{\perp}\right\} $.
We begin by rewriting the equation as:

\begin{equation}
\nabla_{T}\cdot\left(\vec{\dot{T}}\bar{f}\right)\equiv\partial_{t}\left(\bar{f}\right)+\partial_{r}\left(v_{||}\bar{f}\right)+\partial_{v_{||}}\left(\underbrace{\left[\frac{q}{m}E_{||}+\frac{v_{\perp}^{2}}{r}\right]}_{\bar{a}_{||}}\bar{f}\right)+\partial_{v_{\perp}}\left(\underbrace{-\frac{v_{||}v_{\perp}}{r}}_{\bar{a}_{\perp}}\bar{f}\right)=0,
\end{equation}
where $\bar{f}=J_{Sv_{\perp}}f$ and $\vec{\dot{T}}=\left\{ 1,v_{||},\bar{a}_{||},\bar{a}_{\perp}\right\} $.
From steps similar to those in \ref{app:CC_2_SC_transformation},
we obtain:

\begin{equation}
\nabla_{T}\cdot\left(\vec{\dot{T}}\bar{f}\right)=\frac{1}{J_{\xi\widehat{v}}}\frac{\partial}{\partial\tau^{i}}\left(J_{\xi\widehat{v}}\dot{\tau}^{i}\bar{f}\right)=0,\label{eq:vlasov_phys2logi_intermediate}
\end{equation}
where, $J_{\xi\widehat{v}}=\frac{\partial r}{\partial\xi}v^{*^{2}}$
is the Jacobian of transformation for our system. By multiplying by
$J_{\xi\widehat{v}}$, we obtain:

\begin{equation}
\frac{\partial}{\partial\tau^{i}}\left(J_{\xi\widehat{v}}\dot{\tau}^{i}\bar{f}\right)=0.
\end{equation}

The advection velocity in the transformed coordinate system is given
by:

\begin{equation}
\dot{\tau}^{i}=\vec{\dot{T}}\cdot\vec{g}^{i}=\frac{\partial\tau^{i}}{\partial t}+v_{||}\frac{\partial\tau^{i}}{\partial r}+\bar{a}_{||}\frac{\partial\tau^{i}}{\partial v_{||}}+\bar{a}_{\perp}\frac{\partial\tau^{i}}{\partial v_{\perp}},
\end{equation}
and the components are given as:

\begin{equation}
\dot{\bm{\tau}}=\left[\begin{array}{c}
1\\
\frac{\partial\xi}{\partial t}+v_{||}\frac{\partial\xi}{\partial r}\\
\frac{\partial\widetilde{v}_{||}}{\partial t}+v_{||}\frac{\partial\widetilde{v}_{||}}{\partial r}+\bar{a}_{||}\frac{\partial\widetilde{v}_{||}}{\partial v_{||}}\\
\frac{\partial\widetilde{v}_{\perp}}{\partial t}+v_{||}\frac{\partial\widetilde{v}_{\perp}}{\partial r}+\bar{a}_{\perp}\frac{\partial\widetilde{v}_{\perp}}{\partial v_{\perp}}
\end{array}\right].
\end{equation}

We derive the expression for each terms individually, beginning with
the temporal derivatives. Recall that 
\begin{equation}
\frac{\partial}{\partial t}\equiv\left(\frac{\partial}{\partial t}\right)_{r,v_{||},v_{\perp}}
\end{equation}
and that 
\begin{equation}
\left(\frac{\partial\tau^{i\neq t}}{\partial t}\right)_{T^{j}\neq t}=-\left(\frac{\partial\vec{T}}{\partial t}\right)_{\tau^{k}\neq t}\cdot\vec{g}^{i\neq t}.
\end{equation}
We obtain, 
\begin{equation}
\left[\begin{array}{c}
\left(\partial_{t}t\right)_{r,v_{||},v_{\perp}}\\
\left(\partial_{t}\xi\right)_{r,v_{||},v_{\perp}}\\
\left(\partial_{t}\widetilde{v}_{||}\right)_{r,v_{||},v_{\perp}}\\
\left(\partial_{t}\widetilde{v}_{\perp}\right)_{r,v_{||},v_{\perp}}
\end{array}\right]=\left[\begin{array}{c}
1\\
-\left(\frac{\partial r}{\partial t}\right)_{\bm{\tau}^{k\neq t}}\left(\frac{\partial r}{\partial\xi}\right)_{\bm{\tau}^{k\neq\xi}}^{-1}\\
-\left[\left(\frac{\partial v_{||}}{\partial t}\right)_{\bm{\tau}^{k\neq t}}-\left(\frac{\partial r}{\partial t}\right)_{\bm{\tau}^{k\neq t}}\left(\frac{\partial r}{\partial\xi}\right)_{\bm{\tau}^{k\neq\xi}}^{-1}\left(\frac{\partial v_{||}}{\partial\xi}\right)_{\bm{\tau}^{k\neq\xi}}\right]\left(\frac{\partial v_{||}}{\partial\widetilde{v}_{||}}\right)_{\bm{\tau}^{k\neq\widetilde{v}_{||}}}^{-1}\\
-\left[\left(\frac{\partial v_{\perp}}{\partial t}\right)_{\bm{\tau}^{k\neq t}}-\left(\frac{\partial r}{\partial t}\right)_{\bm{\tau}^{k\neq t}}\left(\frac{\partial r}{\partial\xi}\right)_{\bm{\tau}^{k\neq\xi}}^{-1}\left(\frac{\partial v_{\perp}}{\partial\xi}\right)_{\bm{\tau}^{k\neq\xi}}\right]\left(\frac{\partial v_{\perp}}{\partial\widetilde{v}_{\perp}}\right)_{\bm{\tau}^{k\neq\widetilde{v}_{\perp}}}^{-1}
\end{array}\right].
\end{equation}
Here, recall that 
\begin{equation}
v_{||}=\widetilde{v}_{||}v^{*}+u_{||}^{*}\;\;\textnormal{and}\;\;v_{\perp}=\widetilde{v}_{\perp}v^{*},
\end{equation}
therefore 
\begin{equation}
\left(\frac{\partial v_{||}}{\partial t}\right)_{\bm{\tau}^{k\neq t}}=\widetilde{v}_{||}\frac{\partial v^{*}}{\partial t}+\frac{\partial u^{*}}{\partial t},
\end{equation}
\begin{equation}
\left(\frac{\partial v_{\perp}}{\partial t}\right)_{\bm{\tau}^{k\neq t}}=\widetilde{v}_{\perp}\frac{\partial v^{*}}{\partial t},
\end{equation}
\begin{equation}
\left(\frac{\partial v_{||}}{\partial\widetilde{v}_{||}}\right)_{\bm{\tau}^{k\neq\widetilde{v}_{||}}}=\left(\frac{\partial v_{\perp}}{\partial\widetilde{v}_{\perp}}\right)_{\bm{\tau}^{k\neq\widetilde{v}_{\perp}}}=v^{*}.
\end{equation}
The rest of the terms are trivially evaluated as: 
\begin{equation}
v_{||}\frac{\partial\xi}{\partial r}=v^{*}\left(\widetilde{v}_{||}+\widehat{u}_{||}^{*}\right)\frac{\partial\xi}{\partial r},
\end{equation}

\begin{equation}
\frac{\partial\widetilde{v}_{||}}{\partial r}=\frac{\partial\widetilde{v}_{||}}{\partial v^{*}}\frac{\partial v^{*}}{\partial\xi}\frac{\partial\xi}{\partial r}+\frac{\partial\widetilde{v}_{||}}{\partial u_{||}^{*}}\frac{\partial u_{||}^{*}}{\partial\xi}\frac{\partial\xi}{\partial r},
\end{equation}
\begin{equation}
\frac{\partial\widetilde{v}_{\perp}}{\partial r}=\frac{\partial\widetilde{v}_{||}}{\partial v^{*}}\frac{\partial v^{*}}{\partial\xi}\frac{\partial\xi}{\partial r},
\end{equation}

\begin{equation}
\frac{\partial\widetilde{v}_{||}}{\partial v_{||}}=\frac{1}{v^{*}},\;\;\textnormal{and}\;\;\frac{\partial\widetilde{v}_{\perp}}{\partial v_{\perp}}=\frac{1}{v^{*}},
\end{equation}

\begin{equation}
\frac{\partial\widetilde{v}_{||}}{\partial v^{*}}=-\frac{v_{||}-u_{||}^{*}}{v^{*^{2}}}=-\frac{\widetilde{v}_{||}-\widehat{u}_{||}^{*}}{v^{*}},\;\;\frac{\partial\widetilde{v}_{||}}{\partial u_{||}^{*}}=\frac{1}{v^{*}},
\end{equation}
and 
\begin{equation}
\frac{\partial\widetilde{v}_{\perp}}{\partial v^{*}}=-\frac{v_{\perp}}{v^{*^{2}}}=-\frac{\widehat{v}_{\perp}}{v^{*}}.
\end{equation}

Since $\bar{f}=J_{Sv_{\perp}}f$, $\widetilde{f}=v^{*^{3}}f$, by
defining $J_{S\widetilde{v}_{\perp}}=r^{2}\widetilde{v}_{\perp}$,\textcolor{black}{{}
$J_{\xi}=\frac{\partial r}{\partial\xi}$; and noting that $\vec{\widetilde{v}}\frac{\partial v^{*}}{\partial t}+\frac{\partial\vec{u}^{*}}{\partial t}=\frac{\partial}{\partial t}\left(v^{*}\vec{\widetilde{v}}+u^{*}\vec{e}_{||}\right)=\frac{\partial\vec{v}}{\partial t},$}
$\vec{\widetilde{v}}\frac{\partial v^{*}}{\partial\xi}+\frac{\partial\vec{u}^{*}}{\partial\xi}=\frac{\partial}{\partial\xi}\left(v^{*}\vec{\widetilde{v}}+u^{*}\vec{e}_{||}\right)=\frac{\partial\vec{v}}{\partial t},$
where $\vec{e}_{||}$ is the unit vector in the parallel-velocity
direction; $v_{||}=v^{*}\widetilde{v}_{||}+u_{||}^{*}$, $v_{\perp}=v^{*}\widetilde{v}_{\perp}$;\textcolor{black}{{}
and substituting the above results into the Eq. \ref{eq:vlasov_phys2logi_intermediate},
the following transformed equation for $\widetilde{f}\left(\xi,\widetilde{\vec{v}},t\right)$
is obtained:} 
\begin{eqnarray}
\frac{\partial\left(J_{S\widetilde{v}_{\perp}}J_{\xi}\widetilde{f}\right)}{\partial t}+\frac{\partial}{\partial\xi}\left[J_{S\widetilde{v}_{\perp}}\left(v_{||}-\partial_{t}r\right)\widetilde{f}\right]-\frac{1}{v^{*}}\frac{\partial}{\partial\widetilde{\vec{v}}}\cdot\left[\frac{\partial\vec{v}}{\partial t}J_{S\widetilde{v}_{\perp}}J_{\xi}\widetilde{f}\right]\nonumber \\
-\frac{1}{v^{*}}\frac{\partial}{\partial\widetilde{\vec{v}}}\cdot\left[\frac{\partial\vec{v}}{\partial\xi}\left(v_{||}-\partial_{t}r\right)J_{S\widetilde{v}_{\perp}}\widetilde{f}\right]+\frac{1}{v^{*}}\frac{\partial}{\partial\widetilde{v}_{||}}\left(\frac{q}{m}E_{||}J_{S\widetilde{v}_{\perp}}J_{\xi}\widetilde{f}\right)-\frac{1}{v^{*}}\frac{\partial}{\partial\widetilde{\vec{v}}}\cdot\left[\frac{J_{S\widetilde{v}_{\perp}}J_{\xi}\widetilde{\vec{a}}}{r}\widetilde{f}\right].
\end{eqnarray}
Finally by defining $\dot{J}_{r^{3}}=\frac{1}{3}\frac{\partial r^{3}}{\partial t}=J_{S}\frac{\partial r}{\partial t}$,
$J_{S\xi}=J_{S}J_{\xi}$, and dividing by $\widetilde{v}_{\perp}$,
we obtain:

\begin{eqnarray}
\frac{\partial\left(J_{S\xi}\widetilde{f}\right)}{\partial t}+\frac{\partial}{\partial\xi}\left[\left(J_{S}v_{||}-\dot{J}_{r^{3}}\right)\widetilde{f}\right]-\frac{1}{v^{*}}\frac{\partial}{\partial\widetilde{\vec{v}}}\cdot\left[\frac{\partial\vec{v}}{\partial t}J_{S\xi}\widetilde{f}\right]\nonumber \\
-\frac{1}{v^{*}}\frac{\partial}{\partial\widetilde{\vec{v}}}\cdot\left[\frac{\partial\vec{v}}{\partial\xi}\left(J_{S}v_{||}-\dot{J}_{r^{3}}\right)\widetilde{f}_{\alpha}\right]+\frac{1}{v^{*}}\frac{\partial}{\partial\widetilde{v}_{||}}\left(J_{S\xi}\frac{q}{m}E_{||}\widetilde{f}\right)+\frac{1}{v^{*}}\frac{\partial}{\partial\widetilde{\vec{v}}}\cdot\left[\frac{J_{S\xi}\widetilde{\vec{a}}}{r}\widetilde{f}\right]=0.
\end{eqnarray}
Here, $\partial_{\widetilde{\vec{v}}}\cdot\vec{A}=\partial_{\widetilde{v}_{||}}\left(A_{v_{||}}\right)+\widetilde{v}_{\perp}^{-1}\partial_{\widetilde{v}_{\perp}}\left(\widetilde{v}_{\perp}A_{v_{\perp}}\right)$
is the velocity-space divergence operator acting on a vector $\vec{A}=\left[A_{v_{||}},A_{v_{\perp}}\right]^{T}$
.

\section{Vlasov Conservation Symmetries for the Transformed Coordinate System\label{app:symmetries_for_transformed_coordinates}}

Following similar procedures outlined in Ref. \citep{taitano_cpc_2020_1d2v_cartesian_phase_space},
we derive the conservation symmetries for the Vlasov equation in a
1D spherically symmetric configuration space and 2V cylindrically
symmetric velocity-space coordinate system. \textcolor{black}{We note
that the conservation properties associated with the temporal and
spatial variation of velocity-space metrics, as well as the inertial
terms introduced by the spherical configuration space transformation
can all be shown independently. We begin by developing discretization
for exact mass, linear momentum, and energy conservation in a spatially
homogeneous system (0D), and then a spatially inhomogeneous case (1D)
in a periodic spatial domain }\textit{\textcolor{black}{without}}\textcolor{black}{{}
any background field (conservation with the electric field can be
shown separately, as demonstrated in Ref. \citep{Taitano_2018_vrfp_1d2v_implicit}).}

\subsection{\textcolor{black}{Temporal variation of $v^{*}$ and $\widehat{u}_{||}^{*}$\label{subsec:temporal_variation_conservation}}}

\textcolor{black}{For simplicity, we drop the subscript denoting ion
species. By consider only the terms associated with the temporally
varying velocity-space metrics in Eq. (\ref{eq:composite_transformed_vfp}),
we obtain the following simplified form of the Vlasov equation:}

\textcolor{black}{
\begin{equation}
\partial_{t}\left(J_{r^{2}\xi}\widetilde{f}\right)-\frac{J_{r^{2}\xi}}{v^{*}}\frac{\partial}{\partial\widetilde{\vec{v}}}\cdot\left(\widetilde{f}\partial_{t}\vec{v}\right)=0,\label{eq:simplified_temporal_only_vlasov}
\end{equation}
where $\partial_{t}\vec{v}=\partial_{t}\left[v^{*}\left(\vec{\widetilde{v}}+\widehat{u}_{||}^{*}\vec{e}_{||}\right)\right]$.
In the continuum, mass conservation is defined as: 
\begin{equation}
\int_{0}^{1}d\xi\left\langle m,\partial_{t}\left(J_{r^{2}\xi}\widetilde{f}\right)\right\rangle _{\widetilde{\vec{v}}}=0.
\end{equation}
This can be shown trivially due to the divergence form of the inertial
terms (with velocity space fluxes, $\vec{J}$, at the boundaries set
to zero, i.e., $\vec{J}\cdot\vec{n}_{v,B}=0$, where $\vec{n}_{v,B}$
is the velocity space boundary normal vector). }

\textcolor{black}{Momentum and energy conservation is defined by following
Ref. \citep{taitano_cpc_2020_1d2v_cartesian_phase_space} and using
the chain rule and taking the $\left\langle mv_{||},\left(\cdot\right)\right\rangle _{\widetilde{\vec{v}}}$
moment of Eq. (\ref{eq:simplified_temporal_only_vlasov}) as: 
\[
\left\langle 1,m\left\{ \partial_{t}\left(v_{||}J_{r^{2}\xi}\ftilde\right)-\underbrace{\left[J_{r^{2}\xi}\ftilde\partial_{t}v_{||}+\frac{J_{r^{2}\xi}v_{||}}{v^{*}}\frac{\partial}{\partial\widetilde{\vec{v}}}\cdot\left(\partial_{t}\vec{v}\widetilde{f}\right)\right]}_{\textcircled a}\right\} \right\rangle _{\widetilde{\vec{v}}}=0,
\]
and 
\[
\left\langle 1,m\left\{ \partial_{t}\left(\frac{v^{2}}{2}J_{r^{2}\xi}\ftilde\right)-\underbrace{\left[J_{r^{2}\xi}\ftilde\partial_{t}\frac{v^{2}}{2}+\frac{J_{r^{2}\xi}v^{2}}{2v^{*}}\frac{\partial}{\partial\widetilde{\vec{v}}}\cdot\left(\widetilde{f}\partial_{t}\vec{v}\right)\right]}_{\textcircled b}\right\} \right\rangle _{\widetilde{\vec{v}}}=0.
\]
It can be shown by using integration by parts that, $\left\langle m,\textcircled a\right\rangle _{\widetilde{\vec{v}}}=0$
and $\left\langle m,\textcircled b\right\rangle _{\widetilde{\vec{v}}}=0$,
and the key is to ensure these symmetries discretely. These symmetries
are ensure by multiplying the inertial term by a velocity-space dependent
function, $\gamma_{t}\left(\vec{v}\right)$, such that, at a time-step
$p$,}

\textcolor{black}{
\begin{eqnarray}
\left\langle v_{||,i}^{(p)},\delta_{\widetilde{\vec{v}}}\cdot\left(\gamma_{t,i}^{(p+1)}\underbrace{\vec{J}_{t,i}^{(p+1)}}_{\textcircled c}\right)\right\rangle _{\delta\vec{\widetilde{v}}}-\nonumber \\
\left[\left\langle v_{||,i}^{(p)},\delta_{t}\left(J_{r^{2}\xi}\widetilde{f}\right)_{i}^{(p+1)}\right\rangle _{\delta\vec{\widetilde{v}}}-\left\langle 1,\delta_{t}\left(v_{||}J_{r^{2}\xi}\widetilde{f}\right)_{i}^{(p+1)}\right\rangle _{\delta\vec{\widetilde{v}}}\right]=0\label{eq:temporal_semi_discrete_momentum_constraint}
\end{eqnarray}
and 
\begin{eqnarray}
\left\langle \frac{\left[v_{i}^{(p)}\right]^{2}}{2},\delta_{\widetilde{\vec{v}}}\cdot\left(\gamma_{t,i}^{(p+1)}\vec{J}_{t,i}^{(p+1)}\right)\right\rangle _{\delta\widetilde{\vec{v}}}-\nonumber \\
\left[\left\langle \frac{\left[v_{i}^{(p)}\right]^{2}}{2},\delta_{t}\left(J_{r^{2}\xi}\widetilde{f}\right)_{i}^{(p+1)}\right\rangle _{\delta\widetilde{\vec{v}}}-\left\langle 1,\delta_{t}\left(\frac{v^{2}}{2}J_{r^{2}\xi}\widetilde{f}\right)_{i}^{(p+1)}\right\rangle _{\delta\widetilde{\vec{v}}}\right]=0.\label{eq:temporal_semi_discrete_energy_constraint}
\end{eqnarray}
Here, $i$ is the spatial cell index (velocity-space indices are dropped
for brevity), $\delta_{\widetilde{\vec{v}}}\cdot$ is the discrete
velocity-space divergence operator, $\delta_{t}$ is the discrete
temporal derivative, $\gamma_{t}\left(\vec{\widetilde{v}}\right)=1+{\cal O}\left(\Delta_{v}^{\beta},\Delta_{t}^{\zeta}\right)$
is the discrete-nonlinear-constraint function \citep{Taitano_2018_vrfp_1d2v_implicit},
where $\beta$ and $\zeta$ are the velocity-space and temporal discretization
truncation order, the term $\textcircled c$ is the discrete flux
for the inertial term in Eq. \ref{eq:vfp_discrete_eqn} , due to the
temporal variation of the velocity-space transformation metrics {[}details
shown in Eqs. (\ref{eq:discrete_temporal_flux_parallel})-(\ref{eq:temporal_inertial_coeff_perpendicular}){]},
and 
\[
\vec{v}_{i}^{(p)}=v_{i}^{*,(p)}\left[\widetilde{\vec{v}}+\widehat{u}_{||}^{*,(p)}\vec{e}_{||}\right],
\]
\[
\delta_{t}\vec{v}_{i}^{(p+1)}=\frac{c^{(p+1)}\vec{v}_{i}^{(p)}+c^{(p)}\vec{v}_{i}^{(p-1)}+c^{(p-1)}\vec{v}_{i}^{(p-2)}}{\Delta t^{(p)}},
\]
\[
\delta_{t}\left(J_{S\xi}\widetilde{f}\right)_{i}^{(p+1)}=\frac{c^{(p+1)}J_{S\xi,i}^{(p+1)}\widetilde{f}_{i}^{(p+1)}+c^{(p)}J_{S\xi,i}^{(p)}\widetilde{f}_{i}^{(p)}+c^{(p-1)}J_{S\xi,i}^{(p-1)}\widetilde{f}_{i}^{(p-1)}}{\Delta t^{(p)}},
\]
\[
\delta_{t}\left(v_{||}J_{S\xi}\widetilde{f}\right)_{i}^{(p+1)}=\frac{c^{(p+1)}v_{||,i}^{(p)}J_{S\xi,i}^{(p+1)}\widetilde{f}_{i}^{(p+1)}+c^{(p)}v_{||,i}^{(p-1)}J_{S\xi,i}^{(p)}\widetilde{f}_{i}^{(p)}+c^{(p-1)}v_{||,i}^{(p-2)}J_{S\xi,i}^{(p-1)}\widetilde{f}_{i}^{(p-1)}}{\Delta t^{(p)}},
\]
\[
\delta_{t}\left(\frac{v^{2}}{2}J_{S\xi}\widetilde{f}\right)_{i}^{(p+1)}=\frac{c^{(p+1)}\left[v_{i}^{(p)}\right]^{2}J_{S\xi,i}^{(p+1)}\widetilde{f}_{i}^{(p+1)}+c^{(p)}\left[v_{i}^{(p-1)}\right]^{2}J_{S\xi,i}^{(p)}\widetilde{f}_{i}^{(p)}+c^{(p-1)}\left[v_{i}^{(p-2)}\right]^{2}J_{S\xi,i}^{(p-1)}\widetilde{f}_{i}^{(p-1)}}{2\Delta t^{(p)}}.
\]
To evaluate $\gamma_{t}$, we follow Refs. \citep{Taitano_2018_vrfp_1d2v_implicit,taitano_cpc_2020_1d2v_cartesian_phase_space}
and begin by assuming a velocity space dependent local (in configuration
space) functional representation:}

\textcolor{black}{{} 
\begin{equation}
\gamma_{t}\left(\widetilde{v}_{||},\widetilde{v}_{\perp}\right)=1+\sum_{l=0}^{P}C_{l}B_{l}\left(\widetilde{v}_{||},\widetilde{v}_{\perp}\right),\label{eq:temporal_discrete_nonlinear_constraint_function}
\end{equation}
where 
\[
\sum_{l=0}^{P}C_{l}B_{l}\left(\widetilde{v}_{||},\widetilde{v}_{\perp}\right)=\sum_{r=0}^{P_{||}}\sum_{s=0}^{P_{\perp}}C_{rs}B_{||,r}\left(\widetilde{v}_{||}\right)B_{\perp,s}\left(\widetilde{v}_{\perp}\right),
\]
$B_{||,r}$ is a $r^{th}$ representation in the parallel velocity
component, $B_{\perp,s}$ is a similar quantity in the perpendicular
velocity component, and $C_{rs}$ is the coefficient corresponding
to the respective functions. In this study, we chose a Fourier representation
where: 
\[
B_{||,r}=\begin{cases}
1 & \textnormal{if}\;r=0\\
\textnormal{sin}\left[rk_{||}\left(\widetilde{v}_{||}+\widehat{u}_{||}^{*}-u_{||}/v^{*}\right)\right] & \textnormal{if}\;\textnormal{mod}\left(r,2\right)=0\\
\textnormal{cos}\left[\left(r-1\right)k_{||}\left(\widetilde{v}_{||}+\widehat{u}_{||}^{*}-u_{||}/v^{*}\right)\right] & \textnormal{if}\;\textnormal{mod}\left(r,2\right)=1
\end{cases},\;\;\;B_{\perp,s}=\begin{cases}
1 & \textnormal{if}\;s=0\\
\textnormal{sin}\left[sk_{\perp}\widetilde{v}_{\perp}\right] & \textnormal{if}\;\textnormal{mod}\left(s,2\right)=0\\
\textnormal{cos}\left[\left(s-1\right)k_{\perp}\widetilde{v}_{\perp}\right] & \textnormal{if}\;\textnormal{mod}\left(s,2\right)=1
\end{cases},
\]
and $k_{||}=2\pi/L_{||}$, $k_{\perp}=2\pi/\widetilde{L}_{\perp}$
are the wave vectors. We also choose $r=s=\left(0,1,2\right)$. The
coefficients, $C_{l}$, are obtained by minimizing their amplitude
while satisfying the discrete symmetry constraints as given by Eqs.
(\ref{eq:temporal_semi_discrete_momentum_constraint}) and (\ref{eq:temporal_semi_discrete_energy_constraint}).
This is done by solving a constrained-minimization problem for the
following cost function:}

\textcolor{black}{
\begin{equation}
{\cal F}\left(\vec{C},\bm{\lambda}\right)=\frac{1}{2}\sum_{l=0}^{P}C_{l}^{2}-\bm{\lambda}^{T}\cdot\vec{M}.\label{eq:cost_function_t}
\end{equation}
Here, $\vec{C}=\left[C_{1},C_{2},\cdots,C_{P}\right]^{T}$, $\bm{\lambda}$
is a vector of Lagrange multipliers and $\vec{M}$ is the vector of
constraints {[}Eqs. (\ref{eq:temporal_semi_discrete_momentum_constraint})
and (\ref{eq:temporal_semi_discrete_energy_constraint}){]}; and $\vec{C}$
is obtained from the linear system:}

\textcolor{black}{
\begin{equation}
\left[\begin{array}{c}
\partial_{\bm{C}}{\cal F}\\
\partial_{\bm{\lambda}}{\cal F}
\end{array}\right]=\vec{0}.\label{eq:minimization_problem}
\end{equation}
We note that, similarly to previous studies employing discrete nonlinear
constraints \citep{taitano-jcp-15-vfp,Taitano_2015_jcp_cmec_va,Taitano_2015_rfp_0d2v_implicit,Taitano_2016_rfp_0d2v_implicit,Taitano_2018_vrfp_1d2v_implicit},
since $\gamma_{t}$ is an implicit function of the solution, for nonlinearly
implicit system such as ours the quality of discrete conservation
properties depends on the prescribed nonlinear convergence tolerance
of our solver (as demonstrated in Sec. \ref{subsec:doubly_reflective_bc_problem}).}

\subsection{\textcolor{black}{Spatial variation of $v^{*}$ and $\widehat{u}_{||}^{*}$\label{subsec:spatial_variation_conservation}}}

\textcolor{black}{Similarly to the temporal variation, conservation
symmetries for the case of spatial variation of the velocity-space
metrics can be shown independently. Consider only the spatial gradient
terms in the Vlasov equation, Eq. (\ref{eq:composite_transformed_vfp}),
to obtain the following expression:}

\textcolor{black}{{} 
\begin{equation}
\partial_{\xi}\left(v_{||,eff}\widetilde{f}\right)_{\vec{\widetilde{v}},t}-\frac{1}{v^{*}}\frac{\partial}{\partial\widetilde{\vec{v}}}\cdot\left(\left.\partial_{\xi}\vec{v}\right|_{\vec{\widetilde{v}},t}v_{||,eff}\ftilde\right).\label{eq:continuum_spatial_inertial_vlasov}
\end{equation}
Here, $v_{||,eff}=J_{S}v^{*}\left(\widetilde{v}_{||}+\widehat{u}_{||}^{*}\right)-\dot{J}_{r^{3}}$
and $\left.\partial_{\xi}\vec{v}\right|_{\vec{\widetilde{v}},t}=\partial_{\xi}\left[v^{*}\left(\vec{v}+\widehat{u}_{||}^{*}\vec{e}_{||}\right)\right]$,
and the mass conservation theorem is revealed by taking the $mv^{0}$
moment, assuming a periodic boundary condition, and integrating in
$\xi$ to find 
\begin{equation}
\int_{0}^{1}\left\langle 1,\partial_{\xi}\left(mv_{||,eff}\widetilde{f}\right)\right\rangle _{\widetilde{\vec{v}}}d\xi=0.
\end{equation}
Note that the inertial term is in a divergence form in the velocity
space, and therefore its $mv^{0}$ moment trivially vanishes both
in the continuum and discretely.}

\textcolor{black}{In the transformed coordinate, the momentum and
energy conservations are defined as:}

\textcolor{black}{{} 
\begin{equation}
\int_{0}^{1}d\xi\left\{ \left\langle v_{||},\partial_{\xi}\left(v_{||,eff}\widetilde{f}\right)_{\vec{\widetilde{v}},t}\right\rangle _{\widetilde{\vec{v}}}-\left\langle v_{||},\frac{\partial}{\partial\widetilde{\vec{v}}}\cdot\left(\frac{\left.\partial_{\xi}\vec{v}\right|_{\vec{\widetilde{v}},t}}{v^{*}}v_{||,eff}\ftilde\right)_{\xi,t}\right\rangle _{\widetilde{\vec{v}}}\right\} =0.\label{eq:spatial_momentum_conservation_symmetry_2}
\end{equation}
and 
\begin{equation}
\int_{0}^{1}d\xi\left\{ \left\langle \frac{v^{2}}{2},\partial_{\xi}\left(v_{||,eff}\widetilde{f}\right)_{\widetilde{\vec{v}},t}\right\rangle _{\widetilde{\vec{v}}}-\left\langle \frac{v^{2}}{2},\frac{1}{v^{*}}\frac{\partial}{\partial\widetilde{\vec{v}}}\cdot\left(\left.\partial_{\xi}\vec{v}\right|_{\widetilde{v},t}v_{||,eff}\ftilde\right)_{\widetilde{\vec{v}},t}\right\rangle _{\widetilde{\vec{v}}}\right\} =0.\label{eq:spatial_energy_conservation_symmetry_2}
\end{equation}
These symmetries are ensured in the discrete by modifying the inertial
term by a velocity-space dependent function, $\gamma_{r}\left(\vec{\widetilde{v}}\right)$,
such that the following relationships are satisfied:}

\textcolor{black}{
\begin{eqnarray}
\sum_{i=1}^{N_{\xi}}\Delta\xi\left\{ \left\langle v_{||,i}^{(p)},\delta_{\xi}\left(F_{r}+F_{\dot{r}}\right)_{i}^{(p+1)}\right\rangle _{\delta\widetilde{\vec{v}}}\right.\nonumber \\
\left.-\left\langle v_{||,i}^{(p)},\frac{1}{2}\delta_{\widetilde{\vec{v}}}\cdot\left(\gamma_{r,i+1/2}^{(p+1)}\underbrace{\vec{J}_{r}^{-,(p+1)}}_{\textcircled a}+\gamma_{r,i-1/2}^{(p+1)}\underbrace{\vec{J}_{r}^{+,(p+1)}}_{\textcircled b}\right)_{i}\right\rangle _{\delta\widetilde{\vec{v}}}\right\} =0\label{eq:spatial_semi_discrete_momentum_constraint}
\end{eqnarray}
and 
\begin{eqnarray}
\sum_{i=1}^{N_{\xi}}\Delta\xi\left\{ \left\langle \frac{\left[v_{i}^{(p)}\right]^{2}}{2},\delta_{\xi}\left(F_{r}+F_{\dot{r}}\right)_{i}^{(p+1)}\right\rangle _{\delta\widetilde{\vec{v}}}\right.\nonumber \\
\left.-\left\langle \frac{\left[v_{i}^{(p)}\right]^{2}}{2},\frac{1}{2}\delta_{\widetilde{\vec{v}}}\cdot\left(\gamma_{r,i+1/2}^{(p+1)}\vec{J}_{r}^{-,(p+1)}+\gamma_{r,i-1/2}^{(p+1)}\vec{J}_{r}^{+,(p+1)}\right)_{i}\right\rangle _{\delta\widetilde{\vec{v}}}\right\} =0.\label{eq:spatial_semi_discrete_energy_constraint}
\end{eqnarray}
Here, $\vec{v}_{i}^{(p)}=v_{i}^{*,(p)}\left[\widetilde{\vec{v}}+\widehat{u}_{||,i}^{*,(p)}\vec{e}_{||}\right]$,
$\gamma_{x}\left(\vec{\widetilde{v}}\right)=1+{\cal O}\left(\Delta_{v}^{\beta},\Delta_{x}^{\eta}\right)$
is the spatial discrete-nonlinear-constraint function, with the functional
form defined similarly to $\gamma_{t}$ {[}Eq. (\ref{eq:temporal_discrete_nonlinear_constraint_function}){]},
and terms $\textcircled a$ and $\textcircled b$ are the discrete
fluxes for the inertial term in Eq. (\ref{eq:vfp_discrete_eqn}) {[}details
shown in Eqs. (\ref{eq:discrete_spatial_inertial_flux_paralle_ipm})-(\ref{eq:discrete_spatial_inertial_flux_perpendicular_coefficient-1}){]}.
Performing a discrete integration by parts (i.e., telescoping the
summation) on Eqs. (\ref{eq:spatial_semi_discrete_momentum_constraint})
and (\ref{eq:spatial_semi_discrete_energy_constraint}), we obtain
the following constraints that relate $\gamma_{r,i+1/2}$, the discrete
configuration-space flux, and the velocity-space inertial terms for
momentum conservation: 
\begin{eqnarray}
\left\langle v_{||,i}^{(p)}-v_{||,i+1}^{(p)},\left(F_{r}-F_{\dot{r}}\right)_{i+1/2}^{(p+1)}\right\rangle _{\delta\widetilde{\vec{v}}}\nonumber \\
-\frac{\Delta\xi}{2}\left[\left\langle v_{||,i}^{(p)},\delta_{\widetilde{\vec{v}}}\cdot\left(\gamma_{r,i+1/2}^{(p+1)}\vec{J}_{r}^{-,(p+1)}\right)_{i}\right\rangle _{\delta\widetilde{\vec{v}}}+\left\langle v_{||,i+1}^{(p)},\delta_{\widetilde{\vec{v}}}\cdot\left(\gamma_{r,i+1/2}^{(p+1)}\vec{J}_{r}^{+,(p+1)}\right)_{i+1}\right\rangle _{\delta\widetilde{\vec{v}}}\right]=0,\label{eq:semi_discrete_spatial_momentum_constraint_3}
\end{eqnarray}
and for energy conservation: 
\begin{eqnarray}
\left\langle \frac{\left[v_{i}^{(p)}\right]^{2}}{2}-\frac{\left[v_{i+1}^{(p)}\right]^{2}}{2},\left(F_{r}-F_{\dot{r}}\right)_{i+1/2}^{(p+1)}\right\rangle _{\delta\widetilde{\vec{v}}}\nonumber \\
-\frac{\Delta\xi}{2}\left[\left\langle \frac{\left[v_{i}^{(p)}\right]^{2}}{2},\delta_{\widetilde{\vec{v}}}\cdot\left(\gamma_{r,i+1/2}^{(p+1)}\vec{J}_{r}^{-,(p+1)}\right)_{i}\right\rangle _{\delta\widetilde{\vec{v}}}+\left\langle \frac{\left[v_{i+1}^{(p)}\right]^{2}}{2},\delta_{\widetilde{\vec{v}}}\cdot\left(\gamma_{r,i+1/2}^{(p+1)}\vec{J}_{r}^{+,(p+1)}\right)_{i+1}\right\rangle _{\delta\widetilde{\vec{v}}}\right]=0.\label{eq:semi_discrete_spatial_energy_constraint_3}
\end{eqnarray}
The vector of coefficients, $\vec{C}$, for $\gamma_{r,i+1/2}$ is
evaluated by solving a constrained minimization problem as in Eq.
(\ref{eq:minimization_problem}) with the vector of vanishing constraints,
$\vec{M}$, being Eqs. (\ref{eq:semi_discrete_spatial_momentum_constraint_3})
and (\ref{eq:semi_discrete_spatial_energy_constraint_3}). We note
that at configuration-space boundaries we set $\gamma_{r,i+1/2}=1$
(and $\partial_{\xi}v^{*}=\partial_{\xi}\widehat{u}_{||}=\partial_{\xi}\vec{v}=0$),
as the (non-periodic) boundary conditions violate the continuum conservation
principle. Further, in regions in configuration space where the truncation
error is large (e.g., under-resolved shock fronts) such that $\sqrt{\frac{1}{2P_{||}P_{|perp}}\sum_{r=0}^{P_{||}}\sum_{s=0}^{P_{\perp}}C_{rs}^{2}}>\gamma_{tol}$
where $\gamma_{r}=1+\sum_{r=0}^{P_{||}}\sum_{s=0}^{P_{\perp}}C_{rs}B_{||,r}B_{\perp,s}$,
we set $\gamma_{r}=0\;\forall\;\vec{\widetilde{v}}$ for numerical
robustness purposes ($\gamma_{tol}=0.5$ in this study). However,
the phase-space grid adaptation strategy and the Winslow smoothing
of the velocity space metrics, $(v^{*},u_{||}^{*})$, provide sufficient
resolution at most times, and such situations are rare.}

\subsection{Spherical inertial term\label{subsubsec:spherical_inertial_term}}

Consider the inertial term due to the transformation from Cartesian
to spherical geometry: 
\[
v^{*^{-1}}r^{-1}\partial_{\vec{\widetilde{v}}}\cdot\left(J_{S}\vec{\widetilde{a}}f\right).
\]
The mass conservation is trivially ensured due to the divergence form
of the operator while conservation of linear momentum is automatically
satisfied via spherical symmetry. The energy conservation theorem
is shown by, 
\begin{equation}
r^{-1}v^{*^{-1}}\left\langle \frac{v^{2}}{2}\partial_{\vec{\widetilde{v}}}\cdot\left(J_{S}\vec{\widetilde{a}}f\right)\right\rangle _{\vec{v}}=-J_{S}r^{-1}v^{*^{-1}}\left[\underbrace{\left\langle v_{||},v_{\perp}^{2}\widetilde{f}\right\rangle _{\vec{\widetilde{v}}}}_{\textcircled a}-\underbrace{\left\langle v_{\perp},v_{||}v_{\perp}\widetilde{f}\right\rangle _{\vec{\widetilde{v}}}}_{\textcircled b}\right]=0.\label{eq:spherical_inertial_energy_symmetry}
\end{equation}
In the discrete, we ensure the cancellation between terms $\textcircled a$
and $\textcircled b$\textcolor{black}{{} by modifying the parallel-velocity
component of spherical geometry inertial term by a constant, $\gamma_{S}$,
such that: 
\begin{eqnarray}
\left\langle \frac{\left[v_{i}^{(p)}\right]^{2}}{2},\gamma_{S,i}\delta_{\widetilde{v}_{||}}\left(J_{S,||}^{(p+1)}\right)_{i}+\widetilde{v}_{\perp}^{-1}\delta_{\widetilde{v}_{\perp}}\left(\widetilde{v}_{\perp}J_{S,\perp}^{(p+1)}\right)_{i}\right\rangle _{\delta\widetilde{\vec{v}}}=0.\label{eq:spherical_inertial_energy_symmetry_discrete}
\end{eqnarray}
Here, $\textcircled a$ and $\textcircled b$ are the discrete fluxes
for inertial term in Eq. (\ref{eq:vfp_discrete_eqn}) {[}details shown
in Eqs. (\ref{eq:geo_inert_para}) and (\ref{eq:geo_inert_perp}){]},
and $\gamma_{S}$ is trivially evaluated as: 
\begin{equation}
\gamma_{S,i}=-\frac{\left\langle \frac{\left[v_{i}^{(p)}\right]^{2}}{2},\delta_{\widetilde{v}_{\perp}}\left(J_{S,\perp}^{(p+1)}\right)_{i}\right\rangle _{\delta\widetilde{\vec{v}}}}{\left\langle \frac{\left[v_{i}^{(p)}\right]^{2}}{2},\delta_{\widetilde{v}_{||}}\left(J_{S,||}^{(p+1)}\right)_{i}\right\rangle _{\delta\widetilde{\vec{v}}}}.\label{eq:gamma_S_definition}
\end{equation}
We close by noting that the so-called }\textcolor{black}{\emph{skew
bracket }}\textcolor{black}{formalism \citep{quispel_pra_1996_solving_odes_preserving_first_integral,Gasteiger_JOPP_2016_ADI_type_PC_for_SS_Vlasov,kormann_jcp_2021_energy_conserving_time_propagation}
may be an alternative approach to ensure the conservation symmetries
when considering an adaptive phase-space grid. Investigation to such
an approach is left for future consideration.}

\section{Discretization of the Vlasov Equation Phase-Space Fluxes \label{app:discretization_of_fluxes}}

The individual terms in Eq. (\ref{eq:vfp_discrete_eqn}) are elaborated.
The term $\textcircled a$ corresponds to the discrete representation
of the spatial streaming term, with

\begin{equation}
F_{r,\alpha,i+1/2,j,k}^{(p+1)}=J_{S,i+1/2}v_{\alpha,i+1/2}^{*,(p)}\left(\widetilde{v}_{||,j}+\widehat{u}_{||,\alpha,i+1/2}^{*,(p)}\right)\textnormal{SMART}\left(\widetilde{v}_{||,j}+\widehat{u}_{||,\alpha,i+1/2}^{*,(p)},\widetilde{f}_{\alpha}^{(p+1)}\right)_{i+1/2,j,k},\label{eq:discrete_streaming_flux}
\end{equation}

\[
v_{\alpha,i+1/2}^{*,(p)}=\frac{v_{\alpha,i+1}^{*,(p)}+v_{\alpha,i}^{*,(p)}}{2},\;\;\textnormal{and}\;\;\widehat{u}_{||,\alpha,i+1/2}^{*,(p)}=\frac{\widehat{u}_{||,\alpha,i}^{*,(p)}+\widehat{u}_{||,\alpha,i+1}^{*,(p)}}{2},
\]
where $\textnormal{SMART}\left(a,\phi\right)_{face}$ \textcolor{black}{denotes
a SMART cell-face interpolation operation \citep{smart_limiter_gaskel_lau_1988}
of a scalar $\phi$ at a cell face with a given velocity $a$, 
\[
\textnormal{SMART}\left(a,\phi\right)_{face}=\sum_{i'=1}^{N}c_{face,i'}\left(a,\phi\right)\phi_{i'}.
\]
Here, coefficients $c_{face,i'}$ are the interpolation weights for
cell index $i'$ surrounding the cell face. }

The term $\textcircled b$ corresponds to the inertial term in the
configuration space, arising from the moving grid, with

\begin{equation}
F_{\dot{r},\alpha,i+1/2,j,k}^{(p+1)}=-\dot{J}_{r^{3},i+1/2}^{(p+1)}\textnormal{SMART}\left(-\dot{J}_{r^{3},i+1/2}^{(p+1)},\widetilde{f}_{\alpha}^{(p+1)}\right)_{i+1/2,j,k},\label{eq:discrete_grid_motion_flux}
\end{equation}
where the definition of $\dot{J}_{r^{3},i+1/2}$ is given in Sec.
\ref{subsec:nonlinear_mmpde}.

The term $\textcircled c$ corresponds to the electrostatic-acceleration
term with

\begin{equation}
J_{acc,\alpha,i,j+1/2,k}^{(p+1)}=J_{S\xi,i}^{(p+1)}\frac{q_{\alpha}}{m_{\alpha}}\frac{E_{||,i}^{(p+1)}}{v_{\alpha,i}^{*,p}}\textnormal{SMART}\left(q_{\alpha}E_{||,i}^{(p+1)},\widetilde{f}_{\alpha}^{(p+1)}\right)_{i,j+1/2,k}.\label{eq:discrete_acceleration_flux}
\end{equation}

The term $\textcircled d$ corresponds to the inertial terms due to
temporal variation of the velocity-space metrics (i.e., $v_{\alpha}^{*}$
and $\widehat{u}_{||,\alpha}^{*}$) with

\begin{equation}
J_{t,||,\alpha,i,j+1/2,k}^{(p+1)}={\cal I}_{||,t,\alpha,i,j+1/2,k}^{(p)}\textnormal{SMART}\left({\cal I}_{||,t,\alpha,i+1/2,j+1/2,k}^{(p+1)},\widetilde{f}_{\alpha}^{(p+1)}\right)_{i,j+1/2,k}\label{eq:discrete_temporal_flux_parallel}
\end{equation}
and

\begin{equation}
J_{t,\perp,\alpha,i,j,k+1/2}^{(p+1)}={\cal I}_{\perp,t,\alpha,i,j,k+1/2}^{(p+1)}\textnormal{SMART}\left({\cal I}_{\perp,t,\alpha,i,j,k+1/2}^{(p+1)},\widetilde{f}_{\alpha}^{(p+1)}\right)_{i,j,k+1/2},\label{eq:discrete_temporal_flux_perpendicular}
\end{equation}

\begin{eqnarray}
{\cal I}_{t,||,\alpha,i,j+1/2,k}^{(p+1)}=-\frac{J_{S\xi,i}^{(p+1)}}{v_{\alpha,i}^{*,(p)}}\delta_{t}\left(v_{\alpha}^{*}\left[\widetilde{v}_{||,j+1/2}+\widehat{u}_{||,\alpha}^{*}\right]\right)_{i}^{(p+1)},\label{eq:temporal_inertial_coeff_parallel}
\end{eqnarray}
and 
\begin{eqnarray}
{\cal I}_{t,\perp,\alpha,i,j,k+1/2}^{(p+1)}=-\frac{J_{S\xi,i}^{(p+1)}}{v_{\alpha,i}^{*,(p)}}\left[\delta_{t}\left(v_{\alpha}^{*}\widetilde{v}_{\perp}\right)_{i,k+1/2}^{(p+1)}\right].\label{eq:temporal_inertial_coeff_perpendicular}
\end{eqnarray}
As also described in Ref. \citep{Taitano_2016_rfp_0d2v_implicit},
we lag the time level between the BDF2 coefficients and the normalization
speed (and similarly, the shift velocity) to avoid over-constraining
the nonlinear residual (we refer the readers to the reference for
further detail).

The term $\textcircled e$ corresponds to the inertial terms due to
the spatial variation of the metrics with

\begin{equation}
J_{r,||,\alpha,i,j+1/2,k}^{(p+1),-}={\cal I}_{r,||,\alpha,i,j+1/2,k}^{(p+1),-}\textnormal{SMART}\left({\cal I}_{r,||,\alpha,i,j+1/2,k}^{(p+1),-},\widetilde{f}_{\alpha}^{(p+1)}\right)_{i,j+1/2,k}\label{eq:discrete_spatial_inertial_flux_paralle_ipm}
\end{equation}
\begin{equation}
J_{r,||,\alpha,i,j+1/2,k}^{(p+1),+}={\cal I}_{r,||,\alpha,i,j+1/2,k}^{(p+1),+}\textnormal{SMART}\left({\cal I}_{r,||,\alpha,i,j+1/2,k}^{(p+1),+},\widetilde{f}_{\alpha}^{(p+1)}\right)_{i,j+1/2,k}\label{eq:discrete_spatial_inertial_flux_paralle_ipm-1}
\end{equation}
and 
\begin{equation}
J_{r,\perp,\alpha,i,j,k+1/2}^{(p+1),-}={\cal I}_{r,\perp,\alpha,i,j,k+1/2}^{(p+1),-}\textnormal{SMART}\left({\cal I}_{r,\perp,\alpha,i,j,k+1/2}^{(p+1),-},\widetilde{f}_{\alpha}^{(p+1)}\right)_{i,j,k+1/2},\label{eq:discrete_spatial_inertial_flux_perpendicular_ipm}
\end{equation}

\begin{equation}
J_{r,\perp,\alpha,i,j,k+1/2}^{(p+1),+}={\cal I}_{r,\perp,\alpha,i,j,k+1/2}^{(p+1),+}\textnormal{SMART}\left({\cal I}_{r,\perp,\alpha,i,j,k+1/2}^{(p+1),+},\widetilde{f}_{\alpha}^{(p+1)}\right)_{i,j,k+1/2},\label{eq:discrete_spatial_inertial_flux_perpendicular_ipm-1}
\end{equation}
where

\begin{eqnarray}
{\cal I}_{r,||,\alpha,i,j+1/2,k}^{(p+1),-}=-\left\{ \left[J_{S}v_{\alpha}^{*}\left(\widetilde{v}_{||}+\widehat{u}_{||,\alpha}^{*}\right)-\dot{J}_{r^{3}}\right]\frac{\partial}{\partial\xi}\left[v_{\alpha}^{*}\left(\widetilde{v}_{||}+\widehat{u}_{||,\alpha}^{*}\right)\right]\right\} _{i+1/2,j+1/2,k}^{(p+1)}\nonumber \\
\approx-\left[J_{S,i+1/2}v_{\alpha,i+1/2}^{*,(p)}\left(\widetilde{v}_{||,j+1/2}+\widehat{u}_{||,\alpha,i+1/2}^{*,(p)}\right)-\dot{J}_{r^{3},i+1/2}^{(p+1)}\right]\times\nonumber \\
\frac{v_{\alpha,i+1}^{*,(p)}\left(\widetilde{v}_{||,j+1/2}+\widehat{u}_{||,\alpha,i+1}^{*,(p)}\right)-v_{\alpha,i}^{*,(p)}\left(\widetilde{v}_{||,j+1/2}+\widehat{u}_{||,\alpha,i}^{*,(p)}\right)}{\Delta\xi},\label{eq:discrete_spatial_inertial_flux_parallel_coefficient}
\end{eqnarray}
\begin{eqnarray}
{\cal I}_{r,||,\alpha,i,j+1/2,k}^{(p+1),+}=-\left\{ \left[J_{S}v_{\alpha}^{*}\left(\widetilde{v}_{||}+\widehat{u}_{||,\alpha}^{*}\right)-\dot{J}_{r^{3}}\right]\frac{\partial}{\partial\xi}\left[v_{\alpha}^{*}\left(\widetilde{v}_{||}+\widehat{u}_{||,\alpha}^{*}\right)\right]\right\} _{i-1/2,j+1/2,k}^{(p+1)}\nonumber \\
\approx-\left[J_{S,i-1/2}v_{\alpha,i+1/2}^{*,(p)}\left(\widetilde{v}_{||,j+1/2}+\widehat{u}_{||,\alpha,i-1/2}^{*,(p)}\right)-\dot{J}_{r^{3},i-1/2}^{(p+1)}\right]\times\nonumber \\
\frac{v_{\alpha,i}^{*,(p)}\left(\widetilde{v}_{||,j+1/2}+\widehat{u}_{||,\alpha,i}^{*,(p)}\right)-v_{\alpha,i-1}^{*,(p)}\left(\widetilde{v}_{||,j+1/2}+\widehat{u}_{||,\alpha,i-1}^{*,(p)}\right)}{\Delta\xi},\label{eq:discrete_spatial_inertial_flux_parallel_coefficient-1}
\end{eqnarray}

\begin{eqnarray}
{\cal I}_{r,\perp,\alpha,i,j,k+1/2}^{(p),-}=-\left\{ J_{S}\left[v_{\alpha}^{*}\left(\widetilde{v}_{||}+\widehat{u}_{||,\alpha}^{*}\right)-\dot{J}_{r^{3}}\right]\frac{\partial}{\partial\xi}\left(v_{\alpha}^{*}\widetilde{v}_{\perp}\right)\right\} _{i+1/2,j,k+1/2}^{(p+1)}\nonumber \\
\approx-\left[J_{S,i+1/2}v_{\alpha,i+1/2}^{*,(p)}\left(\widetilde{v}_{||,j}+\widehat{u}_{||,\alpha,i+1/2}^{*,(p)}\right)-\dot{J}_{r^{3},i+1/2}^{(p+1)}\right]\frac{v_{\alpha,i+1}^{*,(p)}\widetilde{v}_{\perp,k+1/2}-v_{\alpha,i}^{*,(p)}\widetilde{v}_{\perp,k+1/2}}{\Delta\xi},\label{eq:discrete_spatial_inertial_flux_perpendicular_coefficient}
\end{eqnarray}
and 
\begin{eqnarray}
{\cal I}_{r,\perp,\alpha,i,j,k+1/2}^{(p),+}=-\left\{ J_{S}\left[v_{\alpha}^{*}\left(\widetilde{v}_{||}+\widehat{u}_{||,\alpha}^{*}\right)-\dot{J}_{r^{3}}\right]\frac{\partial}{\partial\xi}\left(v_{\alpha}^{*}\widetilde{v}_{\perp}\right)\right\} _{i-1/2,j,k+1/2}^{(p+1)}\nonumber \\
\approx-\left[J_{S,i-1/2}v_{\alpha,i-1/2}^{*,(p)}\left(\widetilde{v}_{||,j}+\widehat{u}_{||,\alpha,i-1/2}^{*,(p)}\right)-\dot{J}_{r^{3},i-1/2}^{(p+1)}\right]\frac{v_{\alpha,i}^{*,(p)}\widetilde{v}_{\perp,k+1/2}-v_{\alpha,i-1}^{*,(p)}\widetilde{v}_{\perp,k+1/2}}{\Delta\xi}.\label{eq:discrete_spatial_inertial_flux_perpendicular_coefficient-1}
\end{eqnarray}

The term $\textcircled f$ corresponds to the velocity-space inertial
term due to the transformation of VFP equation from Cartesian to spherical
geometry with

\begin{equation}
J_{S,||,\alpha,i,j+1/2,k}^{(p+1)}=r_{i}^{(p+1)}\left[v_{\alpha,i}^{*,(p)}\widetilde{v}_{\perp,k}\right]^{2}\textnormal{SMART}\left(\widetilde{v}_{\perp,k}^{2},\widetilde{f}_{\alpha}^{(p+1)}\right)_{i,j+1/2,k}\label{eq:geo_inert_para}
\end{equation}
and

\begin{equation}
J_{S,\perp,\alpha,i,j,k+1/2}^{(p+1)}=r_{i}^{(p+1)}\left[v_{\alpha,i}^{*,(p)}\right]^{2}\widetilde{v}_{\perp,k+1/2}\left(\widetilde{v}_{||,j}+\widehat{u}_{||,i}^{*,(p)}\right)\textnormal{SMART}\left(\widetilde{v}_{\perp,k+1/2}\left(\widetilde{v}_{||,j}+\widehat{u}_{||,i}^{*,(p)}\right),\widetilde{f}_{\alpha}^{(p+1)}\right)_{i,j,k+1/2}.\label{eq:geo_inert_perp}
\end{equation}

\section{\textcolor{black}{Intuitions on Various Hyperparameters and Normalization
for Grid Adaptation \label{app:intuition_on_ns_mmpde}}}

\begin{figure}
\begin{centering}
\textcolor{black}{\includegraphics[scale=0.6]{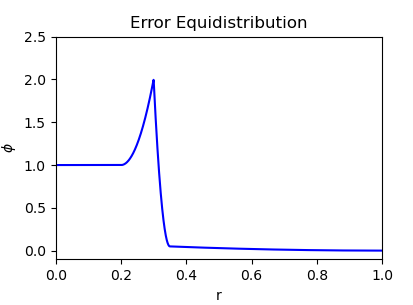}}
\par\end{centering}
\textcolor{black}{\caption{\textcolor{black}{Profile of a reference function, $\phi_{ref}$,
for grid adaptation. \label{fig:reference_solution_profile}}}
}
\end{figure}

\begin{figure}[t]
\begin{centering}
\textcolor{black}{\includegraphics[scale=0.55]{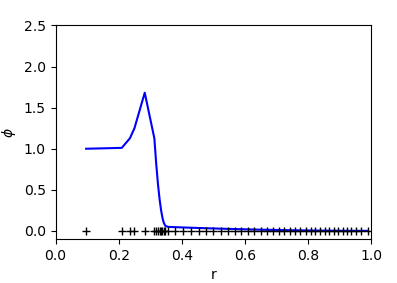}\includegraphics[scale=0.55]{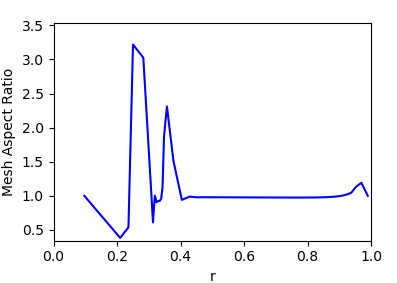}\includegraphics[scale=0.55]{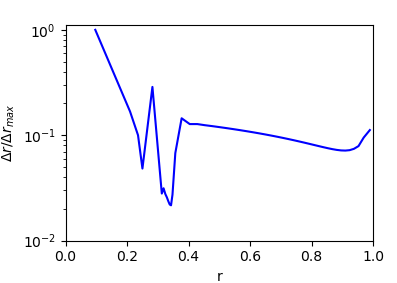}}
\par\end{centering}
\textcolor{black}{\caption{\textcolor{black}{Profiles of the interpolated solution and grid points
(left), the aspect ratio (center), and the cell sizes (right) for
the plain Poisson grid generator without specification of hyperparameters
$\left(\delta_{min}=0,\eta=0,\lambda_{\omega}=0\right)$.\label{fig:plain_mmpde}}}
}
\end{figure}

\textcolor{black}{Herein, we provide intuition on how different grid
hyperparameters affect its evolution. Consider the following reference
function, $\phi$, 
\begin{equation}
\phi=\begin{cases}
1 & \textnormal{if}\;0\le r<0.2\\
1+100\left(r-0.2\right)^{2} & \textnormal{if}\;0.2\le r<0.3\\
0.05+780\left(r-0.35\right)^{2} & \textnormal{if}\;0.3\le r<0.35\\
0.001+\frac{0.049}{0.065^{2}}\left(r-1\right)^{2} & \textnormal{otherwise}
\end{cases}
\end{equation}
 for $r\in\left[0,1\right]$, to which we wish to adapt the grid;
refer to Fig. \ref{fig:reference_solution_profile}.}\textcolor{black}{}
\textcolor{black}{For all generated grids considered in this section,
we employ $N_{\xi}=48$ for the number of grid points and a pre-smoothed
and pre-limited monitor function defined as $\omega^{*}=\delta_{\xi}\ln\phi+\delta_{min}$.
We use the grid aspect ratio,
\begin{equation}
AR_{i}=\frac{1}{2}\left(\frac{\Delta r_{i+1}}{\Delta r_{i}}+\frac{\Delta r_{i}}{\Delta r_{i-1}}\right),\label{eq:grid_aspect_ratio}
\end{equation}
as the metric for the grid quality. As a rule-of-thumb, an $0.8\le AR\le1.2$
will be regarded as a smooth, high-quality grid. The grid and solution
generated from a plain definition of monitor function without the
hyperparameters $\left(i.e.,\delta_{min}=0,\eta=0,\lambda_{\omega}=0\right)$
are shown in Fig. \ref{fig:plain_mmpde}.}
\textcolor{black}{As can be seen, without any processing of the monitor
function, the grid generator attempts to cluster as many points near
sharp gradients as possible, but neglects regions with a smoother
variation (e.g., $0\le r\le0.2$). This leads to large variations
in the grid aspect ratio, which often also results in poor solution
quality \citep{thompson1985numerical,thompson_anm_1985_survey_of_dag,veldman_j_engrg_math_1992_playing_with_mm}.
Further, such grids are brittle to sudden changes in boundary values
(e.g., shock being launched into the system). }

\begin{figure}[t]
\begin{centering}
\textcolor{black}{\includegraphics[scale=0.55]{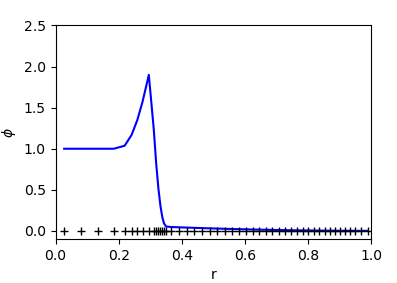}\includegraphics[scale=0.55]{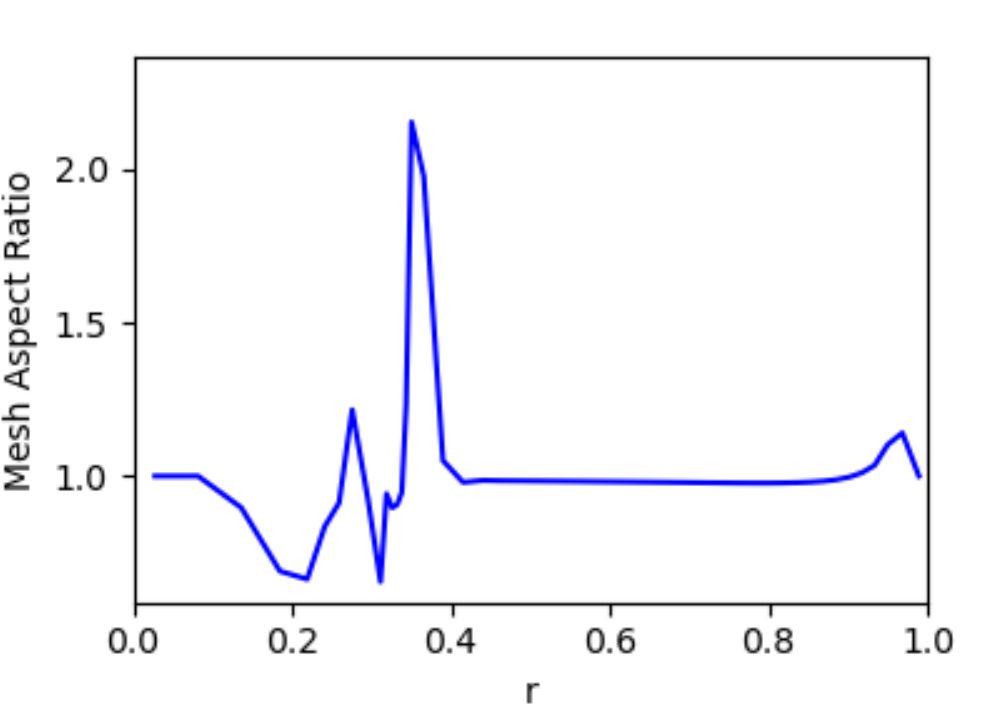}\includegraphics[scale=0.55]{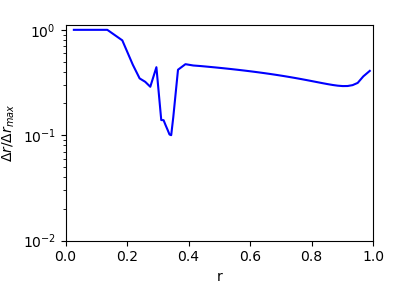}}
\par\end{centering}
\textcolor{black}{\caption{\textcolor{black}{Profiles of the interpolated solution and grid points
(left), the aspect ratio (center), and the cell sizes (right) for
the grid with $\left(\delta_{min}=0,\eta=10^{-1},\lambda_{\omega}=0\right)$.\label{fig:finite_eta_mmpde}}}
}
\end{figure}

\begin{figure}[t]
\begin{centering}
\textcolor{black}{\includegraphics[scale=0.55]{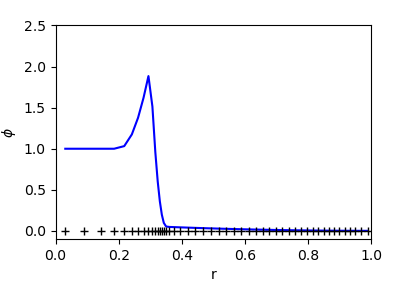}\includegraphics[scale=0.55]{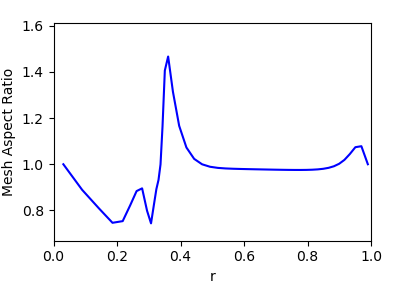}\includegraphics[scale=0.55]{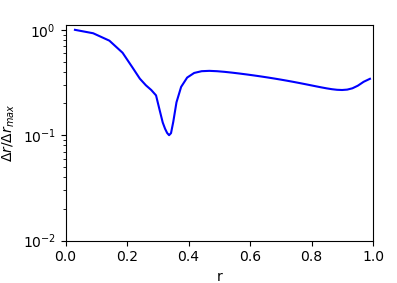}}
\par\end{centering}
\textcolor{black}{\caption{\textcolor{black}{Profiles of the interpolated solution and grid points
(left), the aspect ratio (center), and the cell sizes (right) for
the grid with $\left(\delta_{min}=0,\eta=10^{-1},\lambda_{\omega}=10^{-3}\right)$.\label{fig:finite_eta_smooth_mmpde}}}
}
\end{figure}

\begin{figure}[t]
\begin{centering}
\textcolor{black}{\includegraphics[scale=0.55]{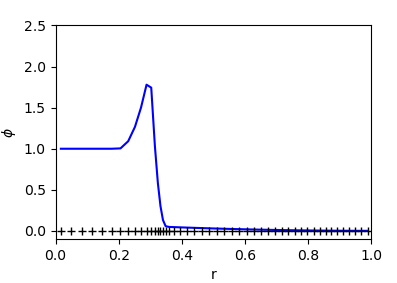}\includegraphics[scale=0.55]{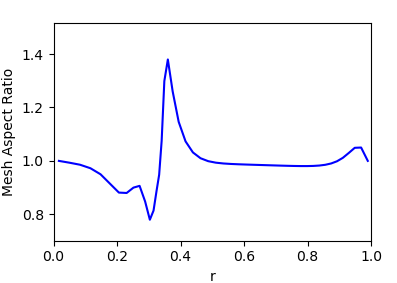}\includegraphics[scale=0.55]{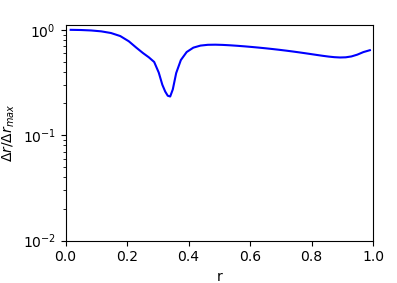}}
\par\end{centering}
\textcolor{black}{\caption{\textcolor{black}{Profiles of the interpolated solution and grid points
(left), the aspect ratio (center), and the cell sizes (right) for
the grid with $\left(\delta_{min}=10^{-1},\eta=10^{-1},\lambda_{\omega}=10^{-3}\right)$.\label{fig:finite_eta_smooth_delta_mmpde}}}
}
\end{figure}

\textcolor{black}{Next, we define a finite cut-off for $\omega_{min}/\omega_{max}\ge\eta=10^{-1}$.
In Fig. \ref{fig:finite_eta_mmpde}, we show the resulting grid with
$(\delta_{min}=0,\eta=10^{-1},\lambda_{\omega}=0)$.}\textcolor{black}{}black
\textcolor{black}{As can be seen, limiting $\omega_{min}/\omega_{max}$
has an immediate effect in improving the grid's quality, as more grid
points are situated in regions with smoother variation in $\phi$.
The grid quality, however, is still unsatisfactory, as observed from
the large aspect ratio near sharp gradients. Further, we note that
since in 1D, the monitor function is proportional to the grid size,
limiting $\omega_{min}/\omega_{max}$ leads to limiting $\Delta r/\Delta r_{max}$
as can be seen in Fig. \ref{fig:finite_eta_mmpde}-right. Next, we
smooth the grid with the Winslow smoothing factor of $\lambda_{\omega}=10^{-3}$.
In Fig. \ref{fig:finite_eta_smooth_mmpde}, we show the resulting
grid with $(\delta_{min}=0,\eta=10^{-1},\lambda_{\omega}=10^{-3})$.
}\textcolor{black}{}
\textcolor{black}{Winslow smoothing leads to a much smoother grid,
as can be seen from both the cell size profile and the smaller variation
in the mesh aspect ratio. Finally, the effect of floor for the monitor
function, $\delta_{min}$, is demonstrated by setting $\delta_{min}=10^{-1}$.
In Fig. \ref{fig:finite_eta_smooth_delta_mmpde}, we show the resulting
grid with $(\delta_{min}=10^{-1},\eta=10^{-1},\lambda_{\omega}=10^{-3})$.
}\textcolor{black}{}
\textcolor{black}{As seen, $\delta_{min}=0.1$ effectively detects
if the grid is trying to resolve variations in the solution below
10\% and redistributes the grid to the rest of the domain. Such a
technique can be useful, particularly when higher order discretization
is used and smooth variations do not require excessive resolution.}

\begin{figure}[t]
\begin{centering}
\textcolor{black}{\includegraphics[scale=0.6]{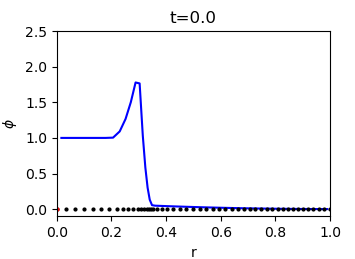}\includegraphics[scale=0.6]{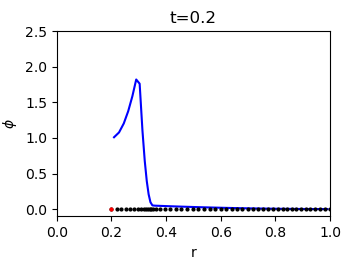}\includegraphics[scale=0.6]{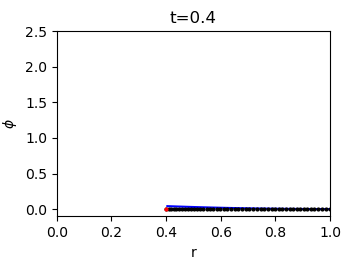}}
\par\end{centering}
\textcolor{black}{\caption{MMPDE evolution of the grid at $t=0,0.2,0.4$ (left to right) for
$\tau_{g}=0.1$ and $(\delta_{min}=10^{-1},\eta=10^{-1},\lambda_{\omega}=10^{-3})$.
The red circles denote the left boundary location. \label{fig:ns_mmpde_taug=00003D0.1}}
}
\end{figure}

\begin{figure}[t]
\begin{centering}
\textcolor{black}{\includegraphics[scale=0.55]{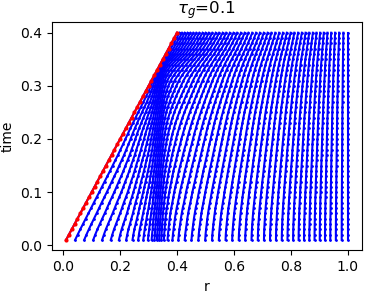}\includegraphics[scale=0.55]{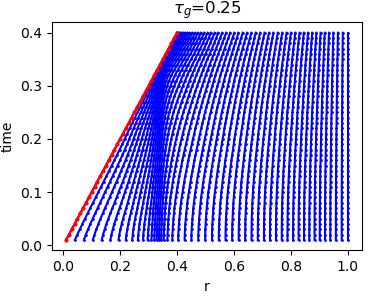}\includegraphics[scale=0.55]{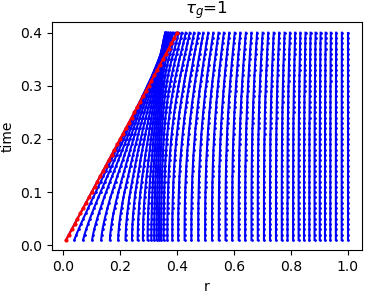}}
\par\end{centering}
\textcolor{black}{\caption{\textcolor{black}{The grid trajectories from MMPDE with $\text{\ensuremath{\tau_{g}=}0.1}$
(left), $0.25$ (center), and $1$ (right). The red circles denote
the left boundary location. \label{fig:ns_mmpde_various_taug}}}
}
\end{figure}

\begin{figure}[t]
\begin{centering}
\textcolor{black}{\includegraphics[scale=0.55]{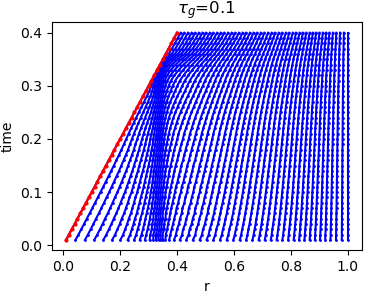}\includegraphics[scale=0.55]{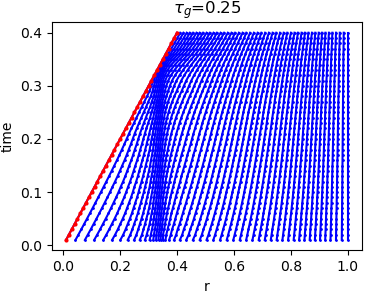}\includegraphics[scale=0.55]{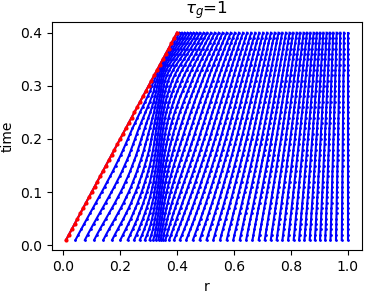}}
\par\end{centering}
\textcolor{black}{\caption{\textcolor{black}{The grid trajectories from normalized MMPDE with
$\tau_{g}=0.1$ (left), $0.25$ (center), and $1$ (right). The red
circles denote the left boundary location. \label{fig:ns_mmpde_normalized_various_taug}}}
}
\end{figure}

\textcolor{black}{Next, we provide an intuitive picture of boundary
grid tangling when one does not follow the procedure of normalizing
the MMPDE equation, as outlined in Sec. \ref{subsec:nonlinear_mmpde}.
Consider the MMPDE equation with a first-order backward Euler time
discretization with staggered finite differencing : 
\begin{equation}
\frac{r_{i+1/2}^{n+1}-r_{i+1/2}^{n}}{\Delta t}=\frac{1}{\tau_{g}}\frac{\omega_{i+1}^{n}\left(r_{i+3/2}^{n+1}-r_{i+1/2}^{n+1}\right)-\omega_{i}^{n}\left(r_{i+1/2}^{n+1}-r_{i-1/2}^{n+1}\right)}{\Delta\xi^{2}},\label{eq:mmpde_appendix}
\end{equation}
where $\omega_{i}^{n}$ is the smoothed and limited monitor function,
and $r_{i}^{n}=\frac{r_{i+1/2}^{n+1}+r_{i-1/2}^{n+1}}{2}$. We begin
by providing a contrived scenario to realize how the boundary grid
point can run ahead of the inner grid points (i.e., grid tangling).
Consider $\tau_{g}^{-1}=0$ such that $r_{i+1/2}^{n+1}=r_{i+1/2}^{n}$.
Together with the boundary condition, $r_{1/2}^{n+1}=L^{n+1}>L^{n}$,
where $L^{n}$ is the left boundary position at time step $n$, (i.e.,
an imploding system) and $L^{n+1}>r_{3/2}^{n}$ immediately leads
to grid tangling at the boundary. Such situations leads to a negative
Jacobian and the catastrophic failure of the underlying solvers for
the VFP and fluid electron equations. For finite values of $\tau_{g}^{-1}$,
it is possible to prevent the tangling, but small enough values for
$\tau_{g}^{-1}$ can lead to undesired }\textcolor{black}{\emph{bunching
}}\textcolor{black}{of grids near the boundaries; boundary can move
faster than internal grids can readjust itself. In Fig. \ref{fig:ns_mmpde_taug=00003D0.1}
we show the MMPDE evolution of the grid for a static background $\phi$,
$\tau_{g}=0.1$, $(\delta_{min}=10^{-1},\eta=10^{-1},\lambda_{\omega}=10^{-3})$,
and $L\left(t\right)=t$. }\textcolor{black}{}
\textcolor{black}{{} }\textcolor{black}{As can be seen, for our particular
choice of $\tau_{g}=0.1$ the grid adapts to $\phi$ with the imploding
left boundary while maintaining grid regularity. In Fig. \ref{fig:ns_mmpde_various_taug},
we show the grid trajectories for various choices of $\tau_{g}$.
}\textcolor{black}{}
\textcolor{black}{As can be seen, for $\tau_{g}=0.1$, the grid is
regular, while $\tau_{g}=0.25$ starts to exhibit bunching of grid
points near the left boundary, and finally $\tau_{g}=1$ leads to
the left boundary point running ahead of the internal computational
grid points. The normalization procedure of the MMPDE equation permanently
addresses the issue of boundary grid tangling, as described in Sec.
\ref{subsec:nonlinear_mmpde}. In Fig. \ref{fig:ns_mmpde_normalized_various_taug},
we show the grid trajectory obtained from the normalized MMPDE formulation.
}\textcolor{black}{}
\textcolor{black}{{} }\textcolor{black}{It is clearly demonstrated, that
the normalization procedure ensures the regularity of the grid at
all time, independent of the $\tau_{g}$ choice. We close by noting
that, in an asymptotic limit (i.e., $\delta_{min}\rightarrow\infty$,
$\eta\rightarrow1$, or $\lambda_{\omega}\rightarrow\infty$), each
hyperparameter leads to a uniform grid. However, as we have demonstrated,
their roles and motivations are unique for finite values.}

\section{Velocity-Space Grid Adaptivity\label{app:velocity_space_grid_adaptivity}}

\textcolor{black}{For completeness, we briefly review the adaptivity
strategy for the velocity space grid metrics from Ref. \citep{taitano_cpc_2020_1d2v_cartesian_phase_space}.
The velocity-space grid metrics, are updated as:}

\textcolor{black}{
\begin{equation}
v_{\alpha}^{*,(p+1)}=\begin{cases}
v_{\alpha}^{*,(p)}+\Delta t^{(p)}\dot{v}_{\alpha}^{*,(p+1)} & \textnormal{if}\;\Delta t^{(p)}\frac{\left|\dot{v}_{\alpha}^{*,(p+1)}\right|}{v_{\alpha}^{*,(p)}}\le0.025\\
v_{\alpha}^{*,(p)}\left[1+0.025\Delta t\textnormal{sign}\left(\dot{v}_{\alpha}^{*,(p+1)}\right)\right] & \textnormal{otherwise}
\end{cases},\label{eq:vstar_def}
\end{equation}
and 
\begin{equation}
u_{||,\alpha}^{*,(p+1)}=\begin{cases}
u_{||,\alpha}^{*,(p)}+\Delta t^{(p)}\dot{u}_{||,\alpha}^{*,(p+1)} & \textnormal{if}\;\Delta t^{(p)}\left|\frac{\dot{u}_{||,\alpha}^{*,(p+1)}}{v_{\alpha}^{*,(p)}}\right|\le0.025\\
u_{||,\alpha}^{*,(p)}\left[1+0.025\Delta t\textnormal{sign}\left(\dot{u}_{||,\alpha}^{*,(p+1)}\right)\right] & \textnormal{otherwise}
\end{cases},\label{eq:urstar_def}
\end{equation}
where}

\textcolor{black}{{} 
\begin{equation}
\dot{v}_{\alpha}^{*,(p+1)}=\frac{\left[v_{\alpha}^{*,(p+1)}\right]^{\dagger}-v_{\alpha}^{*,(p)}}{\Delta t^{(p)}},\label{eq:vstar_time_derivative}
\end{equation}
\begin{equation}
\dot{u}_{||,\alpha}^{*,(p+1)}=\frac{\left[u_{||,\alpha}^{*,(p+1)}\right]^{\dagger}-u_{||,\alpha}^{*,(p)}}{\Delta t^{(p)}},\label{eq:ustar_Time_derivative}
\end{equation}
}

\textcolor{black}{
\begin{eqnarray}
\left[v_{\alpha}^{*,(p+1)}\right]^{\dagger}=\sqrt{\frac{2T_{\alpha}^{(p+1)}}{m_{\alpha}}},\label{eq:intermediate_vstar}
\end{eqnarray}
\begin{equation}
\left[u_{||,\alpha}^{*,(p+1)}\right]^{\dagger}=u_{||,\alpha}^{(p+1)}+\Delta w_{||,\alpha}^{(p+1)},\label{eq:itermediate_ustar}
\end{equation}
and 
\begin{equation}
\Delta w_{||,\alpha}^{(p+1)}=\frac{\left\langle \left(v_{||}^{(p)}-u_{||,\alpha}^{(p+1)}\right)\left(\vec{v}^{(p)}-\vec{u}_{\alpha}^{(p+1)}\right)^{2},f_{\alpha}^{(p+1)}\right\rangle _{v}}{\left\langle \left(\vec{v}^{(p)}-\vec{u}_{\alpha}^{(p+1)}\right)^{2},f_{\alpha}^{(p+1)}\right\rangle _{v}}.\label{eq:pdf_skewness_measure}
\end{equation}
To ensure that the profiles of $v^{*}$ and $u_{||}^{*}$ are smooth
in space, we employ a Winslow smoother. The grid adaptivity algorithm
is summarized in Alg. \ref{alg:vel_grid_update}.} 
\begin{algorithm}[h]
1. Compute the velocity-space grid metrics, $v^{*}$ and $u_{||}^{*}$
from Eqs. (\ref{eq:vstar_def})-(\ref{eq:pdf_skewness_measure}).

2. Winslow smooth the $v^{*}$ and $u_{||}^{*}$ by solving for $\left[1-\lambda_{\omega}\partial_{\xi\xi}^{2}\right]v^{*}=v^{**}$
and $\left[1-\lambda_{\omega}\partial_{\xi\xi}^{2}\right]u_{||}^{*}=u_{||}^{**}$
where $\lambda_{\omega}=10^{-2}$ in this study and the superscript,
$**$, denotes the non-smoothed quantities.

\caption{Update procedure for the velocity-space grid metrics. \label{alg:vel_grid_update}}
\end{algorithm}

\section{Poisson Grid Generation and Initial Optimization\label{app:init_grid_optimization}}

It is critical to ensure an initial high quality grid. Severe numerical
pollution of the solution can occur if one fails at this task and
solely relies on MMPDE to transiently resolve the initial gradients
as is demonstrated in Sec. (\ref{subsec:gudurley}). To address this
issue, we employ an initial Poisson mesh generation and grid optimization
to resolve initial gradients.

Consider a reference solution, $F^{*}\left(r^{*}\right)\in\mathbb{R}^{N_{\xi}^{*}}$
and $N_{\xi}^{*}\gg N_{\xi}$, defined on a highly resolved reference
grid, $r^{*}\in\mathbb{R}^{N_{\xi}^{*}}$. We begin by interpolating
this solution onto the coarse, uniform computational grid, $r^{0}\left(\xi\right)\in\mathbb{R}^{N_{\xi}}$,
such that ${\cal F}^{0}=\mathbb{M}^{0}F^{*}$, where ${\cal F}^{0}\in\mathbb{R}^{N_{\xi}}$
is the reference solution interpolated onto the initial coarse-computational
grid and $\mathbb{M}^{0}=\mathbb{M}^{0}\left(r^{0},F^{*}\right)\in\mathbb{R}^{N_{\xi}\times N_{\xi}^{*}}$
is the interpolation map. Note that, to ensure the monotonicity of
the interpolated quantity, we restrict ourselves to linear interpolation
in this study. The grid optimization strategy is shown in Algorithm
\ref{alg:grid_optimization_algorithm}. 
\begin{algorithm}[h]
1. Initialize iteration index, $k=0$.

2. Increment $k=k+1$.

3. Compute monitor function, $\omega\left({\cal F}^{k-1}\right)$,
given ${\cal F}^{k-1}$, and perform limiting and smoothing as necessary.

4. Solve the Poisson equation, $\frac{\partial}{\partial\xi}\left[\omega\left({\cal F}^{k-1}\right)\frac{\partial r^{\dagger}}{\partial\xi}\right]=0,$
for $r^{\dagger}$.

5. Update the grid location: $r^{k}=r^{k-1}+w_{relax}\left(r^{\dagger}-r^{k-1}\right).$
Here $w_{relax}=0.75$ is the under-relaxation factor used in this
study.

6. Interpolate reference solution onto the new grid: ${\cal F}^{k}=\mathbb{M}^{k}F^{*}.$

7. If $\sqrt{\frac{1}{N_{\xi}+1}\sum_{i=0}^{N_{\xi}}\left(\frac{r_{i}^{k}-r_{i}^{k-1}}{r_{i}^{k-1}+10^{-12}}\right)^{2}}\le10^{-10}$
exit iteration, else return to 2.

\caption{Grid optimization procedure.\label{alg:grid_optimization_algorithm}}
\end{algorithm}

\begin{figure}[h]
\centering{}\centering{}\includegraphics[scale=0.5]{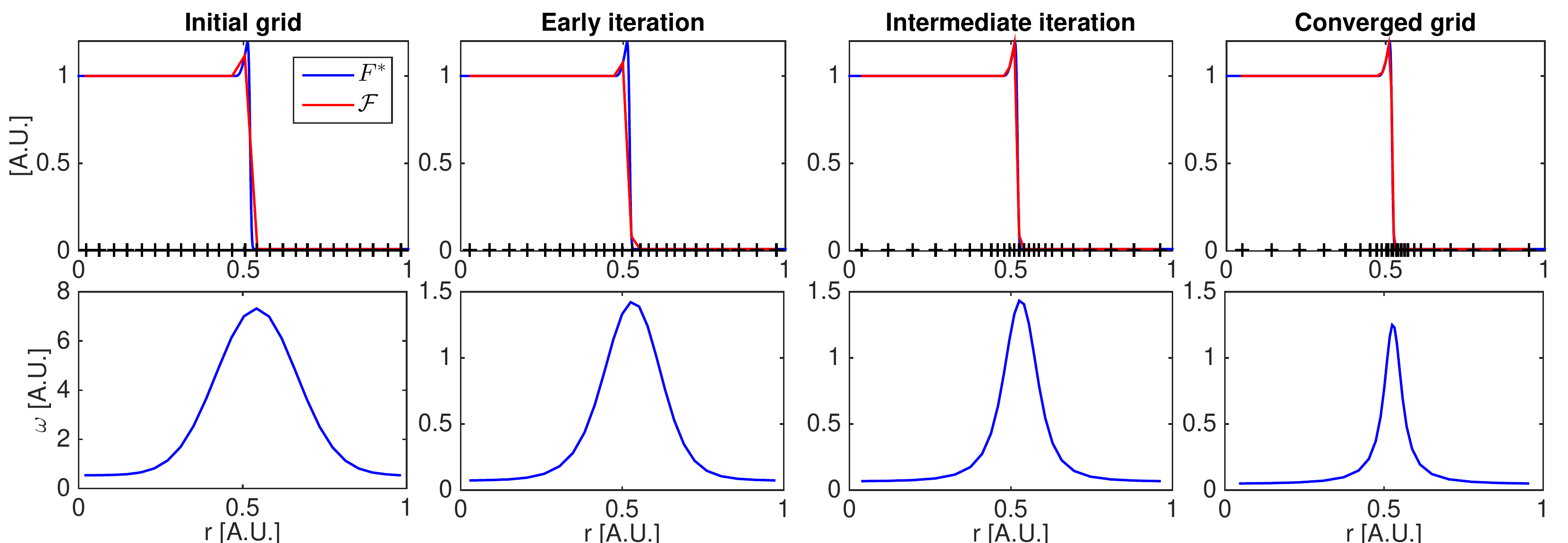}\caption{Illustration of the iterative procedure for the initial grid generation.
The top row shows the reference solution, $F^{*}$ (blue) and the
interpolated solution, ${\cal F}$ (red), while the bottom row shows
the corresponding monitor function. The black markers denote the grid
points. As can be seen, the grid adapts to resolve the initial gradients.
\label{fig:illustration_of_grid_optimization}}
\end{figure}

In Fig. \ref{fig:illustration_of_grid_optimization}, we illustrate
the iterative procedure for the grid optimization.

\section{Guderley and Van-Dyke Problem\label{app:gudurley_problem}}

The Guderley solution is the self-similar solution to the Euler equations
for a strong, spherically converging/diverging shock. In 1D spherical
geometry, we may write the Euler equations as \citep{murakami_pop_2015_stability_of_spherical_shock}:
\begin{equation}
\partial_{t}{\rho}+\frac{1}{r^{2}}{\partial_{r}}\left(r^{2}\rho u_{r}\right)=0,\label{eq:euler_cont}
\end{equation}
\begin{equation}
\left(\partial_{t}+u_{r}\partial_{r}\right)u_{r}=-\frac{1}{\rho}\partial_{r}P,\label{eq:euler_mom}
\end{equation}
\begin{equation}
\left(\partial_{t}+u_{r}\partial_{r}\right)\text{ln}\frac{P}{\rho^{\gamma}}=0,\label{eq:euler_ener}
\end{equation}
where $P$ is the gas pressure, $u_{r}$ is the radial drift velocity,
and $\rho$ is the mass density. Guderley was the first to show that
Eqs.\ (\ref{eq:euler_cont}-\ref{eq:euler_ener}) admits a self-similar
solution \citep{guderley_luftfahrtf_1942_guderley_problem,vallet_pop_2013_finite_mach_number_guderley,murakami_pop_2015_stability_of_spherical_shock,lazarus_siam_jna_guderley,ramsey_2017_JVVUQ_guderley}
with the shock position: 
\begin{equation}
R_{s}(t)\propto|t|^{\alpha},\label{eq:radius_solution-1}
\end{equation}
where $\alpha$ is a constant which differs between the converging
and diverging phases. This constant must be obtained by numerical
integration of the Euler equations, with the coordinate transformation:
$r\rightarrow\xi\equiv r/R_{s}(t)$. In this self-similar variable,
$\xi$, the fluid quantities follow as \citep{murakami_pop_2015_stability_of_spherical_shock}:
\begin{equation}
u_{r}=\frac{r}{t}\ V(\xi),
\end{equation}
\begin{equation}
\rho=\rho_{0}\ G(\xi),
\end{equation}
\begin{equation}
c_{s}^{2}=\left(\frac{r}{t}\right)^{2}Z(\xi),
\end{equation}
where $c_{s}^{2}\equiv\gamma P/{\rho}$ is the sound speed, $\gamma$
is the adiabatic index (which is $5/3$ for an ideal plasma), $\rho_{0}$
is the initial (upstream) density, and the functions V, G, and Z are
determined via numerical integration.

Whereas the Guderley solution is self-similar (i.e., it assumes a
converging shock that has ``forgotten'' its initial conditions),
Van Dyke \& Guttmann \citep{van_dyke_JFM_1982_van_dyke_problem} consider
the creation of a shock from the surface of a spherically converging
piston (moving at a constant velocity). The key assumption is that
the shock, after emerging from the piston face, is approximately planar.
The Euler equations are then solved with a Taylor series expansion
of the shock position: 
\begin{equation}
R_{s}(t)=\sum_{n=1}^{\infty}R_{n}t^{n},\label{eq:radius_solution-2}
\end{equation}
where each subsequent term represents a spherical correction to the
planar trajectory (which is given by the Rankine-Hugoniot conditions).
Likewise, Taylor series expansions were employed for the pressure,
density, and drift velocities: 
\begin{equation}
u_{r}=\sum_{n=1}^{\infty}U_{n}(\xi)t^{n-1},
\end{equation}
\begin{equation}
\rho_{r}=\sum_{n=1}^{\infty}R_{n}(\xi)t^{n-1},
\end{equation}
\begin{equation}
P=\sum_{n=1}^{\infty}P_{n}(\xi)t^{n-1},
\end{equation}
where the similarity variable is $\xi\equiv\frac{2}{\gamma-1}\left(\frac{r}{u_{p}t}-1\right)$,
and $u_{p}$ is the piston velocity. Van Dyke \& Guttmann \citep{van_dyke_JFM_1982_van_dyke_problem}
calculated the shock trajectory to 40-terms in Eq.\ (\ref{eq:radius_solution-1}).
This is the ``Van Dyke'' solution that appears in Sec. \ref{subsec:gudurley_moving_wall}. 
\end{document}